\documentclass[fleqn,usenatbib]{mnras}

\usepackage{newtxtext,newtxmath}

\usepackage[T1]{fontenc}

\DeclareRobustCommand{\VAN}[3]{#2}
\let\VANthebibliography\thebibliography
\def\thebibliography{\DeclareRobustCommand{\VAN}[3]{##3}\VANthebibliography}

\usepackage{graphicx}	
\usepackage{amsmath}	
\usepackage{booktabs}   
\usepackage{siunitx}[print-unity-mantissa=false]    
\usepackage[labelfont=bf]{caption}    
\usepackage{url}      
\usepackage{bm}
\usepackage{fancyhdr}
\usepackage{textcomp}
\usepackage{multicol} 
\usepackage{mathtools}
\usepackage{nicefrac}
\usepackage{xcolor}
\usepackage{float}
\usepackage{sidecap}
\usepackage{enumitem}
\usepackage{geometry}
\usepackage{svg}
\usepackage[version=4]{mhchem}
\usepackage{multirow}
\usepackage{wasysym}      
\usepackage[title]{appendix}
\usepackage{titlesec}
\usepackage{hyperref}
\usepackage[capitalise]{cleveref}

\DeclareSIUnit\bar{bar}
\DeclareSIUnit\AU{AU}
\DeclareSIUnit\dex{dex}
\DeclareSIUnit\erg{erg}
\def\code#1{\texttt{#1}}

\newcommand{\comment}[1]{}

\title[Flare-driven self-consistent gas giant atmospheres]{Temperature-chemistry coupling in the evolution of gas giant atmospheres driven by stellar flares}

\author[H. Nicholls et al.]{
    Harrison Nicholls,$^{1,2}\thanks{harrison.nicholls@physics.ox.ac.uk}$
    Eric H\'ebrard,$^{2}$
    Olivia Venot,$^{3}$
    Benjamin Drummond,$^{4}$
    Elise Evans$^{5}$
    \\
    $^{1}$Atmospheric, Oceanic and Planetary Physics, Department of Physics, University of Oxford, Oxford, OX1 3PU, UK \\
    $^{2}$Physics and Astronomy, Faculty of Environment, Science and Economy, University of Exeter, Exeter EX4 4QL, UK\\
    $^{3}$Universit\'e de Paris Cit\'e and Univ Paris Est Creteil, CNRS, LISA, F-75013 Paris, France\\
    $^{4}$Met Office, Fitzroy Road, Exeter EX1 3PB, UK\\
    $^{5}$Institute for Astronomy, University of Edinburgh, Royal Observatory, Blackford Hill, Edinburgh, EH9 3HJ, UK
}

\date{Accepted XXX. Received YYY; in original form ZZZ}

\pubyear{2023}

\begin{document}
    \label{firstpage}
    \pagerange{\pageref{firstpage}--\pageref{lastpage}}
    \maketitle
    
    \begin{abstract}
        The effect of enhanced UV irradiation associated with stellar flares on the atmospheric composition and temperature of gas giant exoplanets was investigated. This was done using a 1D radiative-convective-chemical model with self-consistent feedback between the temperature and the non-equilibrium chemistry. 
        
        It was found that flare-driven changes to chemical composition and temperature give rise to prolonged trends in evolution across a broad range of pressure levels and species. Allowing feedback between chemistry and temperature plays an important role in establishing the quiescent structure of these atmospheres, and determines their evolution due to flares. It was found that cooler planets are more susceptible to flares than warmer ones, seeing larger changes in composition and temperature, and that temperature-chemistry feedback modifies their evolution. 
        
        Long-term exposure to flares changes the transmission spectra of gas giant atmospheres; these changes differed when the temperature structure was allowed to evolve self-consistently with the chemistry. Changes in spectral features due to the effects of flares on these atmospheres can be associated with changes in composition. The effects of flares on the atmospheres of sufficiently cool planets will impact observations made with \textit{JWST}. It is necessary to use self-consistent models of temperature and chemistry in order to accurately capture the effects of flares on features in the transmission spectra of cooler gas giants, but this depends heavily on the radiation environment of the planet.
    \end{abstract}
    
    
    

    \section{Introduction}
    \label{sec:introduction} 
    The relationship between stellar activity and exoplanet atmospheres is not well understood. It is thought that activity from active stars can affect orbiting exoplanets through various means including UV-driven photochemistry and escape, enhanced ionisation and upper atmosphere particle flux, and internal Joule heating \citep{Chen2020, Grayver2022}. 
    \par 
    Various flare and atmospheric models have been coupled, but with inconsistent findings \citep{Segura2010, Venot2016, Chen2020, Louca2022, Konings2022, Ridgway2022}. While much of the existing literature has focused on implications on the habitability of terrestrial exoplanets with nitrogen dominated atmospheres, only a few have considered planets which could feasibly be observed with current telescopes. There has been demand for an assessment of the role of self-consistent modelling of atmospheres exposed to repeated flares, where temperature and composition are allowed to evolve together, although such an assessment is yet to be performed \citep{Venot2016, Louca2022, Konings2022}. The atmospheres of gas giants are easier to observe than those of terrestrial and rocky exoplanets, so determining the effect of stellar flares could be important for accurate retrievals in the near-future, as we do not yet know the full extent of their impact on exoplanet atmospheres \citep{Fortney2021}. Initial Early Release Science from \textit{JWST} has found direct evidence of sulphur photochemistry in Hot Jupiter (HJ) atmospheres, demonstrating the relevance of these photochemical processes \citep{tsai2022}.
    
    \par 
    AD Leonis (AD Leo) is a high-activity M-dwarf star with luminosity $\sim 0.024 L_\odot$ and mass $\sim 0.42 M_\odot$ \citep{Pettersen1981, Reiners2009}. \citet{Segura2010} used observations of the `Great Flare of AD Leo' in 1985\footnote{This particular flare is henceforth referred to as GF85} and an atmospheric model to simulate the effect of enhanced UV irradiation and proton flux due to flares on an atmosphere. This particular flare was exceptional because of its large integrated energy \mbox{$>$ \SI{e34}{\erg}}, most of which is attributed to enhanced continuum emission \citep{Hawley1991}. The aim of \cite{Segura2010} was to better understand the depletion of ozone species on a hypothetical Earth-like exoplanet orbiting this star, and assess the potential risk to life. The authors found that the flare had little effect on the ozone distribution in the atmosphere, and thus there was little risk to life on the planet, as ozone is necessary for shielding the surface from UV high fluxes \citep{Kim2013}. They did not consider the effect of repeated flares, but suggested that there could be compounding effects.
    \par 
    \citet{Chen2020} used a coupled chemistry-climate model originally developed for Earth science to simulate enhanced UV and proton flux due to flares from G-, K-, and M-type stars and their impact on rocky planets. This 3D model allowed for differences between the substellar and anti-substellar points, as well as the addition of magnetic fields and land/ocean fraction. They found that the behaviour of the atmosphere varies depending on the type of star involved. For planets orbiting K- and M-type stars, recurrent flares drive these atmospheres into a different steady-state to the pre-flare state (PFS; the steady-state of the atmosphere model before any flares were applied), with the abundance of \ce{O3} being significantly depleted. Spectral changes induced by flares included features associated with bio-signature gases such as \ce{CH4} and \ce{O3}, which could reduce their detectability in future observations. Similar work was more recently performed by \citet{Ridgway2022}, which concluded that flares do not induce changes in spectral features observable with current telescopes, and that there could be significant production of \ce{O3}. These two works make opposing conclusions for planets orbiting M-type stars.
    \par 
    \citet{Venot2016} --- henceforth abbreviated as V16 --- simulated a HJ orbiting AD Leo to characterise changes in composition due to enhanced UV irradiation from flares when applied both once and periodically. They used a 1D model implementing the chemical network from \cite{Venot2012} for hydrogen-dominated atmospheres. They considered two cases for the orbital separation, $a = 0.0690 \text{ AU}$ and $a = 0.0069 \text{ AU}$, corresponding to effective temperatures ($T_{\text{eff}}$) of \SI{412}{\kelvin} and \SI{1303}{K}. These planets are relatively cold compared to the wider population of confirmed exoplanets, despite having small $a$, because M-dwarves are relatively dim stars \citep{RodriguezLopez2019, Sing2015}. The use of relatively cool planets is important to note because their chemical abundances are more sensitive to time-dependent processes, while hotter planets remain closer to chemical and thermodynamic equilibrium \citep{Venot2012}. V16 found that a single flare --- GF85, which was also applied in \cite{Segura2010} --- lasting for approximately 3000 seconds could modify the compositions of the atmospheres of the orbiting planets. Additionally, they applied this same flare periodically which caused these effects to compound, significantly affecting species such as \ce{H}, \ce{NH3}, \ce{CO2}, and \ce{CH3}. While this caused abundances to oscillate, for many species these oscillations tended towards a limiting mean abundance which was different to the PFS. V16 concluded that planets around active stars are continuously altered by flares, and resultant changes in transit depth (of up to 1200 ppm) would likely be observable with the \textit{JWST}.
    \par 
    The model of V16 does not account for energetic particles, atmospheric escape, or changes to the temperature structure of the atmosphere, the latter of which they regard as insignificant based on the findings of \cite{Segura2010}. Their simulation results show differences between the PFS and the post-flare states of the atmospheres after the application of a single flare, which are deemed permanent as they persist after \SI{e12}{\second}. Assuming that flares occur periodically at constant amplitude is unrealistic as flares occur stochastically following an energy distribution, the position and shape of which is related to the stellar subtype \citep{Pettersen1981, Hilton2010, Loyd2018}. \citet{Guenther2020} do not observe flares on active M-dwarf stars occurring less frequently than \mbox{0.1 flares/day}, meaning that planets orbiting active M-dwarf stars never experience a significant period of time during which there is little perturbation to the input UV flux. Therefore, it is not informative to discuss the state of an atmosphere a long time (months) after the application of a single flare, as this is a situation which is extremely unlikely to occur naturally.
    \par 
    \citet{Louca2022} --- henceforth abbreviated as L22 --- used a 1D chemical kinetics model (\code{VULCAN}) to simulate three cases of hydrogen-dominated atmospheres. They applied flares stochastically, sampling an energy distribution, over a period of \SI{e6}{\second} (approximately 12 days). Their model included diffusion-limited atmospheric escape, which may be important for species of low molecular weight such as atomic hydrogen. The rate of escape is coupled to UV irradiation, and they attribute many of the post-flare behaviours to enhanced escape. Throughout their simulations the atmospheric temperature profile was held constant in time, as their model is not capable of solving for it self-consistently with the chemistry. The authors highlight that an assessment of the interplay between the chemistry and the temperature has not been performed in this context. Their discussion is limited to cold planets ($T_{\text{eff}} = \{ 395, 428, 479 \} \text{ K}$) in order to avoid the regime where hydrodynamic escape is significant. Two of their conclusions were that: 
    \begin{itemize}
        \item the abundance of some species changes abruptly in response to individual flares while others change cumulatively as flares are applied,
        \item more simulation time ($>10^6 \text{ s}$) would likely be required to determine the observability of compositional change due to flares, as the effects on transmission spectra accumulate over time.
    \end{itemize}
    
    \par 
    More recently, \citet{Konings2022} --- henceforth abbreviated as K22 --- used a pseudo-2D model to simulate the effect of flares on the atmospheric composition. They found that the atmosphere's recovery after the application of a single flare depends highly on the effective temperature of the planet involved, as well as the chemical species considered. This is attributed to differences in chemical time-scales. The conclusions of K22 are significantly different to those of V16. For example, V16 found that the upper atmospheres of their cold planet (\mbox{$T_{\text{eff}} = $\SI{412}{\kelvin}}) case was depleted of \ce{NH3}, while for their warm planet (\mbox{$T_{\text{eff}} = $\SI{1303}{\kelvin}}) it was enriched. In contrast, K22 did not find production of \ce{NH3} in any cases. Additionally, L22 finds continuous trends in the mole fraction of various species (e.g. \ce{CO2}, \ce{H2}, \ce{CH4}); K22 does not see this, which they attribute to the horizontal transport afforded by their model. The simulations performed in K22 show that when flares are repeatedly applied, the composition of the atmospheres rapidly approach a new pseudo steady-state, about which the system fluctuates. The authors of K22 suggest research on the impact of flares on temperature profiles, especially in hot planets, as a further line of inquiry.
    \par 
    Previous research in this area has demonstrated that flares and associated phenomena can lead to permanent changes to the composition of orbiting planets in cases where temperature is held constant with time. The natural next-step would therefore be to evolve the temperature structure alongside the chemistry. This may prove to be a necessary condition, as it was demonstrated by \citet{Drummond2016} that atmospheric models can only conserve energy when the temperature structure is solved for self-consistently  alongside the chemical composition. \cite{Drummond2016} showed this by solving for the steady-state of a test planet with and without coupled temperature and chemistry. The differing converged temperature profiles between the two cases lead to different outward emission fluxes under identical stellar and internal heating; these differing emission fluxes show that the case where temperature was uncoupled does not conserve energy.  \citet{Drummond2016} concluded that neglecting feedback between temperature and composition `can lead to overestimates of the impact of non-equilibrium chemistry' on observations, as it does not allow the atmosphere to re-adjust to radiative-convective equilibrium when composition is perturbed. Instead, the effects of non-equilibrium processes (quenching in particular) are absorbed into the chemistry, rather than being realistically handled by changes in temperature. 
    \par 
    When time-evolution is relevant, such as when a time-series of flares are being applied, energy must be conserved. As there are many chemical species and pressure regimes to consider, it is not possible to predict precisely the effect of energy conservation on these simulations. One effect could be to delay the atmosphere's return to its PFS between flares: chemical processes have shorter timescales than thermodynamic processes, so leaving chemistry uncoupled to temperature would allow the atmosphere to recover rapidly, and possibly inaccurately; allowing feedback between chemistry and temperature could have the effect of damping the chemical evolution while the temperature is evolving.
    \par 
    The aim of this work is to attempt to build upon V16 and L22 by using a self-consistent model of coupled atmospheric chemistry and temperature. V16, L22, and K22 have all called for a self-consistent approach to modelling atmospheric response to flaring. This work will, for the first time, evaluate the role of this feedback, its implications for observables, and whether or not it should be considered in future work. 
    \par 
    We predicted that when simulating repeated flares with CNEQ chemistry, the atmosphere was likely to enter a new steady-state different to the quiescent one, in contrast to solely NEQ cases. We expected that the most significant changes would occur in the upper atmosphere, and that the deep atmosphere would remain mostly unaffected.
    \par 
    The remainder of the paper is structured as follows:
    \begin{itemize}
        \item Methods. We outline of our photochemical model, provide a description of three flare models used in this work, and explain our approach to synthetic observations.
        \item Results and Discussion. We first generate quiescent pre-flare states, from which the effects of flares are simulated using our photochemical model. The effects of flares and the role of self-consistent modelling on various species are discussed and are physically justified. Synthetic observations of these planets before and after they are exposed to flares are used to assess the observability of the effects of flares in different cases.
        \item Conclusion. We summarise our findings and enumerate key points to take away from this work. We also discuss weaknesses of this work and potential future steps.
    \end{itemize}
    
    
    \section{Methods}
    \label{sec:methods} 
    
    \subsection{ATMO}
    \label{ssec:atmo}
    \code{ATMO} is a 1D atmospheric model which simulates a column of atmosphere between two pressure levels \citep{Tremblin2015, Amundsen2014, Drummond2016}. It is similar to the model used in V16 and uses the same chemical network from \cite{Venot2012}. The column is discretised into cells, with energy fluxes balanced at their faces to ensure \textit{radiative-convective equilibrium} (RCE).
    \par
    The radiative transfer (RT) equation is solved in its integral form using the correlated-$k$ approximation, following the method in \cite{Amundsen2014} and references therein. $k$-coefficients are calculated by the radiative-convective scheme at run-time, allowing changes in composition to be reflected in the radiative-transfer calculations \citep{Goyal2017}. The convective flux is non-zero in regions which satisfy the Schwarzschild stability criterion. Solving this radiative-convective scheme to give the pressure-temperature (PT) structure of the atmosphere is done using the Newton-Raphson method, which converges when the maximum relative error in the energy flux balance across all cells is less than \SI{E-4}{}.
    \par
    Opacity data used in RT calculations are adapted from \cite{Amundsen2014}, which include absorption by \ce{H2}, \ce{He}, \ce{H2O}, \ce{CO2}, \ce{CO}, \ce{CH4}, \ce{NH3}, \ce{Na}, \ce{K}, \ce{Li}, \ce{Rb}, \ce{Cs}, \ce{TiO}, \ce{VO}, \ce{FeH}, \ce{PH3}, \ce{H2S}, \ce{HCN}, \ce{C2H2}, \ce{SO2}, and \ce{Fe}. Note the inclusion of TiO and VO, which are known to cause thermal inversions due to absorption of optical radiation \citep{Fortney2021}. Collision-induced absorption due to \ce{H2}-\ce{H2} and \ce{H2}–\ce{He} interactions is also included. Line profiles are pressure-broadened by \ce{H2} and {He}, as well as by the Doppler effect \citep{Stamnes2017,Sharp2007}. The aforementioned broadening processes dominate over turbulent- and self-broadening, so these latter two processes are neglected \citep{Amundsen2014}. Table 4 of \citet{Amundsen2014} lists the correlated-$k$ bands used in \code{ATMO} which cover the range $200 \text{nm} \le \lambda \le 230 \text{ }\mu\text{m}$. Absorption coefficients are tabulated on a pressure-temperature grid, following the method in \citet{Thomas2002}, with 30 pressure points in the interval [\SI{e-1}{\pascal}, \SI{e-8}{\pascal}] and 20 temperature points in the interval [\SI{70}{\kelvin}, \SI{3000}{\kelvin}] uniformly distributed on a logarithmic scale. 
    \par
    The abundances of species at chemical equilibrium (EQ) are determined by following the method in \citet{Gordon1994}, i.e. by minimising the Gibbs free energy \mbox{$G = \sum_j \mu_j n_j$}, where $\mu_j$ is the chemical potential and $n_j$ is the number density of a species $j$ at a given level. This sets the initial abundances of the atmosphere, before non-equilibrium processes are introduced. 
    \par
    \code{ATMO} can also model the composition within each cell over time by solving the rate equation,
    \begin{equation}
        \label{eq:chemCont}
        \frac{\partial n_j}{\partial t} = P_j - n_j L_j - \frac{\partial \phi_j}{\partial z}
    \end{equation}
    for each species $j$ listed in Appendix \ref{app:chemList}. $P_j$ and $L_j$ are the net production and loss within a given cell, and $\phi_j$ is the vertical transport of $j$ out of that cell. $P_j$ and $L_j$ are derived using the chemical reaction network of \citet{Venot2012}. The equation has units of $[\text{molecules}]\text{cm}^{-3}\text{s}^{-1}$. $\phi_j$ includes eddy diffusion, molecular diffusion, self-buoyancy, and thermal diffusion. Data for these diffusion processes are derived from \citet{Poling2001}. Eddy diffusion is parameterised by the coefficient \mbox{$K_{zz} = $\SI{E8}{\centi\meter\tothe{2}\second\tothe{-1}}}. As the model allows for time-dependent processes, it describes the \textit{non-equilibrium chemistry} (NEQ) of the system, as opposed to the \textit{equilibrium chemistry} (EQ) which assumes that the atmosphere is not evolving with time. The system of equations across cells and species defined by Equation \ref{eq:chemCont} is solved using the time-accurate backwards-differentiation formula method from the publicly available FORTRAN library DLSODE \citep{Hindmarsh1983}. Convergence of the NEQ chemical scheme occurs when $\partial n_j/n_j < 10^{-2}$ and $(\partial n_j/n_j)/dt < 10^{-4}$ for three consecutive iterations, for all species with mole-fractions greater than \SI{e-30}{}. $dt$ is the chemical time-step. 
    \par 
    \code{ATMO} uses a different chemical network (\citet{Venot2012}) to the one used in K22 and L22, so a quantitative comparison with these works is not possible, although qualitative conclusions may be drawn. The photodissociation pathways used in this work are available in Appendix D of \citet{Venot2012}.
    \par 
    \code{ATMO} can be set-up to periodically re-converge the temperature profile during NEQ simulations by repeatedly solving for RCE. Solving for RCE happens instantaneously in simulation-time, although it does not occur continuously. See Appendix \ref{app:energyError} for a discussion on the frequency at which the system re-adjusts to RCE. Re-solving for RCE makes \code{ATMO} a \textit{consistent solver}, as temperature and chemistry are capable of bi-directional feedback. Together, this allows us to model the \textit{consistent non-equilibrium} (CNEQ) chemistry of the system. This method is explained in greater detail in \citet{Drummond2016}.
    \par 
    Note that while chemical reaction rates are temperature-dependent, enthalpy changes corresponding to each reaction in the network do not contribute to the temperature of the atmosphere in the temperature range explored by these gas-only simulations \citep{Roth2021}. Temperature changes are solely a result of radiative absorption, scattering, and convective mixing. Convection in HJs usually occurs at pressures greater than \SI{1}{\bar} \citep{Thorngren2019}. 
    \par 
    Changes in UV flux associated with flares could drive formation and removal of hazes, as many haze-precursors (e.g. \ce{CH4}, \ce{HCN}, \ce{C2H2}) are sensitive to radiation in the UV regime \citep{Steinrueck2021,Helling2020,Zhang2015,Sing2015}. Less radiation would be absorbed in the lower atmosphere, instead being absorbed by the hazes at lower pressures. This would most likely change the temperature structure both above and below the haze layer (K22). Changes to the temperature structure could trigger condensation, leading to cloud formation \citep{Wakeford2016}. However, the version of \code{ATMO} used in this work does not include hazes or clouds. \cite{Steinrueck2021} found that hazes cannot currently be correlated to features in transmission spectra of HJs.

    \subsection{Flare models}
    \label{ssec:flaringModels}
    V16 used observational data of GF85 \citep{Pettersen1981,Hawley1991,Segura2010}. They simulated the impacts of both an isolated application of GF85 on atmospheric chemistry, and periodic applications of this same high-energy flare over an interval of \SI{E6}{\second}. They selected a resting time between flare applications of \SI{2e4}{\second}, where UV irradiation was restored to its quiescent value. The same procedure is initially adopted in our work for Sections \Cref{ssec:flaresSglNeq,ssec:flaresSglCneq,ssec:flaresPerNeq,ssec:flaresPerCneq}. In the case where we apply the same flare periodically, we adopt the same resting time of \SI{2e4}{\second} but extend the integration time up to \SI{9E6}{\second}. We do not discuss the behaviour of the atmosphere on long time-scales after flares have stopped because a situation in which there is no flare activity for a long period of time is extremely unlikely (see \Cref{sec:introduction}).
    \par
\begin{figure}
        \centering
        \includegraphics[width=\columnwidth,keepaspectratio]{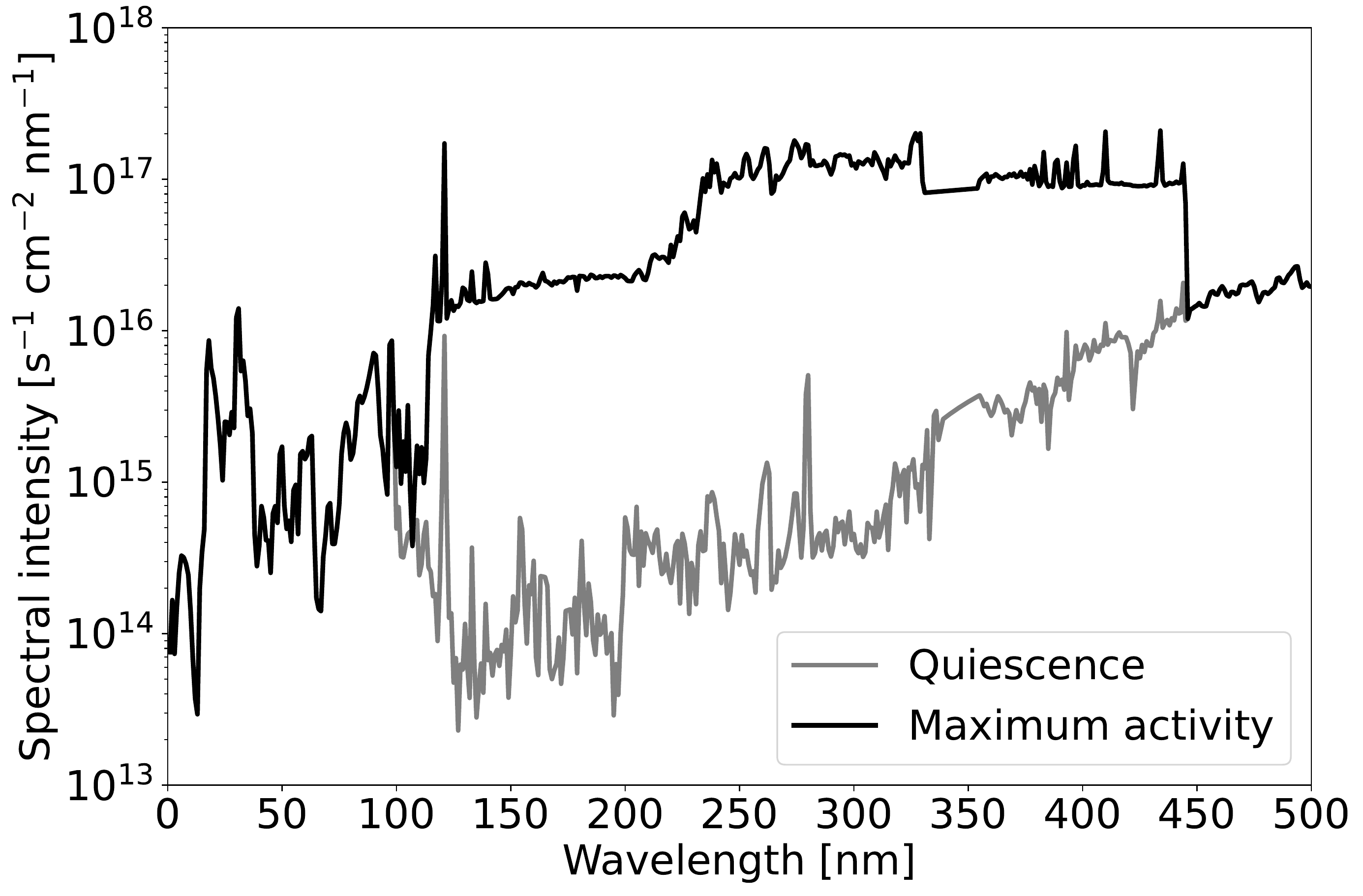}
        \captionof{figure}{Spectral intensity of UV emission from AD Leo at quiescence and maximum activity, which occurs \SI{912}{\second} after a given flare begins.}
        \label{fig:adleoUV}
    \end{figure}

 
    \Cref{fig:adleoUV} compares the intensity of UV emission before and during a flare. GF85 was observed between \SI{100}{\nano\meter} and \SI{444}{\nano\meter}: the \textit{variable interval}. Below \SI{100}{\nano\meter} the intensity is derived from solar observations and scaled by a factor of 100 to fit with the flare spectra, and above \SI{444}{\nano\meter} it is equal to the pre-flare (quiescent) spectrum of AD Leo \citep{Segura2010}.
    \par 
    There is a dearth of sufficient data and emission models in the EUV and X-ray regime for these stars, which makes obtaining a more realistic spectrum difficult. Recent work by \cite{Loyd2018} and \citet{Zhuleku2020} has sought to resolve this by interpolating into the X-ray regime from the FUV regime. As the stellar radiative emission is only variable within our variable interval, some species are not excited and their abundances will not be directly affected by the flares. Most notably, including X-ray emission and subsequent absorption would drive ion chemistry in the thermosphere, as well as enhanced escape \citep{Czesla2013, Odert2020}. In order to make reasonable comparisons with L22 and V16, we have chosen not to include ion chemistry in \code{ATMO}, so neglecting enhanced X-ray irradiation is taken to be a reasonable assumption. Furthermore, radiative-convective models such as \code{ATMO} are not capable of solving for the temperature structure of the thermosphere, where IR opacity is low \citep{Fortney2021}. However, this also means that we do not account for enhanced charged particle fluxes associated with flare events, which could drive chemistry at low pressure levels in the atmosphere \citep{Chadney2017}.
    \par
\begin{figure}
        \centering
        \includegraphics[width=\columnwidth,keepaspectratio]{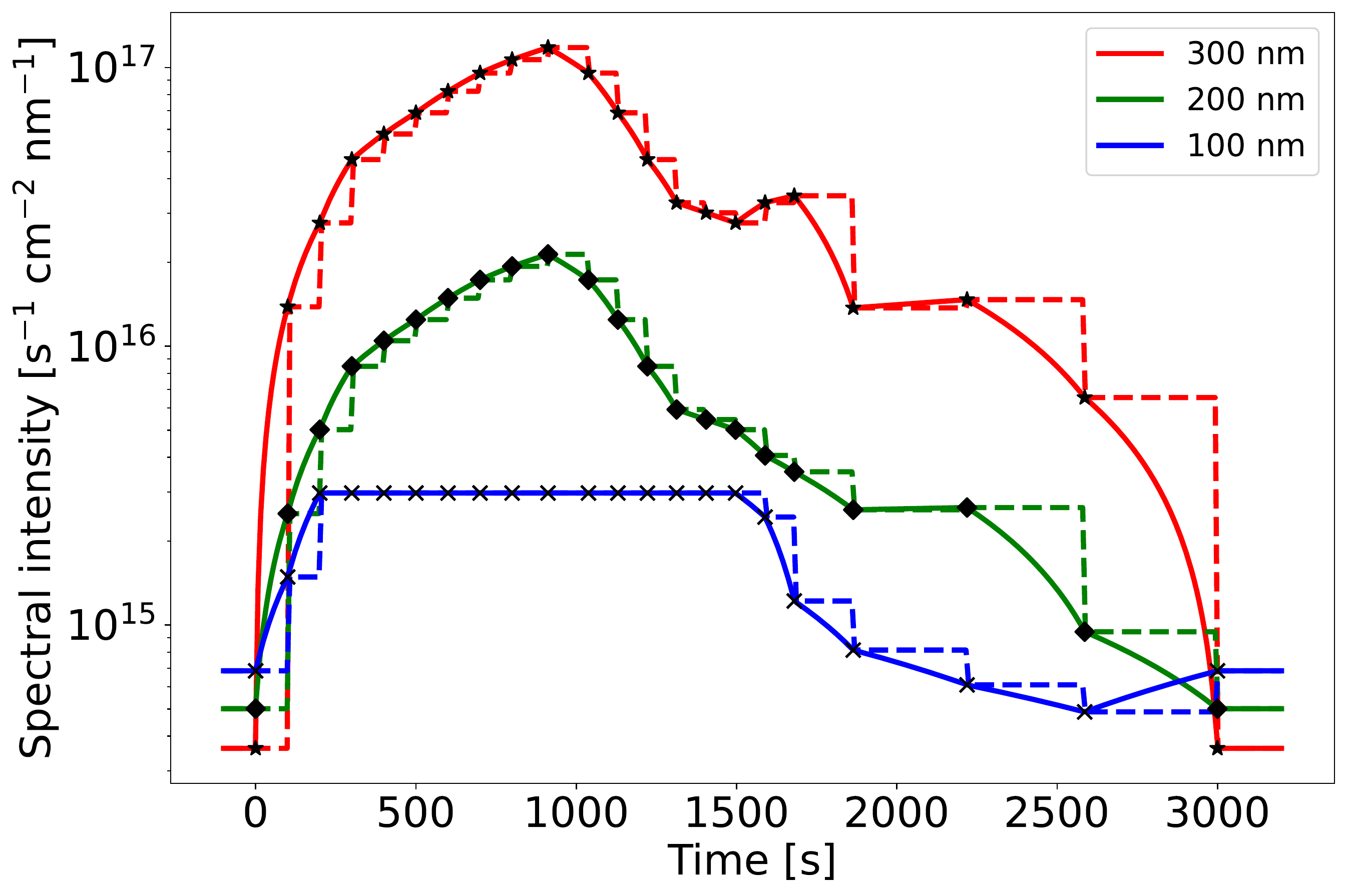}
        \captionof{figure}{Spectral intensity of UV emission vs time for the \SI{300}{}, \SI{200}{}, and \SI{100}{\nano\meter} bins. The solid lines shows linear interpolation between observed data, which are marked in black. The dashed lines demonstrate the interpolation method used in V16.}
        \label{fig:interp}
    \end{figure}

    The 3000 seconds of UV intensity variation for each flare is recorded at 22 points in time across 500 wavelength bins \citep{Hawley1991,Segura2010,Venot2016}. As the input UV flux at the top of the atmosphere $F_{\text{rad}}^0(\lambda, t)$ is required to determine $P_j$ and $L_j$ for photochemical processes at every chemical iteration, it is often necessary to evaluate $F_{\text{rad}}^0(\lambda, t)$ at times not exactly equal to one of these 22 observation points. V16 resolved this by maintaining constant spectral flux across time between these each of these data points. In this work, these data are instead interpolated linearly over time between adjacent points, independently for each wavelength bin, to more accurately represent the time-dependence of the UV flux. \Cref{fig:interp} plots the UV flux versus time with both interpolation methods, for three wavelength bins. Using the method from V16 caused problematic behaviour in the upper atmosphere during time-evolving simulations with \code{ATMO}, where the discontinuous jumps in flux at each of the 22 points forced the solver to choose extremely small time-steps, effectively preventing further evolution of the system.
    \par
    Stellar flares follow a flare-frequency distribution (FFD) over many orders of magnitude, where flare energy and occurrence frequency are inversely related in log-log space \citep{Hawley2014,Loyd2018}. A given flare may have a different energy to the ones preceding and proceeding it, but will have the same characteristic behaviour over time: starting with a linear rise from quiescence to its peak, and then decaying exponentially back to quiescence. \citet{Loyd2018} developed a statistical model\footnote{\url{https://github.com/parkus/fiducial_flare}} for flares originating on M-dwarf stars which samples an FFD derived from observations. This model was also used in \citet{Chen2020}, L22, and K22. Their FFD (plotted in Figure 3 of \citet{Loyd2018}) does not include the most energetic flare events which have been observed --- these would likely drive atmospheric chemistry and escape significantly \citep{MurrayClay2009}. 
    \par 
    It is worth noting that recent observations have shown that most M-dwarf stars are variable in their activity over time in a way which is not correlated with their spectral type \citep{Mignon2023,Melbourne2020}. This contrasts with the canonical `quiet' and `active' classifications applied to these stars, such as in \cite{Loyd2018}. Consequently, these planets may transition between various pseudo-steady states as the stellar activity, and thus their relative exposure to flares, varies over time.
    \par 
    The single and periodic flare models in this work (and in V16) use observational UV data from GF85 \citep{Hawley1991}, which had an energy output of \SI{e34}{\erg}, larger than the $\sim 10^{32} \text{ erg}$ upper-limit of the FFD of \cite{Loyd2018}. As high energy flares occur less frequently, more observational data would allow the FFD to be extended to higher energies, and thus capture `super flare' events, such as GF85.
    \par 
    In our work, Version 1 of the \cite{Loyd2018} model was applied to generate a \SI{9E6}{\second} time-series of flares that follow this energy distribution stochastically. This model was used to produce the results shown in Section \ref{ssec:flaresSto}. \code{ATMO} determines the quiescent flux scale factor (QFSF) $Q(t)$ of the star's UV emission from this time-series at each chemical iteration. From this, the instantaneous UV irradiation at the top of the atmosphere can be calculated by evaluating $F_{\text{rad}}^0(\lambda)$ according to:
    \begin{equation}
        \label{eq:qfsf}
        F_{\text{rad}}^0(\lambda, t) = F_{\text{Quiescent}}^0(\lambda, t) \cdot \big(Q(t)+1\big)
    \end{equation}
    In cases where $Q > Q_{\text{min}}$, \code{ATMO} considers the system in an \textit{active state}, and decreases the time-step for accuracy at the cost of performance. Below this threshold, the increased $Q$ is set to $0$. Neglecting these flares introduces some inaccuracy to this implementation, but is necessary for viable computational performance. However, note that L22 highlighted the importance of including flares of small amplitude, as they occur much more frequently than high-energy flares, and contribute to overall trends in chemical composition. We found that $Q_{\text{min}} = 0.05$ provided good computational performance without compromising the accuracy of our model.
    \par 
    It should be noted that the $Q(t)$ applied in our stochastic flare model does not depend on wavelength $\lambda$. For each wavelength bin, the UV flux across the entire variable interval is scaled relative to its quiescent value by a factor of $Q(t)+1$ at a time $t$. This is the method used in L22, and also follows the philosophy of \cite{Loyd2018} in that `all bands are assumed to follow the same temporal evolution'. This simplified method contrasts with our periodic flare model in which each wavelength bin effectively has a different scale factor compared to other bins at a given time. The difference between these two approaches results from the fact that the periodic model uses observations of GF85 only, while the stochastic model samples a distribution to scale the UV flux relative to its quiescent value. Both models use the same quiescent spectrum, so the differences are solely related to scaling flux over time. This difference is acceptable given that quantitative comparisons are not made between simulations resulting from different flare models. In comparison, K22 used a model which applied flares stochastically with $Q(t)$ also depending on wavelength.
    \par
    The time-step $dt$ of the chemical solver within \code{ATMO} is inversely related to the maximum rate of change of $dn_j/n_j$ across all species $j$, such that faster processes require the solver to take smaller steps. The initial time-step is \SI{e-9}{\second}. It is important to consider this behaviour for stochastic and periodic flare simulations: were $dt$ to reach a length at least equal to a flare duration, individual flares would no longer be temporally resolved by the model, allowing the $dt$ to increase, which increases the probability of missing future flares. To mitigate this, $dt$ is limited to a maximum of \SI{15}{\second} during active periods and \SI{80}{\second} during rest periods, although a time-step of approximately \SI{10}{\second} is often selected by the solver.
    
    \subsection{Transmission spectra and observations}
    \label{ssec:transSpec}
    It is important to consider whether changes in atmospheric composition due to flares could be observable, and whether it requires consideration during retrievals when generating forward models. In HJs, the pressure level which contributes the most to the observed flux generally occurs between \SI{e2}{\milli\bar} and \SI{e3}{\milli\bar}, although this varies moderately between planets and wavelengths \citep{Drummond2017,Parmentier2020,Lee2011}. To have a significant impact on emission observations, changes in composition due to flares would have to occur at or near this pressure range, which is relatively deep in the atmosphere in the context of non-equilibrium processes \citep{Drummond2016,Goyal2017,Venot2012}. Transmission spectroscopy probes regions higher in the atmosphere than emission spectroscopy; changes in composition due to flares at pressures greater than \SI{e2}{\milli\bar} will likely not be observed in transmission, so changes in composition at pressures lower than \SI{e2}{\milli\bar} are of greatest interest.
    \par 
    \code{ATMO} can generate synthetic transmission spectra \citep{Baudino2017}. These can be combined with the \textit{JWST} noise simulator PandExo to generate realistic synthetic observations, in order to assess the observability of the impact of flare-driven chemistry \citep{Phillips2020,Tremblin2019,Batalha2017}. 
    
    Generating transmission spectra before and after flares have been applied, and then simulating the noise on these with PandExo, permits an assessment of the observability of the effects of flares on these atmospheres. The total propagated noise estimated by PandExo $\delta y'$ is a combination of shot-, background-, and read-noise \citep{Batalha2017}.
    \par 
    Quantifying the observability of changes in chemical composition was done by synthesising transmission spectra from several simulations. Relative changes $C$ in wavelength-dependent transit depth $y = (R_p/R_*)^2$ are quantified as
    \begin{equation}
        \label{eq:transDiff}
        C(\lambda,t) =  \frac{y_f(\lambda,t) - y_q(\lambda) }{y_q(\lambda)}
    \end{equation}
    Where $y_q(\lambda)$ is transit depth of the planet immediately before the onset of flares. $y_f = y(\lambda,t)$ represents the average transit depth of the planet between $t-2\tau_c$ and $t$, where $\tau_c$ is the transit duration. Note that all instances of $R_*$ inside the $y$ in \Cref{eq:transDiff} out, so that $C$ does not depend on $R_*$. This formulation for $y_f$ was used to best represent the average state of an atmosphere exposed to flares; it is a metric for typical changes to observables, rather than for a specific flare event. As flares do not occur predictably, it would be unfeasible to accurately time an observation of one with JWST. Transit durations $\tau_c$ for each of the orbital separation $a$ were calculated according the formula
    \begin{equation}
        \label{eq:transit_duration}
        \tau_c = 2 R_{*} \sqrt{\frac{a}{Gm_{*}}}
    \end{equation}
    where $m_{*}$ and $R_{*}$ are the mass and radius of AD Leo, $a$ is the orbital separation of the planet, and $G$ is the gravitational constant. This formula assumes that the planet and orbit are both circular, and that the centre of the planet passes in front of the centre of the star relative to the observer.
    \par 
    To gain insight into the observability of these spectral changes, $y_q$ and $y_f$ were used as input spectra for PandExo, which re-binned and truncated the data, applied the appropriate noise, and estimated the errors. Quantities representing the output of PandExo are denoted with a prime symbol. Overall, for a given simulation, PandExo yielded the wavelength-dependent quantities: $y_f'$ ,$\delta y_f'$, $y_q'$, and $\delta y_q'$. These were analysed using Equation \ref{eq:transDiff} to quantify the effect of (potentially observable) flare-driven changes to the transmission spectra, giving us $C'$ and $\delta C'$. The quantity $C'/\delta C'$ represents the signal-to-noise ratio (SNR) of the flare-driven behaviour. One of the main purposes of transit spectroscopy is to identify specific spectral features, which can inform us about the composition of an observed planet \citep{Mounzer2022}. Therefore, for the effects of flares to have implications for observations, spectral features associated with flare-driven behaviours must have a `significant' SNR: $|\text{SNR}|>1$.

    \subsection{Outline of cases}
    \label{ssec:outlineCases}
    Combinations of the following parameter ranges were modelled, choosing one value per row for each simulation.
    \begin{itemize}[labelwidth=1mm,itemindent=!] 
        \label{itm:grid}
        \item Chemistry: NEQ, \textbf{CNEQ} \\
        \item Flare types: Single, Periodic, \textbf{Stochastic} \\
        \item $T_{\text{eff}}$: 412 K (cold), \textbf{1632 K (hot)} \\
    \end{itemize}
    Cases in \textbf{bold} type were not explored by V16. The planet with \mbox{$T_{\text{eff}} = $\SI{1632}{\kelvin}} was considered because the various processes involved (e.g. photochemistry, mixing) will likely be quite different to in the cold case. This small orbit increases the irradiation on the planet and decreases the chemical time-scale, which may change the relative importance of photochemistry compared to equilibrium and dynamic processes \citep{Fortney2006}. The hot planet was placed at an orbital separation of \SI{0.0044}{\AU}, which is the smallest semi-major axis within the recorded exoplanet population for which both the mass and the radius of the planet are known \citep{Bailes2011}. As in V16, the cold planet -- with an effective temperature of \SI{412}{\kelvin} -- corresponds to an orbital separation of \SI{0.0690}{\AU}. 
    
    \par 
    The effective temperatures of both planets are derived using Equation \ref{eq:teff}, where $\sigma$ is the Stefan-Boltzmann constant and $L_{*}$ is the luminosity of AD Leo.
    \begin{equation}
        \label{eq:teff}
        T_{\text{eff}} = \Big( \frac{ L_{*} }{ 16 \sigma \pi a^2 } \Big)^{1/4}
    \end{equation}
    \par 
    Note that the set of planets explored in our work is not the same as that in V16. While both our work and V16 simulate a cold planet orbiting AD Leo at $a = 0.069 \text{ AU}$, our hotter planet is instead placed at \SI{0.0044}{\AU}, in substitution for V16's placed at \SI{0.0069}{\AU}. This was done in order to determine the relative importance of self-consistent modelling across a broad range of environments (chemical and thermodynamic), while constraining our parameter space in order to enable sufficiently deep discussion.
    \par 
    
    Before any flares are applied, the models need to be initialised from a converged PFS. In generating this initial state, the same physics and parameters must be used as in the simulation which includes the flares. For example, a simulation of the cold planet with NEQ chemistry must be initialised from a simulation which also used NEQ chemistry and had the same effective temperature. This is especially important in the CNEQ cases where we want to capture the feedback between chemistry and $T(p)$ due to flares specifically, so the PT profile for the CNEQ cases must be derived with CNEQ chemistry. In the NEQ cases, $T(p)$ is fixed and unable to evolve; for the cold planet we chose to use the PT profile from V16 (which was derived from \citet{Fortney2013} and is isothermal above \SI{E-3}{\milli\bar}) to enable comparisons with their work, and for the hot planet we generated our own PT profile using \code{ATMO}. Initialising a simulation of flare evolution with an appropriate PFS also improves the computational performance of the analysis.
    
    \par
    The bulk elemental abundances of our model atmospheres can be found in Appendix \ref{app:elemList}. Bulk abundances are used as input parameters for the EQ chemistry, which is itself used to initialise the NEQ and CNEQ models. 
    \par 
    
    All atmospheres are simulated across the pressure and optical-depth regimes applicable for this analysis: \mbox{$10^{-4} \le p/(\text{mbar}) \le 10^{5}$} for the cold planet, and \mbox{$10^{-2} \le p/(\text{mbar}) \le 10^{5}$} for the hot planet. These low-pressure boundaries $p_{\text{min}}$ were selected such that the composition of the mixture at pressures lower than $p_{\text{min}}$ is generally dominated by atomic hydrogen, where flares will not induce photochemistry and UV will not be strongly absorbed. Most of the energy delivered by flares occurs via enhancement to continuum emission, rather than at specific spectral peaks, including those of H \citep{Loyd2018,Hawley1991}. Simulating regions of low opacity and low pressure (such as H-dominated regions) is difficult without the use of specialised models, which come with their own set of compromises \citep{Fortney2008,Lothringer2018,Fossati2021}. With \code{ATMO}, solving for the temperature structure at pressures lower than $p_{\text{min}}$ manifested oscillations in $T(p)$. This spurious oscillatory behaviour interfered with the effects of the time-dependent UV irradiation on the temperature structure, so the simulation region is truncated in each case to the pressure limits stated previously. The primary reason for this behaviour is that the model struggles to maintain radiative equilibrium at such low optical depths, while simultaneously solving for both middle atmosphere and time-dependent changes in composition. As a result, the pressure grid of the hot planet has a smaller range across $p$-space than for the cold planet. This transition to a H-dominated regime at very low pressures is discussed further in Section \ref{ssec:quiescent}. Similarly, regions with greater pressure than \SI{E5}{\milli\bar} are irrelevant for this analysis as UV radiation will not penetrate to such large optical depths, and the atmosphere is dominated by equilibrium processes \citep{Lewis2020,Goyal2017,Drummond2016,Fortney2006}. Including ion chemistry in the model would make neglecting the radiative and chemical effects of the H-dominated region less feasible, as \ce{H-} cations present in HJ atmospheres have been found to generate distinctive spectral features \citep{Lothringer2018,Parmentier2018,Ohmura1961}.
    
    \par
    Data were saved at least every 12 chemical iterations, and for CNEQ simulations the radiative-convective scheme was re-converged every $M$ chemical iterations. We found that $M=15$ provided good model accuracy, while not compromising performance and thus limiting our integration time (see Appendix \ref{app:energyError}). The atmosphere was allowed to evolve for a \textit{delay period} of \SI{8E5}{\second} after reaching a pre-flare steady state, but before flares were applied. This delay period ensured that any numerical effects from initialising the flare-configured simulations could dissipate. It was verified that this delay period was sufficient, and that the atmosphere was in a steady-state before flares were applied in all cases. Flares were applied for up to \SI{9E6}{\second} ($\sim 100 \text{ days}$), at which point the models were stopped. Solving for the time-dependent behaviour of the system self-consistently is computationally expensive due to the extra steps required to ensure RCE, so in some cases it was unfeasible to evolve the model for the whole \SI{9E6}{\second}. Results presented in this work are analysed with this under consideration. Simulations applying the stochastic flare model ran faster than those applying the periodic flare model, primarily because the solver can recover from smaller flares more rapidly than from larger ones. Similarly, simulations of the hot planet were faster than of the cold one, because non-equilibrium effects were less relevant, and thus longer time-steps were feasible at the same error tolerance. In the most computationally expensive case explored in this work (stochastic flares/cold planet/CNEQ), \SI{3e6}{\second} of simulation time required approximately 50 days of real time to integrate, at which point the model was stopped.

    
    \section{Results and Discussion}
    \label{sec:resAndDis} 
    
    This section combines our results and discussion together, following the development of increasingly advanced flare models, and concluding with a discussion on potentially observable effects. Throughout, simulations with coupled temperature and chemistry are compared to those in which the temperature profile was fixed. This is done in order to assess the role of self-consistent solving in this context. Our results and discussion are therefore structured as follows.
    \begin{itemize}
        \item \Cref{ssec:quiescent}: introduction of quiescent states used to initialise atmospheres before any flares were applied.
        \item \Cref{ssec:flaresSglNeq,ssec:flaresSglCneq}: demonstration of the impact of applications of a single instance of GF85, and how the chemistry recovers from such an event.
        \item \Cref{ssec:flaresPerNeq,ssec:flaresPerCneq}: investigation into the chemical and thermal response to periodic applications of GF85, mirroring the work of V16.
        \item \Cref{ssec:flaresSto} and subsections therein: analysis of the complex chemical and thermal response to stochastically applied flares, where seven species are selected as case studies.
        \item \Cref{ssec:synthObs}: discussion on the impact of flare-driven compositional and temperature change in the context of \textit{JWST} observations.
    \end{itemize}

    \subsection{Quiescent states}
    \label{ssec:quiescent}
\begin{figure*}
        \includegraphics[width=2\columnwidth,keepaspectratio]{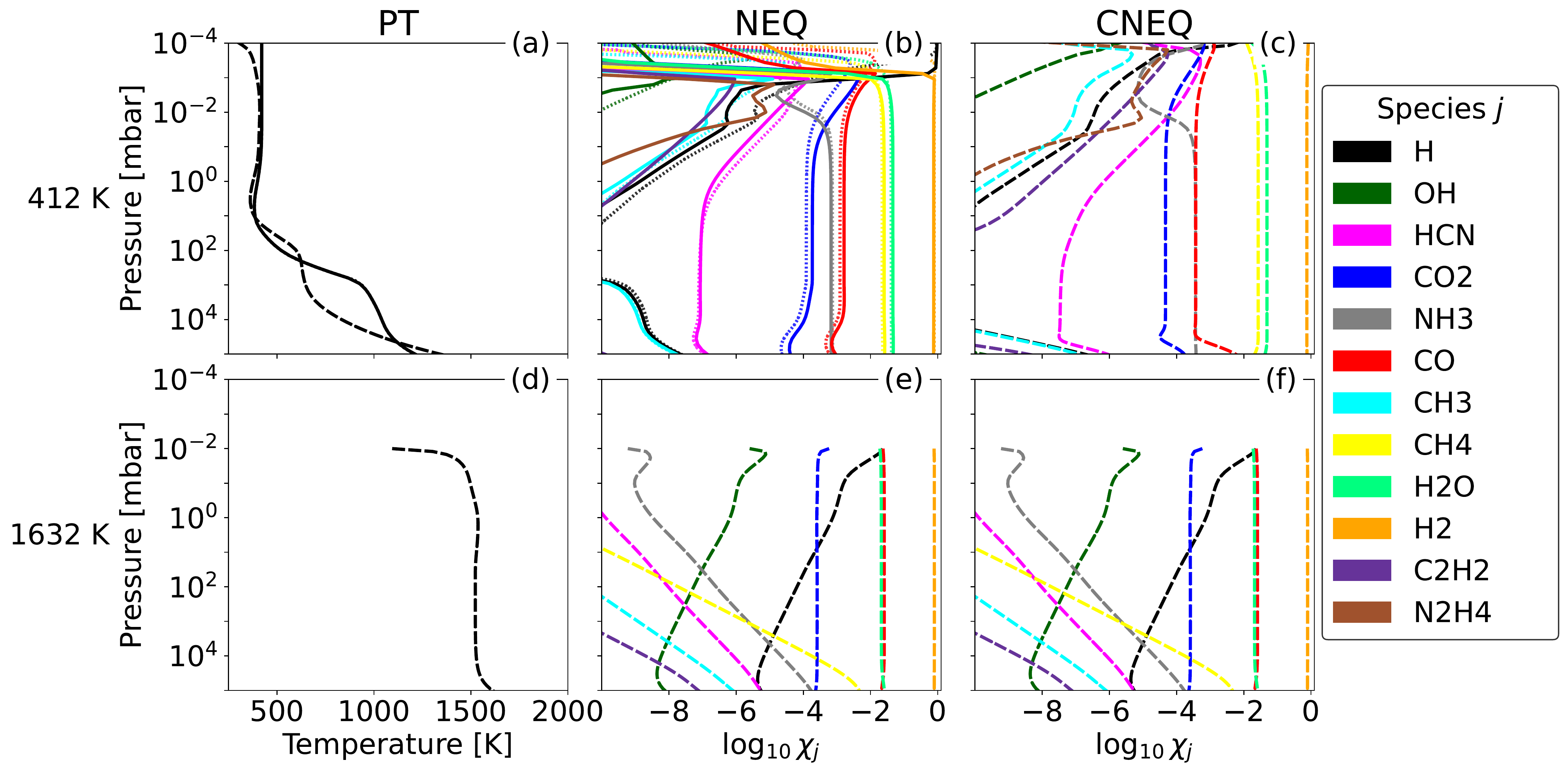}
        \captionof{figure}{Quiescent atmospheres generated by \code{ATMO}. These were used as initial states from which the atmosphere later evolved with flares. Top row: cold planet. Bottom row: hot planet, for which the simulation region is truncated to higher pressures, as discussed in Section \ref{sec:methods}. The chemical abundance of a species $j$ is quantified by its mole fraction $\chi_j$. (a): temperature profiles used for the NEQ cases (solid) and to initialise the CNEQ (dashed) cases; the solid-line is identical to the PT profile used in V16 for $T_{\text{eff}} = 412 \text{ K}$. (b): initial NEQ chemical abundances derived using \code{ATMO} (solid), plotted alongside those used in V16 (dotted). (c): initial CNEQ-derived chemical abundances. (d): the temperature profile used for the NEQ cases and to initialise the CNEQ cases. (e): initial NEQ chemical abundances; this atmosphere was not derived from a temperature profile used in V16, so a consistent simulation is necessary to generate a temperature profile, so it is identical to panel (f). (f): initial CNEQ chemical abundances. }
        \label{fig:tilesQsc}
    \end{figure*}

    \Cref{fig:tilesQsc} shows converged states from which simulations involving various flare models were later initialised (Sections \ref{ssec:flaresSglNeq}-\ref{sssec:stoNO}). As described in \Cref{ssec:outlineCases}, NEQ simulations of the hot planet must be first initialised using CNEQ chemistry because the radiative-convective scheme is required in order to generate a temperature profile. This does not preclude use of the NEQ case of the hot planet, because temperature and chemistry were uncoupled once a PFS was established.
    This therefore does not limit the analysis performed in later sections of this work. \Cref{fig:tilesQsc} shows that the temperature structure and composition differ significantly between the hot and cold planets, enabling an investigation the effect of flares and the role of self-consistent modelling under a range of conditions.
    
    \par 
    Panel (b) of \Cref{fig:tilesQsc} shows that \code{ATMO} can reasonably replicate the quiescent NEQ results of V16, with only minor discrepancies in the upper atmosphere. 
    \par
    At low pressures, the mole fraction $\chi_{\ce{H}}$ of atomic hydrogen is enhanced primarily due to photolysis of \ce{H2}, with some contribution from \ce{H2O} and other hydrogen-bearing species. This means that in each case there exists a pressure level above which \ce{H} is a major component of the atmosphere; for the cold planet this occurs at approximately \SI{E-3}{\milli\bar} and \SI{E-4}{\milli\bar} in the NEQ and CNEQ cases respectively. For the hot planet, $\chi_{\ce{H}}$ increases towards the low-pressure boundary in panels (e) and (f), although in these cases the planet is not simulated to sufficiently low pressures for H to dominate over all other species. Trials where the pressure-grid of the hot planet was extended to lower pressures found that H dominated the composition at pressures less than \SI{2e-3}{\milli\bar}. Nevertheless, the regions for which atomic hydrogen is a significant component of the mixture is not particularly interesting for this work, as the partial pressures of other species are sufficiently low that reaction rates will be negligible. In the wavelength region modulated by the flares, H will absorb some of the UV radiation incident on the top of the atmosphere, however the Balmer series absorption is weak, and strong absorption by the Lyman series \mbox{$\lambda \sim$ \SI{121}{\nano\meter}} does not cover a large portion of the variable interval used in our flare model \citep{Wiese2009}. \cite{Loyd2018} states that enhancement to the continuum component of emission `accounts for the bulk of the flux' associated with flares, so capturing narrow absorption features associated with hydrogen in the uppermost parts of the atmosphere are not of high importance.
    \par 
    The rapid decreases in temperature at the top of the PT profile in panels (a) and (d) of \Cref{fig:tilesQsc} can be attributed to the low opacity and low pressure of the topmost regions of the model columns. It is worth noting that similar behaviour is not seen in the PT profiles of V16 (plotted as a solid line in \ref{fig:tilesQsc}(a)) as they are assumed to be isothermal at pressures less than \SI{E-3}{\milli\bar}.
    States generated using CNEQ chemistry (\Cref{fig:tilesQsc}(c,f)) are broadly similar to NEQ equivalents (\Cref{fig:tilesQsc}(b,e)). 
    \par 
    The absorption of optical radiation by vapours of TiO and VO is thought to cause thermal inversions in hot gas planets, but this was not found to occur in the model atmospheres explores in this work despite \ce{TiO} and \ce{VO} being included in both the chemical and radiative transfer schemes within \code{ATMO}. The reason for this is that the cases explored in this work are not hot enough for vapour phases of \ce{TiO} and \ce{VO} to contribute significantly to the optical opacity, so they do not manifest a thermal inversion. It is known that the transition to the inverted regime occurs for $T_{\text{eff}}$ between \SI{1600}{\kelvin} and \SI{2000}{\kelvin} when \ce{TiO} and \ce{VO} evaporate depending on pressure, metallicity, and C/O \citep{Lothringer2018,Piette2020,Desert2008,Fortney2008}. No further discussion will be made regarding \ce{TiO} and \ce{VO}, as they do not strongly absorb UV radiation, and do not significantly impact the temperature profiles used in this work.
    
    \subsection{Single flares - NEQ}
    \label{ssec:flaresSglNeq}
    Single flare events were simulated using NEQ chemistry starting from the PFSs presented in Section \ref{ssec:quiescent}. 
    \par
\begin{figure}
        \centering
        \includegraphics[width=\columnwidth]{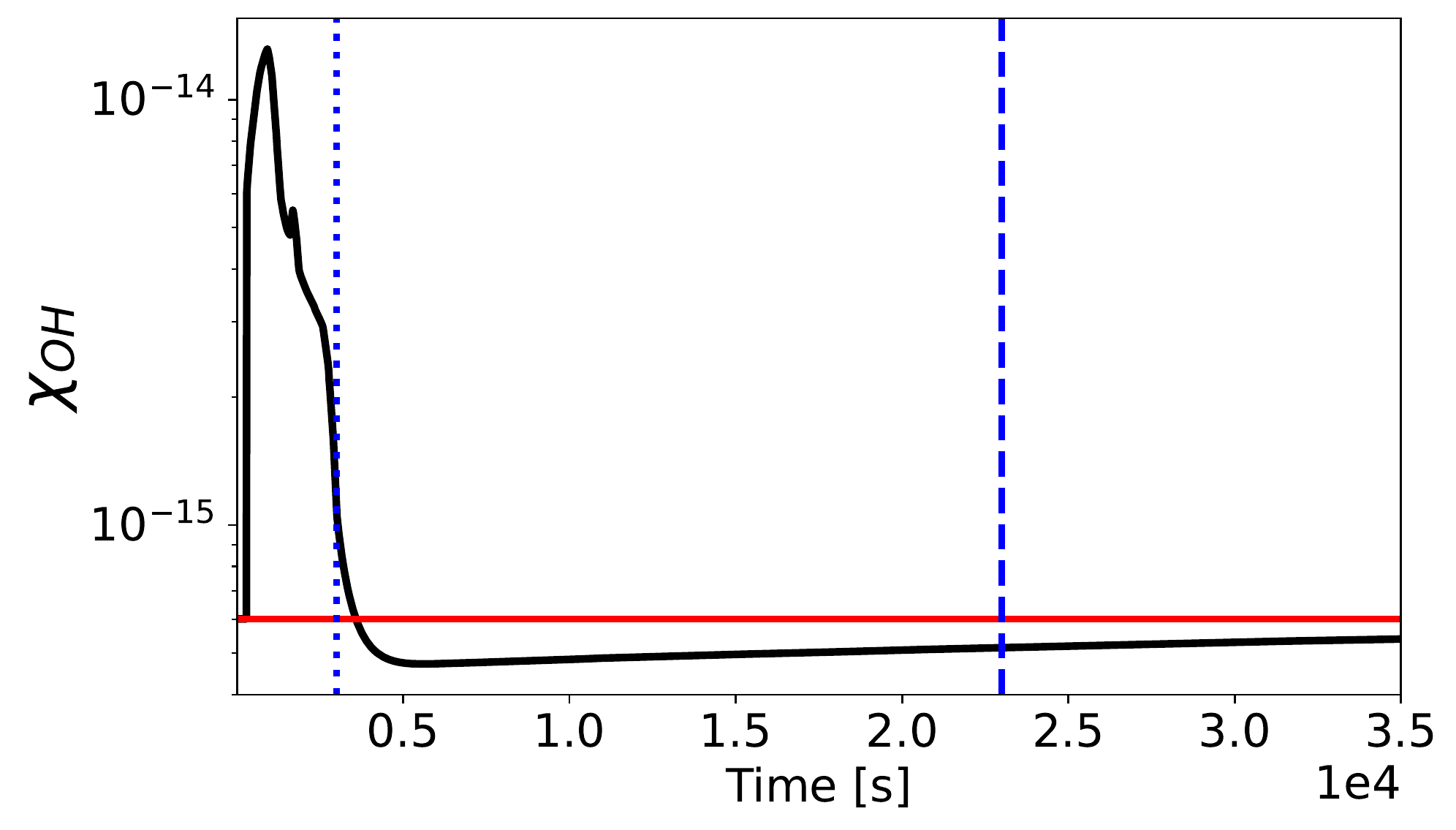}
        
        \vspace{2mm}
        
        \includegraphics[width=\columnwidth]{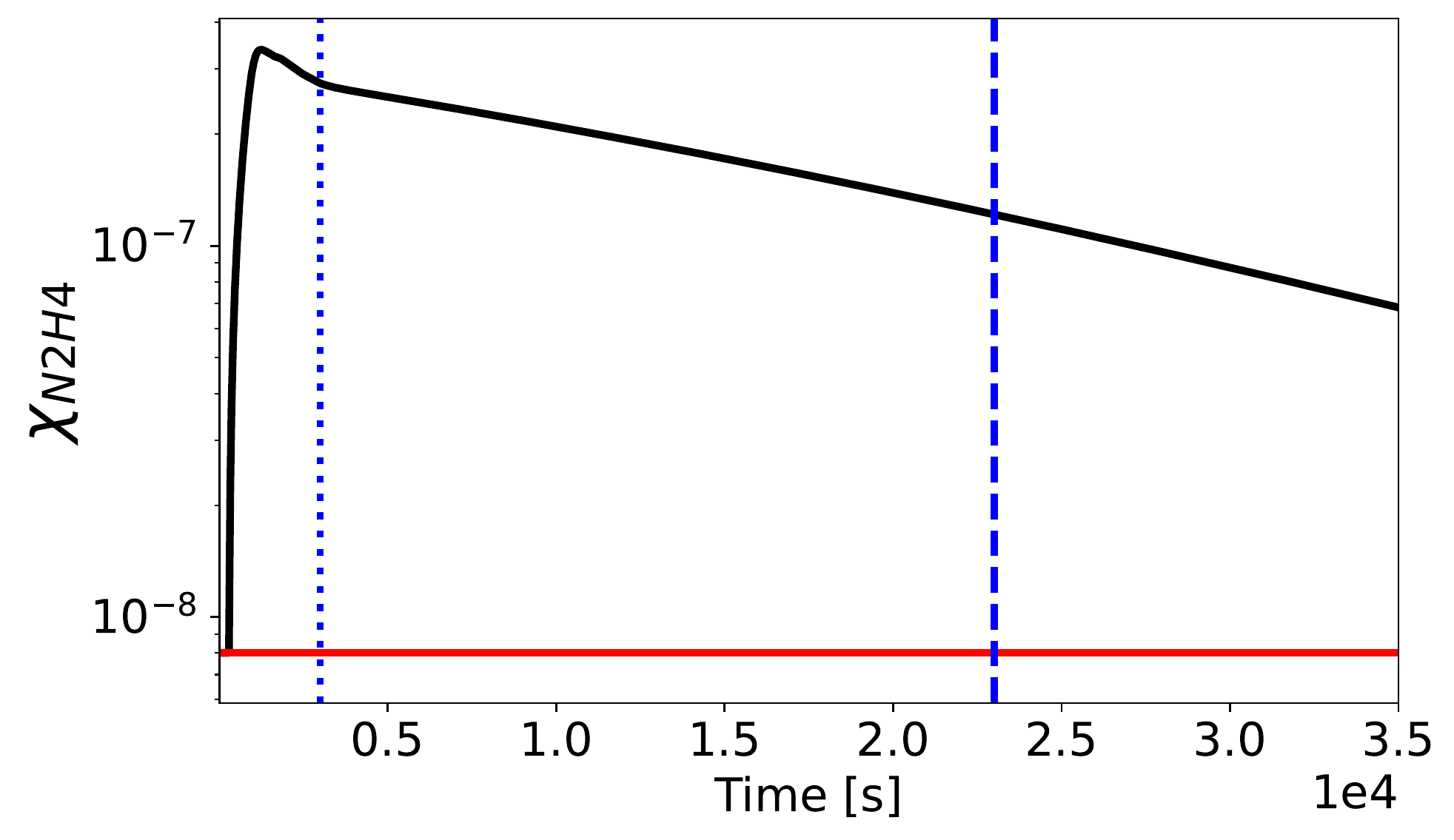}
        
        \captionof{figure}{$\chi_{\text{OH}}$ (top) and $\chi_{\text{N2H4}}$ (bottom) versus time at a pressure of \SI{0.063}{\milli\bar}. These species evolved simultaneously in a simulation of the cold planet using NEQ chemistry. Evolution was driven by a single instance of GF85 applied at $t=0$. The horizontal red lines denote the mole fractions of each species before any flares were applied. The dotted vertical blue line represents the point in time at which the UV flux returns to its quiescent value ($t=3000\text{ s}$), although it decays significantly before this point. The dashed vertical blue line is drawn at the point in time at which a successive flare would be applied using the periodic flaring scheme \mbox{($t=$\SI{2.3e4}{\second})}. }
        
        \label{fig:mftsSglOH-N2H4}
    \end{figure}

 
    Changes in composition measured at this point in time are small and are heavily species dependent. The largest changes are seen for \ce{NH3}, \ce{H}, and \ce{OH}. It is unsurprising to see \ce{H} changed most significantly as it is the product of a number of photolysis pathways, such as \ce{H2O ->[h\nu] H + OH} and \ce{NH3 ->[h\nu] H + NH2} \citep{Liang2003}. The hotter planet's composition is more affected than that of the cooler one, which follows from that fact that it has shorter chemical time-scales, and parallels the trends seen in V16 and K22. As only one flare is applied, the timeframe for which photochemistry is especially relevant is short in comparison to flare models in which many flares are applied. Despite the hotter planet being more greatly affected by a single flare during the active period, its fast recovery once the UV flux is restored to its quiescent state could mean that sequential flares compound less constructively.
    \Cref{fig:mftsSglOH-N2H4} shows the responses of \ce{OH} and \ce{N2H4} at \SI{0.063}{\milli\bar} when a single flare is applied to the cold planet. These species and pressures were chosen because they demonstrate that different species in the atmosphere respond to flares with various behaviours across several time-scales. Some species did not manifest any significant response to the flares; e.g. \ce{CO}, \ce{N2}, \ce{TiO}, \ce{VO}, \ce{Na}, \ce{NH}.
    \par 
    The top panel of \Cref{fig:mftsSglOH-N2H4} shows the response of \ce{OH}. The abundance of \ce{OH} is increased in accordance with the increase UV flux associated with the flare; this behaviour can be attributed to photolytic processes where \ce{OH} is a product. Throughout this process, the abundance of \ce{OH} traces the change in UV flux closely, including the double-peak behaviour shown in \Cref{fig:interp}. At \SI{2.3e4}{\second} --- the point at which a subsequent flare would occur under the periodic flaring scheme --- it is close to its pre-flare abundance, and by \SI{6E4}{\second} it has effectively returned to its PFS. 
    \par 
    The bottom panel of \Cref{fig:mftsSglOH-N2H4} shows that the abundance of \ce{N2H4} initially increases when the flare is applied, and does not evolve significantly away from this state until the flare ends at \SI{3000}{\second}. Once the flare ends, the abundance of \ce{N2H4} slowly decays towards the PFS, but does not reach it by the same a successive flare would be applied. Production of \ce{N2H4} during the active period can be attributed to reactions between \ce{N2H3} molecules, which are themselves produced by reactions between photolysis products (i.e. \ce{NH + NH2 -> N2H3}). The abundance of \ce{N2H4} in the rest period displays semi-permanent changes in composition driven by flares; further discussion of \ce{N2H4} is made in \Cref{sssec:stoN2H4}.
    \par 
    The atmosphere continues to evolve once UV irradiation returns to its quiescent state. Although analysing chemical evolution for times long after the flare has passed is not physically motivated --- as discussed in \Cref{sec:introduction,ssec:flaringModels} --- the short-term behaviour following a single flare event can be used explain the effects of repeated flaring. Across species, it is common for their forward pathways to react strongly to the enhanced UV, but once the UV flux returns to quiescence the atmosphere model does not necessarily tend directly to the PFS. Instead, the composition first overshoots the PFS and only then decays back towards the PFS (see the top panel of \Cref{fig:mftsSglOH-N2H4}). This \textit{chemical inertia} implies that the duration of the inter-flare resting period could play an important role in the overall effect of repeated flares on these atmospheres, as it determines how closely the atmosphere has approached the PFS when a successive flare is applied.
    \par 
    Post-flare, the most significant effects are seen within \mbox{$\sim$\SI{3E4}{\second}}, at which point most of the atmosphere is slowly returning to the PFS. This indicates that, despite the overshooting behaviour, the chemistry still acts to restore species abundances to their PFS after a flare has occurred. This means that for cases where flares occur in series, we should expect species abundances to approach their PFS until another flare occurs.

    \subsection{Single flares - CNEQ}
    \label{ssec:flaresSglCneq}
    
    Allowing the temperature structure of the atmosphere to vary throughout these simulations does not lead to significant temperature changes, although they are non-zero. The changes in temperature seen here are less than 1 K, which is less than that in \cite{Segura2010}. The energy of a single flare (even GF85) is too small to remove a significant fraction of the species in the atmosphere which contribute to the temperature structure by absorbing radiation. Thus, the opacity of the atmosphere is relatively unchanged, and so the temperature changes are small compared to the absolute temperatures of these planets. It is worth noting that solving for RCE iteratively leads to an marginal under-estimate of the peak change in temperature throughout the course of a given flare, compared to the ideal case in which the temperature profile is updated continuously over time (see Appendix \ref{app:energyError}).
    \par 
    Despite changes in composition being small, similarly to the NEQ cases in the previous section, it was found that the atmospheres do not entirely return to their PFS before \SI{2.3e4}{\second}. This indicates that repeated flares could have a cumulative effect. 
    \par 
    When only simulating the effects of a single flare, it is not possible to confidently state whether self-consistent simulations lead to more or less significant changes to the composition compared to cases where $T(p)$ is fixed, as in both cases the changes in composition and temperature are small. 
    
    \subsection{Periodic flares - NEQ}
    \label{ssec:flaresPerNeq}
    With the application of repeated flares, it is common to see dependent variables of the system (such a mole fraction or temperature) fluctuate over time about a stable value. To make an analogy with the Reynolds Decomposition, this `stable' value is termed the `average steady-state' \citep{Adrian2000}. Similarly, V16 makes reference to each variable having a `limiting value', and K22 takes the time-average of mole fraction of particular species to describe this state. As we are interested in characterising the net effect of flares and its relationship with self-consistency, it is the average steady-state of variables such as composition and temperature which are of most interest. We take a variable to be in an `average steady-state' at a time $t$ and pressure $p$ when the difference between its average values for the intervals $[t-t_d,t]$ and $[t,t+t_d]$ are very small ($t_d=$\SI{1e5}{\second}).
    \par 
    For simulations of periodic flares with NEQ chemistry, our models demonstrate similar behaviour to that in K22: the atmosphere tends generally towards an average steady-state different to the PFS, with the most significant changes occurring when the flares begin. At \SI{3e6}{\second}, some species are still evolving at higher pressures , which indicates that more simulation time may be required for the model to closely approach an average steady-state in this regime, if one exists. 
    \par
\begin{figure}
        \centering
        
        \includegraphics[width=\columnwidth]{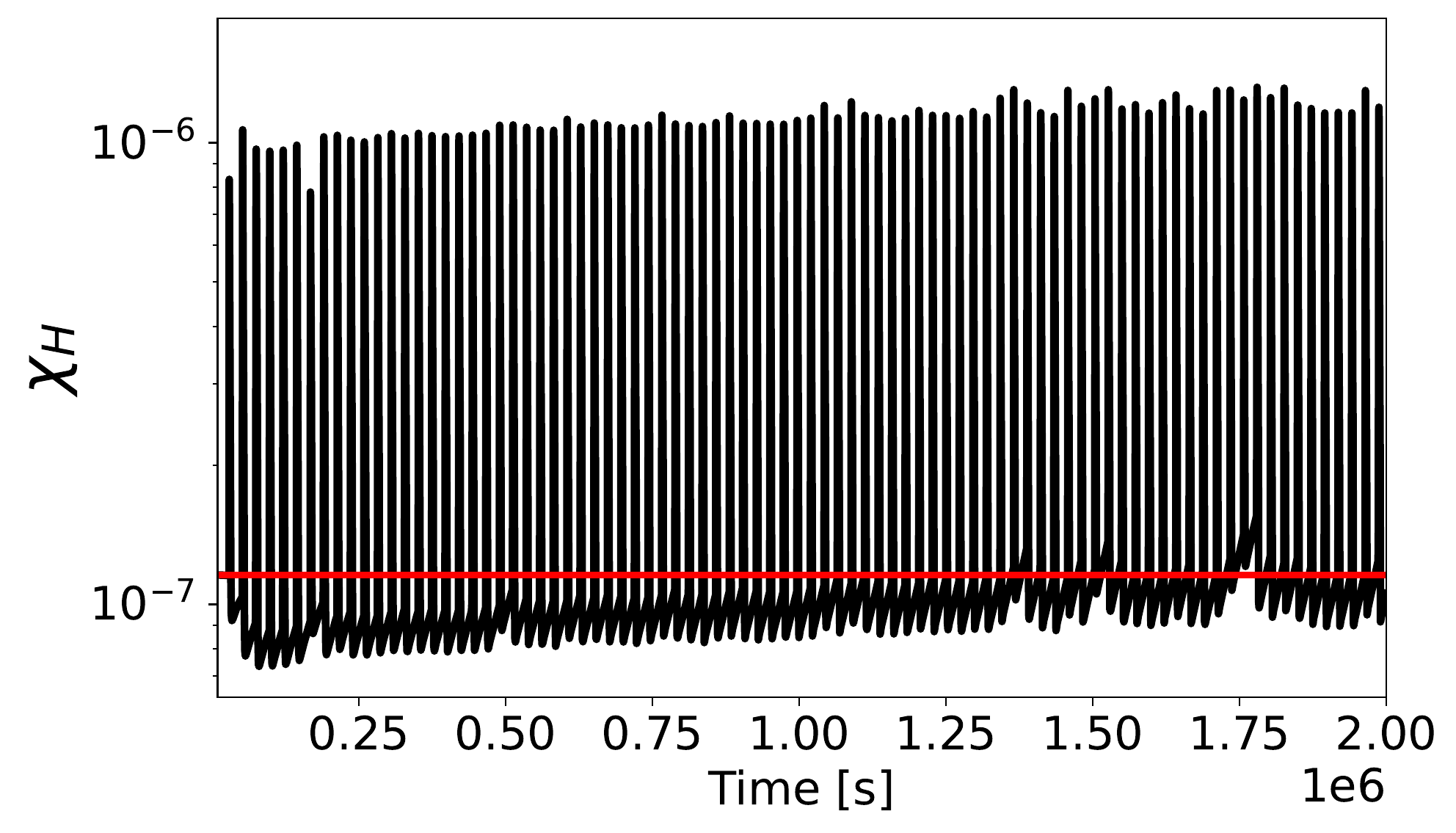}
        \vspace{2mm}
        \includegraphics[width=\columnwidth]{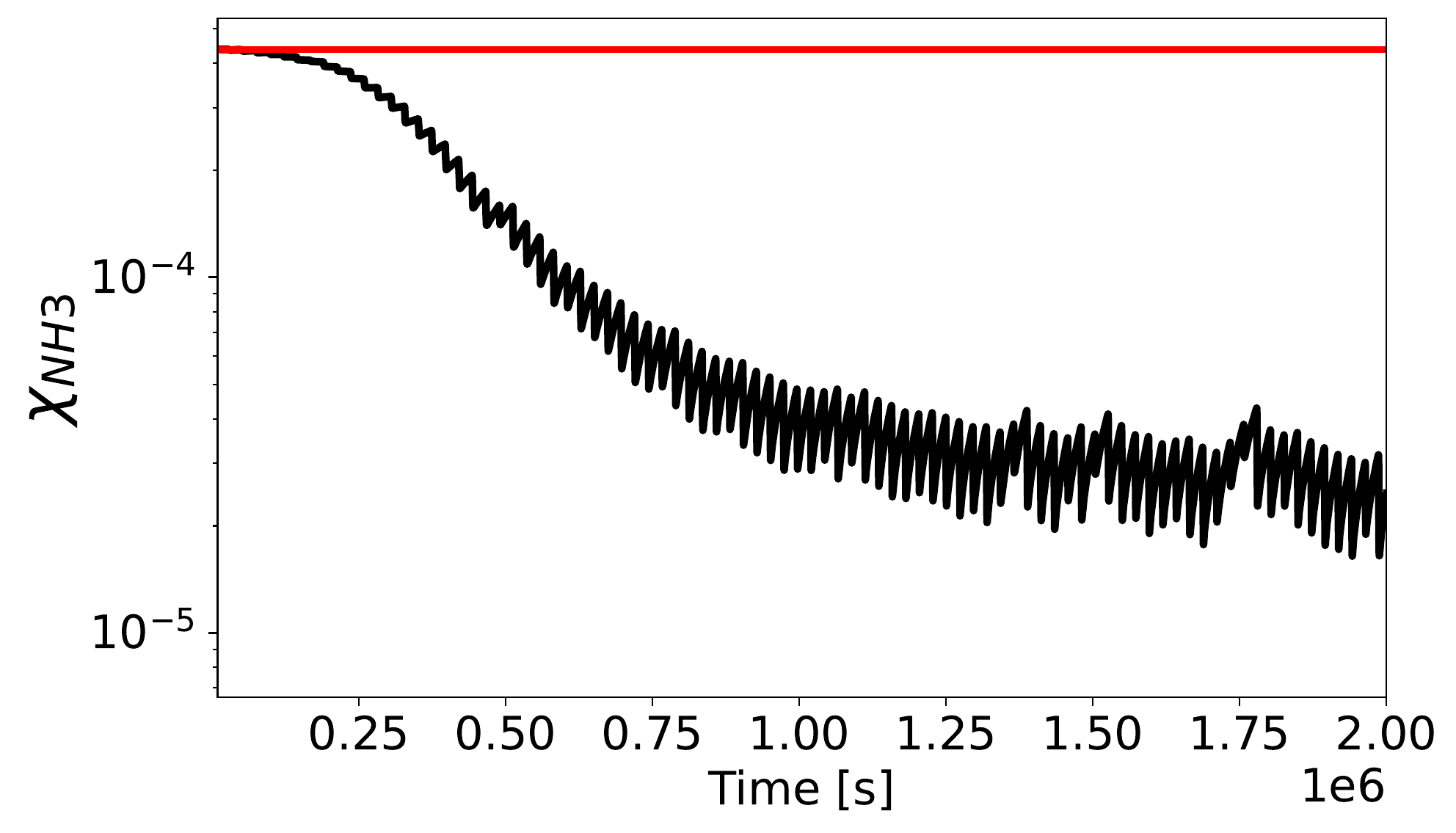}
        
        \captionof{figure}{Top panel: $\chi_{\ce{H}}$ versus time at pressure of \SI{0.063}{\milli\bar}. Bottom panel: $\chi_{\ce{NH3}}$ versus time at pressure of \SI{0.022}{\milli\bar}. These species evolved simultaneously in a simulation of the cold planet using NEQ chemistry. The horizontal red lines denote the mole fractions of each species before any flares were applied.}
        
        \label{fig:mftsIPer412H-NH3}
    \end{figure}

 
    The panels of \Cref{fig:mftsIPer412H-NH3} are analogous to the panels in Figure 11 of V16, although in this case plotted for the colder planet. The trend in the top panel of \Cref{fig:mftsIPer412H-NH3} is similar to the analogous plot in V16, with \ce{H} quickly entering an oscillatory state corresponding to the application of each flare. Similar behaviour is also seen in Figure 12 of K22, where the abundances of some species return to their pre-flare values during inter-flare resting periods. However, in the case of K22 this behaviour varies significantly depending on the duration of this quiet period, which is inversely related to the rate at which flares are applied. The bottom panel of \Cref{fig:mftsIPer412H-NH3} shows that \ce{NH3} initially responds both quickly and dramatically to the enhanced UV flux, but then gently tends towards a new mean abundance, about which it fluctuates. This plot demonstrates the importance of sufficiently long simulations as limiting the time axis to \SI{E6}{\second} in this case would erroneously indicate a trend of decreasing \ce{NH3} abundance without approaching an average steady-state. For a handful of abundant species (\ce{H}, \ce{CH3}, \ce{NH3}, and \ce{HNO}) an integration time of to \SI{E6}{\second} --- as was used in V16 and L22 -- would be sufficient for the abundances to approach an average steady-state across all pressure levels. For both cases plotted in \Cref{fig:mftsIPer412H-NH3}, the magnitude of the changes in mole fraction are large, with each both changing by more than an order of magnitude during this part of the simulation. This is not to say that the mole fractions of every species in the atmosphere were enhanced or depleted to such a degree, but to demonstrate that such changes are possible.
    
    \par 
    For all cases discussed in this section there is no compositional change in the deepest levels of the atmosphere (\mbox{$P>$\SI{1e3}{\milli\bar}}), primarily because the UV radiation does not penetrate to this level. However, it is also due to the relationship that $P_j$ and $L_j$ have with the atmospheric pressure: reaction rates generally increase with pressure as the mean free path between particle collisions decreases. Thus, chemistry in the deep atmosphere (at higher pressures) is dominated by equilibrium processes and responds quickly to reverse flare-driven chemistry, so there is little opportunity for flare-driven effects to accumulate. It is for this same reason that the hot planet recovers faster than the cold one.
    \par 
    The panels in \Cref{fig:mftsIPer412H-NH3} each plot abundance $\chi$ at a single pressure level; they represent some of the most affected parts of the atmosphere, but much of it is unaffected. It is clearer to plot the relative change in the abundance $\chi_s$ for a species $s$ versus time for each pressure level $p$,
    \begin{equation}
        \label{eq:mfRelChange}
        \Delta_{s}(t,p) = \frac{\chi_{s}(t,p) - \chi_{s}(0,p)}{\chi_{s}(0,p)} \cdot 100
    \end{equation}
    With $\chi_{s}(0,p)$ being the abundance immediately before any flares are applied. Plotting changes in chemistry this way requires mapping $\Delta_{s}$ to a colour bar. Throughout the course of a simulation of a planet exposed to flares, there are both short/fast and long/slow changes to composition. A logarithmically normalised colour bar is effective at displaying large changes in composition; a linearly normalised one is good for showing small changes precisely. A colour bar which transitions from linear (near $\Delta_{s}=0$) to logarithmic normalisation ($ |\Delta_{s}| \gg 0$) allows visualisation of both kinds of change, and also handles the case of $\Delta_{s}=0$. We choose a colour bar which is normalised linearly between 0 and the $1 \sigma$ and logarithmically between $1 \sigma$ and $2\sigma$ (the 68th and 95th percentiles of $|\Delta_{s}|$ respectively). $1 \sigma$ is labelled on the colour bars in red text in all figures. Values of $|\Delta_{s}| >  2 \sigma$ are clipped. We found that this type of normalisation is particularly necessary constructive analysis of fast-reacting species, which respond strongly to flares on short time-scales, but may also have a gradually changing trends in abundance. Our adoption of this method for plotting changes in composition reflects one of the conclusions of L22 that some species respond rapidly to individual flares, and some species respond cumulatively over time.
    \par
    Note that the data are not normally distributed across pressure and time, so $\sigma$ does not represent the standard deviation of the data; $1\sigma$ and $2\sigma$ are simply selected to represent percentiles of $\Delta_{s}(t)$.
    \par
    In this way, a value of $\Delta_{s} = +100$ denotes that a species $s$ has doubled in abundance at the pressure level $p$, relative to the PFS. However, this means that when making comparisons with these figures, a relatively greater boldness of colour in one does not necessarily mean a greater $\Delta$, as the potentially different normalisations should be considered.
    \par 
    The $1\sigma$ and $2\sigma$ values can be used to quantitatively compare the behaviour of different cases. The $1\sigma$ value is the 68th percentile of the data. A large $1\sigma$ would therefore indicate that $\Delta_{s}$ is large across most of pressure- and time-space, for a given simulation case and species. From this large $1\sigma$ we could conclude that, even neglecting large changes in composition, the abundance of a species $s$ is significantly affected by flares. Conversely, a small $1\sigma$ value would indicate that the composition is typically unchanged, although there could be briefly extreme peak changes. The $2\sigma$ value is the 95th percentile of the data, which includes both overall trends in relative abundance as well as more extreme and brief changes. 
    \par
    Therefore, cases in which $2\sigma$  is large but $1\sigma$ is comparatively small would indicate that the abundance of $s$ has small gradual trends throughout the course of flare activity as well as brief periods of time where there are very large changes --- this could correspond to the case of where a species reacts strongly and rapidly to enhanced UV flux, but is also gradually produced or removed due to its involvement in other (slower) pathways.
    \par
\begin{figure}
        \includegraphics[width=0.95\columnwidth,keepaspectratio]{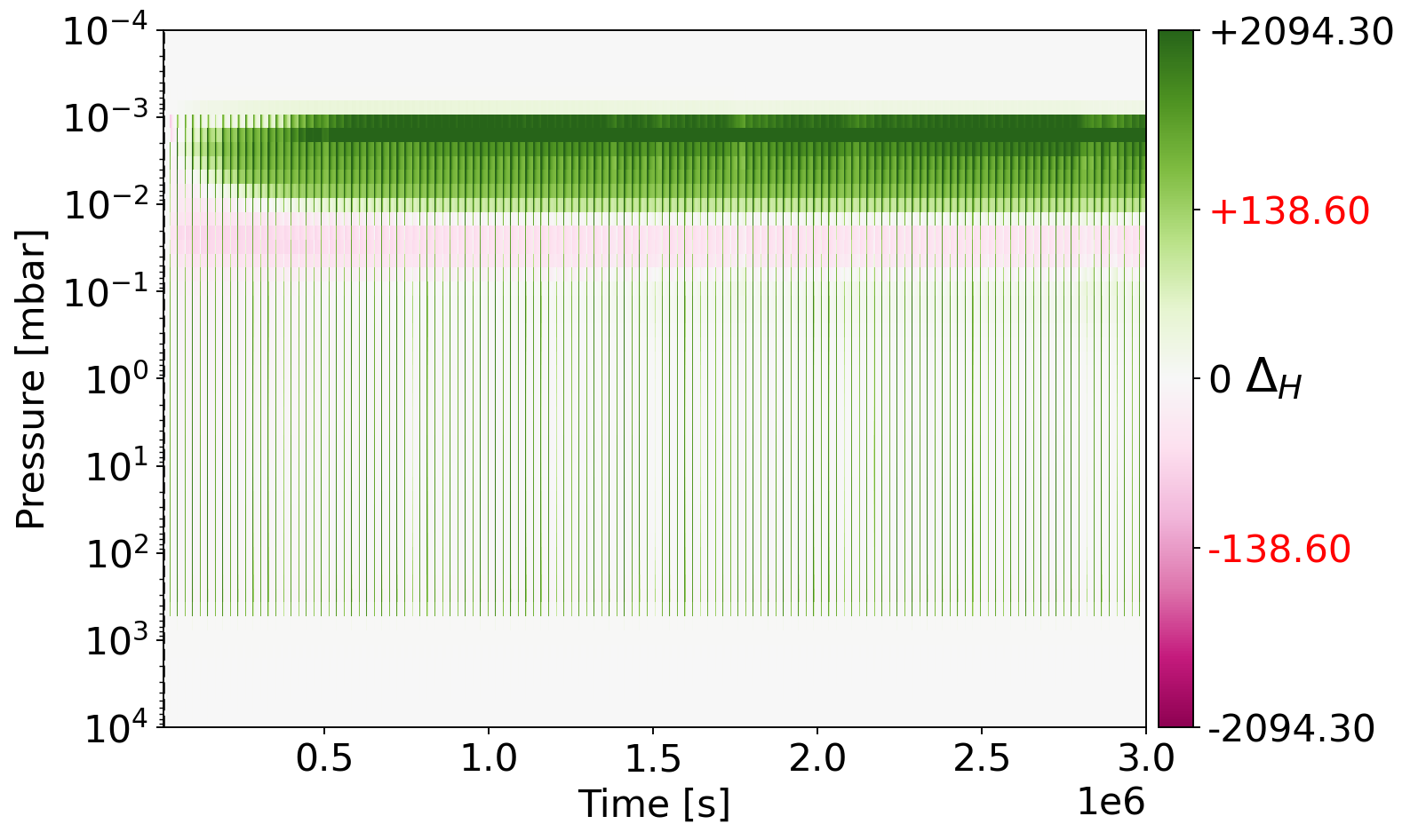}
        \captionof{figure}{$\Delta_{\text{H},p}$ in the cold planet versus time. The system was solved using non-consistently with the periodic flare model. }
        \label{fig:pcmIPer412H}
    \end{figure}

 
    \Cref{fig:pcmIPer412H} plots $\Delta_{s}$ in this way for \ce{H}. Changes in the upper atmosphere persist, and approach a state different to the PFS as indicated by the top panel of \Cref{fig:mftsIPer412H-NH3}. Hydrogen in the deeper atmosphere is briefly affected by flares, but fast reaction rates quickly return it to pre-flare abundances. At pressures greater than \SI{500}{\milli\bar}, changes in the abundance of hydrogen are small due to the large optical depth, and dominating EQ processes.
    \par
\begin{figure}
        \includegraphics[width=0.95\columnwidth,keepaspectratio]{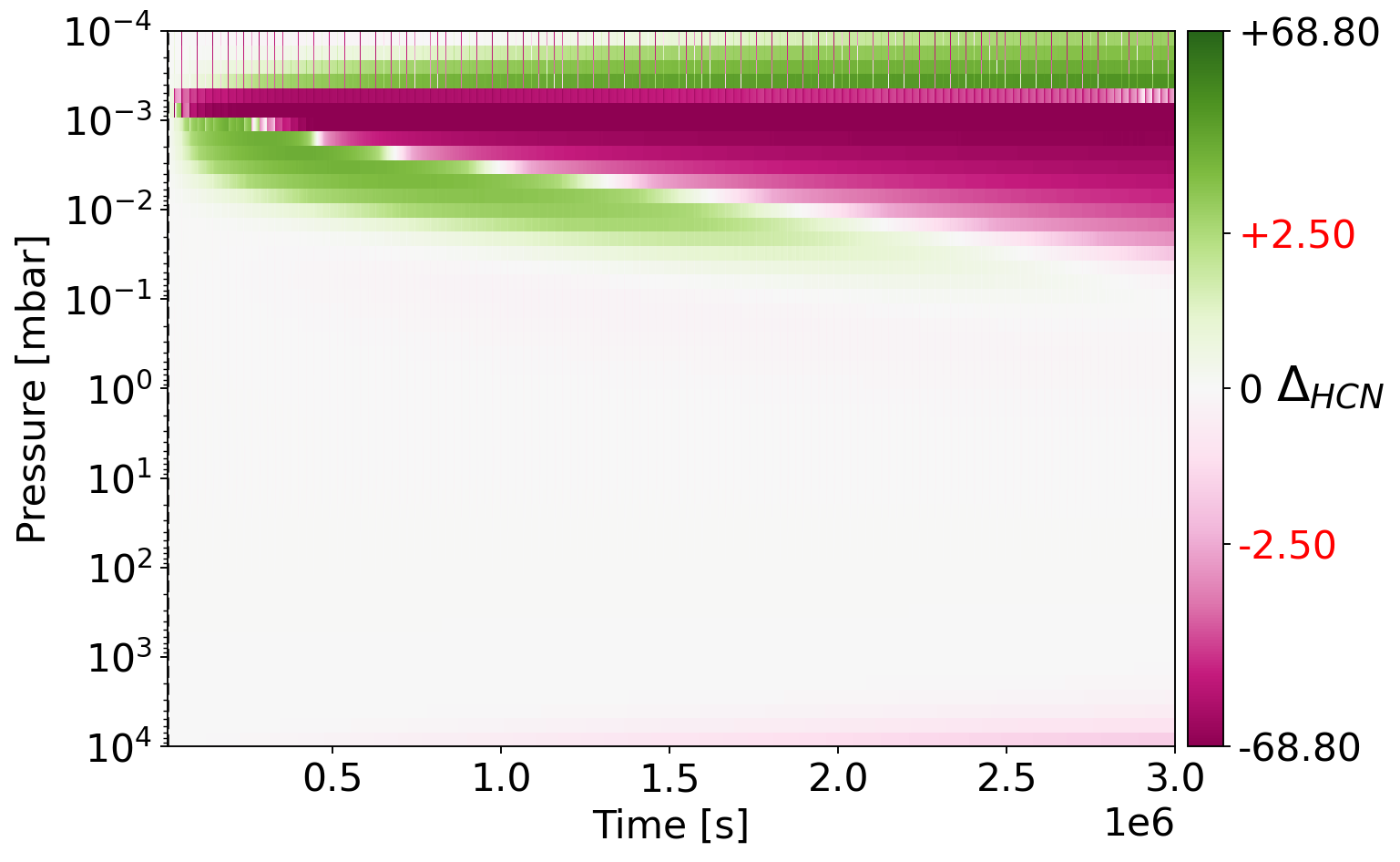}
        \captionof{figure}{$\Delta_{\text{HCN},p}$ in the cold planet versus time. The system was solved non-consistently with the periodic flare model.}
        \label{fig:pcmIPer412HCN}
    \end{figure}

 
    It is common to see depletion of a species in one region and enhancement in an adjacent region. For example, in \Cref{fig:pcmIPer412HCN} one can see that there is initially a decrease in \ce{HCN} abundance at a pressure \mbox{$P_L\sim$\SI{E-3}{\milli\bar}} and a corresponding increase at a higher pressure \mbox{$P_H\sim$\SI{2E-3}{\milli\bar}}. This is explained by the reaction \ce{HCN ->[h\nu] H + CN} \citep{Mizutani1975}. As \ce{HCN} is depleted at $P_L$, H is created (increasing $\chi_{\text{H}}$ at $P_L$), but less radiation of this wavelength is transmitted to deeper levels because it is absorbed at $P_L$ during photolysis. This decreases the photolysis rates of species at deeper levels, and thus increases the abundance of \ce{HCN} at $P_H$. However, gradual depletion of \ce{HCN} at $P_L$ eventually leads to deeper UV penetration, so this behaviour propagates to increasingly higher pressures, explaining the trend in \Cref{fig:pcmIPer412HCN}. Vertical mixing acts to transport species downwards at this point in the atmosphere, but $\partial \phi/\partial z$ is more than a factor of 100 smaller than the other terms in the RHS of Equation \ref{eq:chemCont}, which include photochemical processes. It is possible that increasing the eddy diffusion coefficient $K_{zz}$ would eliminate or modify this banding behaviour. Additionally, the abundance of \ce{HCN} in the atmosphere requires a significant amount of time to adapt to the introduction of flares, demonstrated by the fact that at \mbox{$t=$\SI{3e6}{\second}} it is still evolving.
    \par
\begin{figure}
        \centering
        \includegraphics[width=0.95\columnwidth,keepaspectratio]{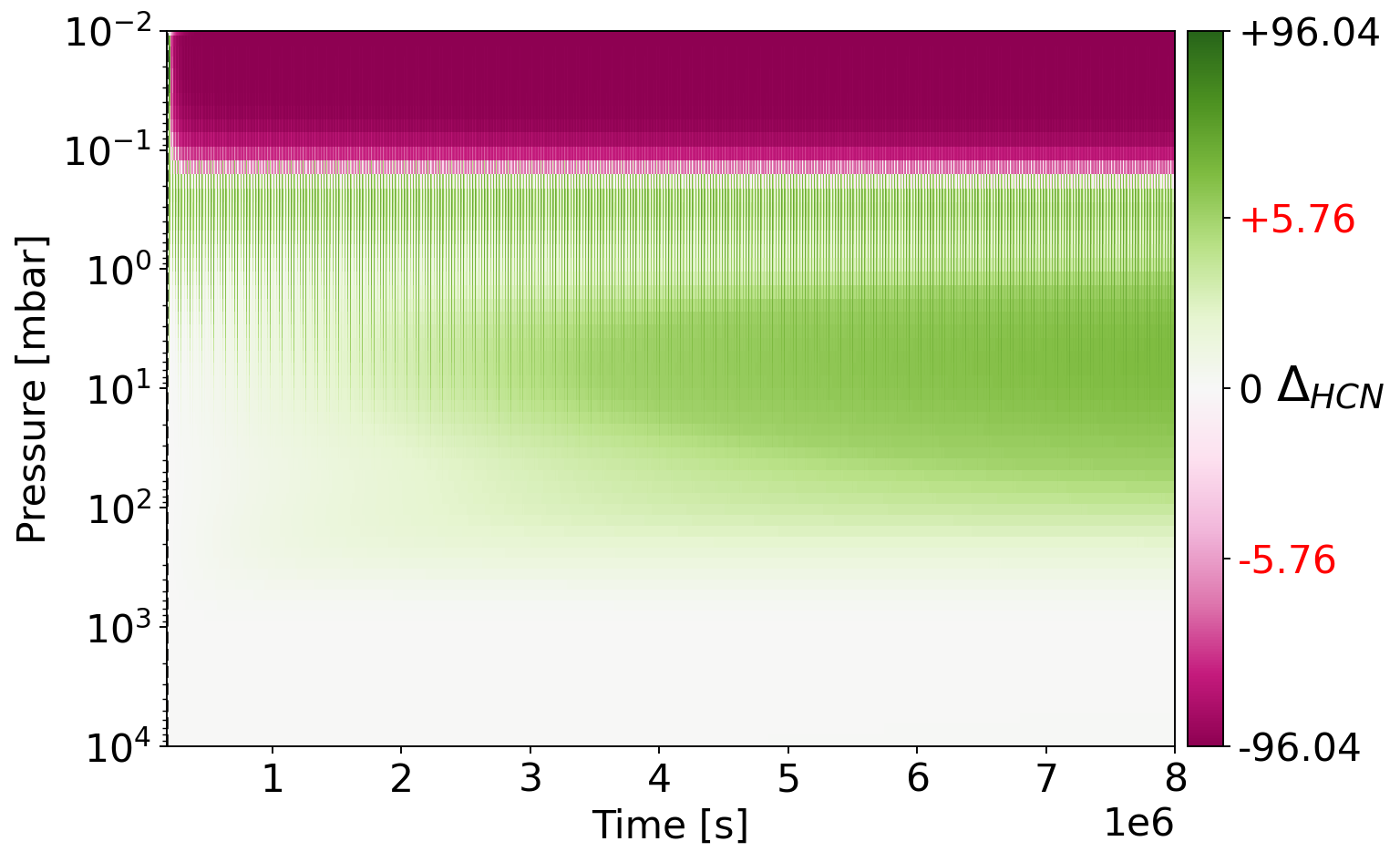}
        \captionof{figure}{$\Delta_{\text{HCN},p}$ in the hot planet versus time. The system was solved non-consistently with the periodic flare model.}
        \label{fig:pcmIPer1632HCN}
    \end{figure}

 
    Although there is similar banding behaviour, the hotter planet approaches an average steady-state in the upper atmosphere much more rapidly than the cooler one, which parallels the findings of K22 (see \Cref{fig:pcmIPer1632HCN}). This is attributed to the significantly reduced chemical time-scales in the high-pressure high-temperature environment. 
    The atmosphere of the hot planet was still evolving at \SI{3e6}{\second}, although with much smaller $1\sigma$ changes compared to the equivalent case of the cold planet ($1\sigma = 0.85$ versus 2.50, up to $t=$\SI{3e6}{\second}). When the simulation stopped at \SI{9e6}{\second} the mole fraction of \ce{HCN} at the top of the atmosphere was no longer evolving with time, but a small high-pressure trend in \ce{HCN} abundance remained.
    
    \subsection{Periodic flares - CNEQ}
    \label{ssec:flaresPerCneq}
    When periodic flares are applied, atmospheres can behave differently when $T(p)$ is repeatedly solved for RCE (CNEQ) compared when $T(p)$ is fixed (NEQ). The exact behaviour is heavily temperature- and species-dependent.
    \par 
    For the purposes of this discussion, we define \textit{temperature anomaly} (TA) as the temperature change of the atmosphere relative to the PFS such that
    \begin{equation}
        \label{eq:temperatureAnom}
        \text{TA}_p(t) = T_p(t) - T_p(0)
    \end{equation}
    Where $T_p(t)$ is the temperature of the atmosphere at pressure $p$ at a time $t$ since flares began (i.e. after any delay period).
    \par
\begin{figure}
        \includegraphics[width=0.95\columnwidth,keepaspectratio]{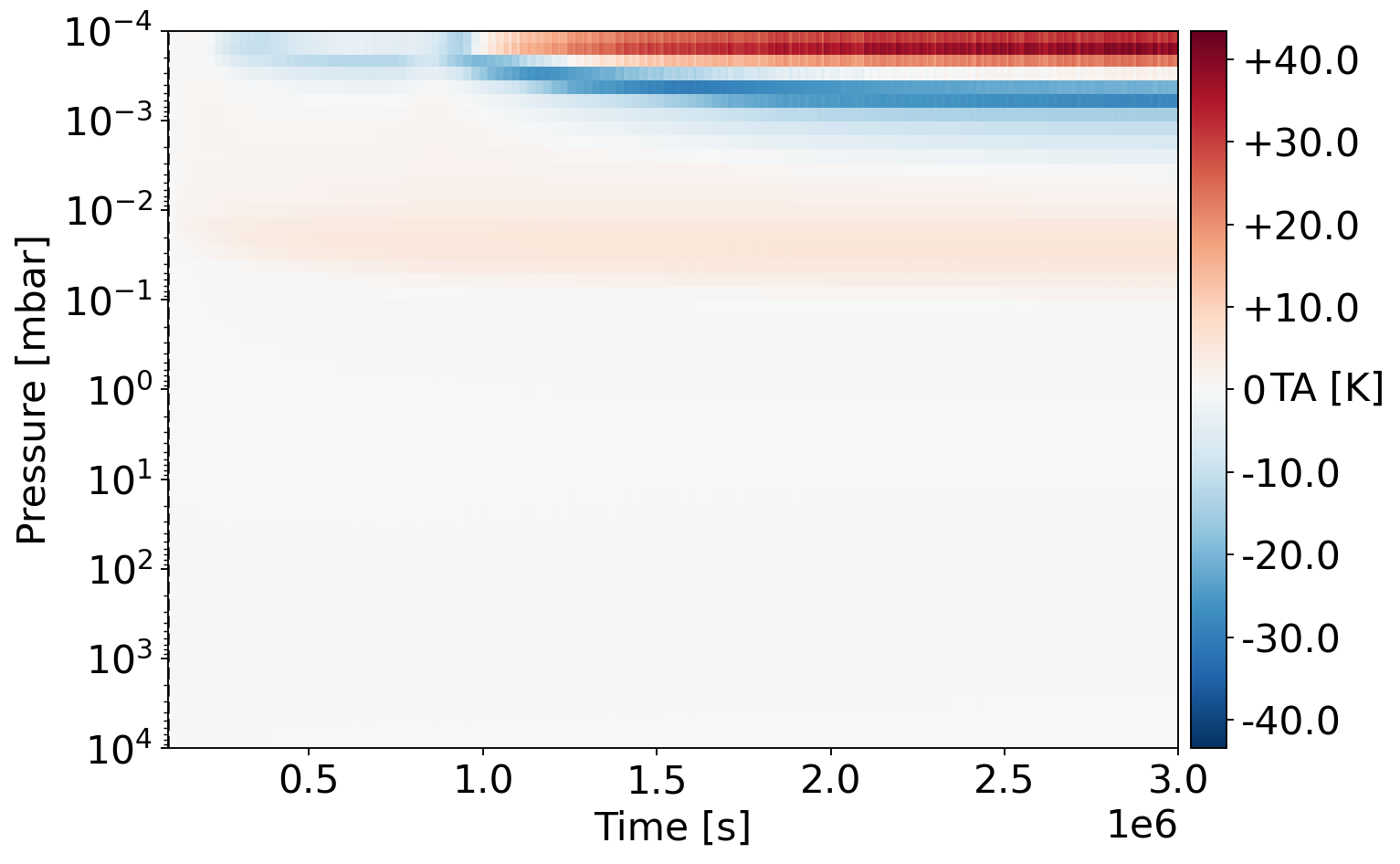}
        \captionof{figure}{Time evolution of temperature anomaly in the cold planet over time. The colour bar is normalised linearly. Solved self-consistently with the periodic flare model.}
        \label{fig:pcmCPer412T}
    \end{figure}

    The time evolution of the TA of the atmosphere of the cold planet is plotted in \Cref{fig:pcmCPer412T}. Shortly after the onset of flares a region of cooling is established at the top of this atmosphere, and a region of heating is established below. The enhanced UV flux acts to quickly change the abundance of species which shape the temperature structure of the atmosphere, which in turn perturbs the radiating temperature. Species which are relevant to this feedback include \ce{H2}, \ce{H2O}, \ce{CO2}, \ce{CH4}, \ce{HCN}, and \ce{C2H2}.
    \par
\begin{figure}
        \includegraphics[width=0.95\columnwidth,keepaspectratio]{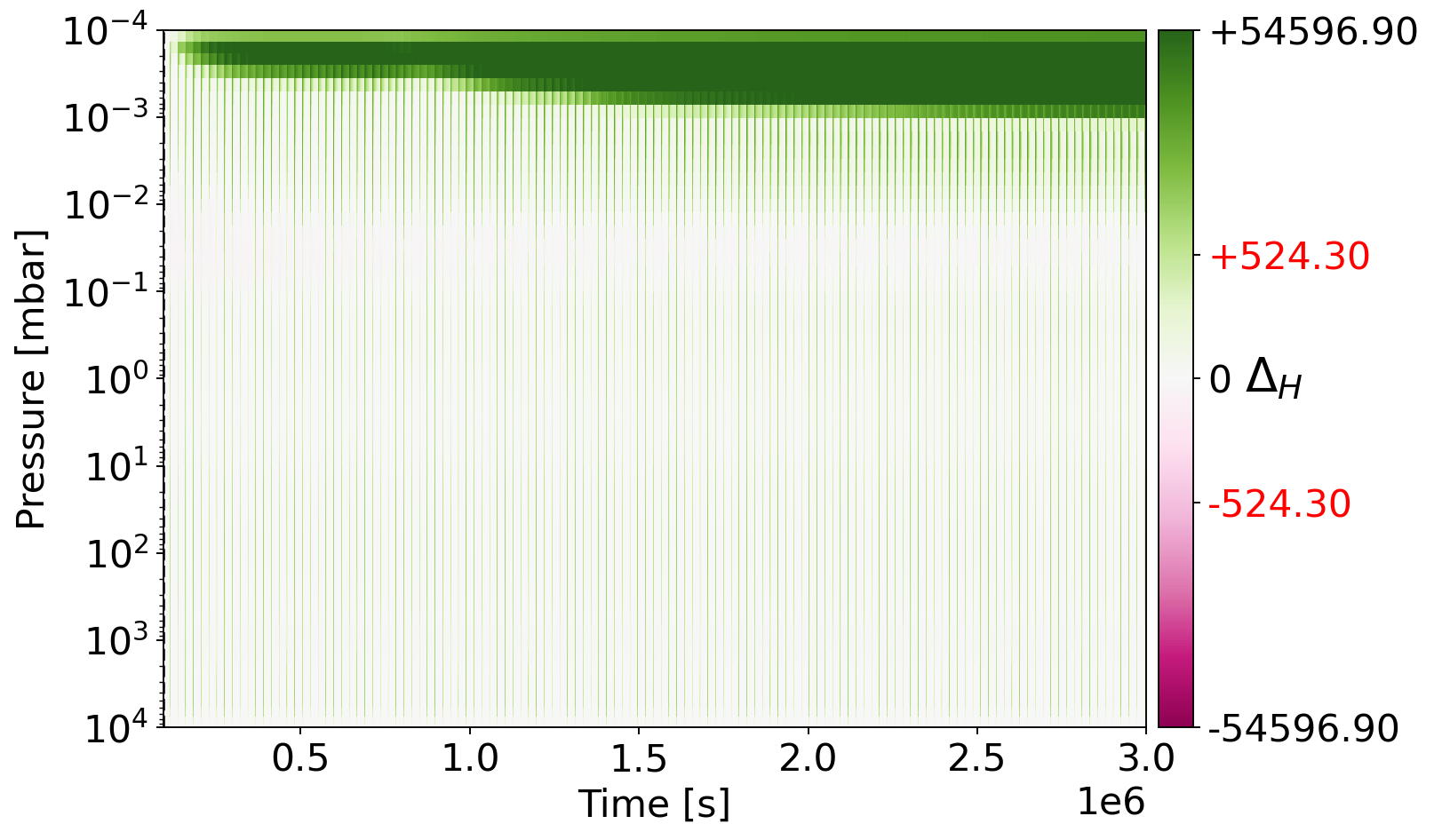}
        \captionof{figure}{$\Delta_{\text{H},p}$ in the cold planet versus time between. Solved self-consistently with the periodic flare model.}
        \label{fig:pcmCPer412H}
    \end{figure}

 
    After some time a tipping point is reached where these two regions of heating and cooling are shifted to higher pressures, with an additional region of heating being formed at the top of the atmosphere. As the TA is driven by the changing opacity of the atmosphere, the emergence of this additional region of heating at \mbox{$t \sim$\SI{e6}{\second}} can be attributed to complex compositional changes at low pressures due to chemical pathways involving photochemically active species. For example, the abundance of \ce{H} (\Cref{fig:pcmCPer412H}) sees a shift in behaviour at the same time that the TA does, indicating that the opacity contribution of both \ce{H} itself and (photo-)chemically related species such as \ce{CH4} and \ce{NH3} contribute to changing behaviour of the TA. The TA in this new heated region initially grows over time as flares are applied. 
    \par
    The temperature of the whole column approaches an average steady-state by \SI{2E6}{\second} where there is a maximum TA on the order of \SI{40}{\kelvin}, which occurs near the lowest pressures of the model. This maximum TA is much larger than the single-flare TA both in this work, and in \citet{Segura2010} ($<8 \text{ K}$), which follows from the prediction in the previous sections that repeated flares could have compounding effects. 
    
    \par 
    The profile of the TA is similar in both cases, with alternating levels of warming and cooling. 
    
    Despite the hot planet receiving more radiant flux from AD Leo, its atmospher absorbs significantly less UV radiation near its upper boundary compared to the cold atmosphere -- in spite of effects such as Doppler line broadening -- due to it being less opaque in the uppermost part of its simulation region. UV-absorbing species (such as \ce{HCN}, \ce{CH4}, and \ce{NH3}) are much less abundant near the upper boundary of the hot planet, while species with smaller UV absorption cross-sections -- such as \ce{H} and \ce{OH} -- are more abundant. This can be seen by comparing panels (c) and (f) of \Cref{fig:tilesQsc}: at \SI{0.01}{\milli\bar} \ce{NH3}, which strongly absorbs UV radiation, differs in abundance by \SI{5}{\dex} between these two planets \citep{Dishoeck2006,Weng2021}. As a result of these differing abundances near the upper boundaries of the atmospheres, the UV radiation is absorbed at higher pressures (and temperatures) in the atmosphere of the hot planet, where the chemical kinetics favours back-reactions which restore photochemically active species, and mitigate the effects of flares on the chemistry. One result of this is that the TA of the atmosphere of the hot planet is negligible ($< 1 \text{ K}$), and the majority of chemical species evolve identically in between the NEQ and CNEQ cases of the hot planet.
    \par
    For \ce{H} in the atmosphere of cold planet, there is a much more dramatic response to periodic flares when the model is evolved self-consistently. Comparison of  \Cref{fig:pcmIPer412H,,fig:pcmCPer412H} shows that in the upper atmosphere the periodicity of flares does not necessarily translate to periodic changes in \ce{H} dominating the chemical response, as it does when the temperature profile is fixed. Due to the chemistry-temperature feedback, we see larger changes in the abundance of H relative to the PFS in the self-consistent case. \ce{H} in this region is produced by photolysis of \ce{C2H2}, \ce{HCN}, and \ce{H2}, which see corresponding decreases in abundance. As a result, the low pressure regions (where H is primarily enhanced) are where the atmosphere's temperature is most significantly affected (\Cref{fig:pcmCPer412T}). 
    \par 
    Many of the differences between the self-consistent and non-consistent results in this work stem from the fact that the quiescent abundance distributions differ between the NEQ and CNEQ cases of a given planet. For example, in the cold planet's self-consistent case the cross-over region to where \ce{H} dominates occurs at lower pressures compared to the NEQ case, meaning that the UV radiation is generally absorbed at lower pressures and compositional changes occur higher-up in the atmosphere. As many reactions are sensitive to partial pressure, this shift plays a role in the evolution associated with flares --- \Cref{fig:pcmIPer412H,,fig:pcmCPer412H} show that \ce{H} sees greatest changes in abundance at differing pressure levels between the NEQ and CNEQ cases.
    \par
\begin{figure}
        \includegraphics[width=0.95\columnwidth,keepaspectratio]{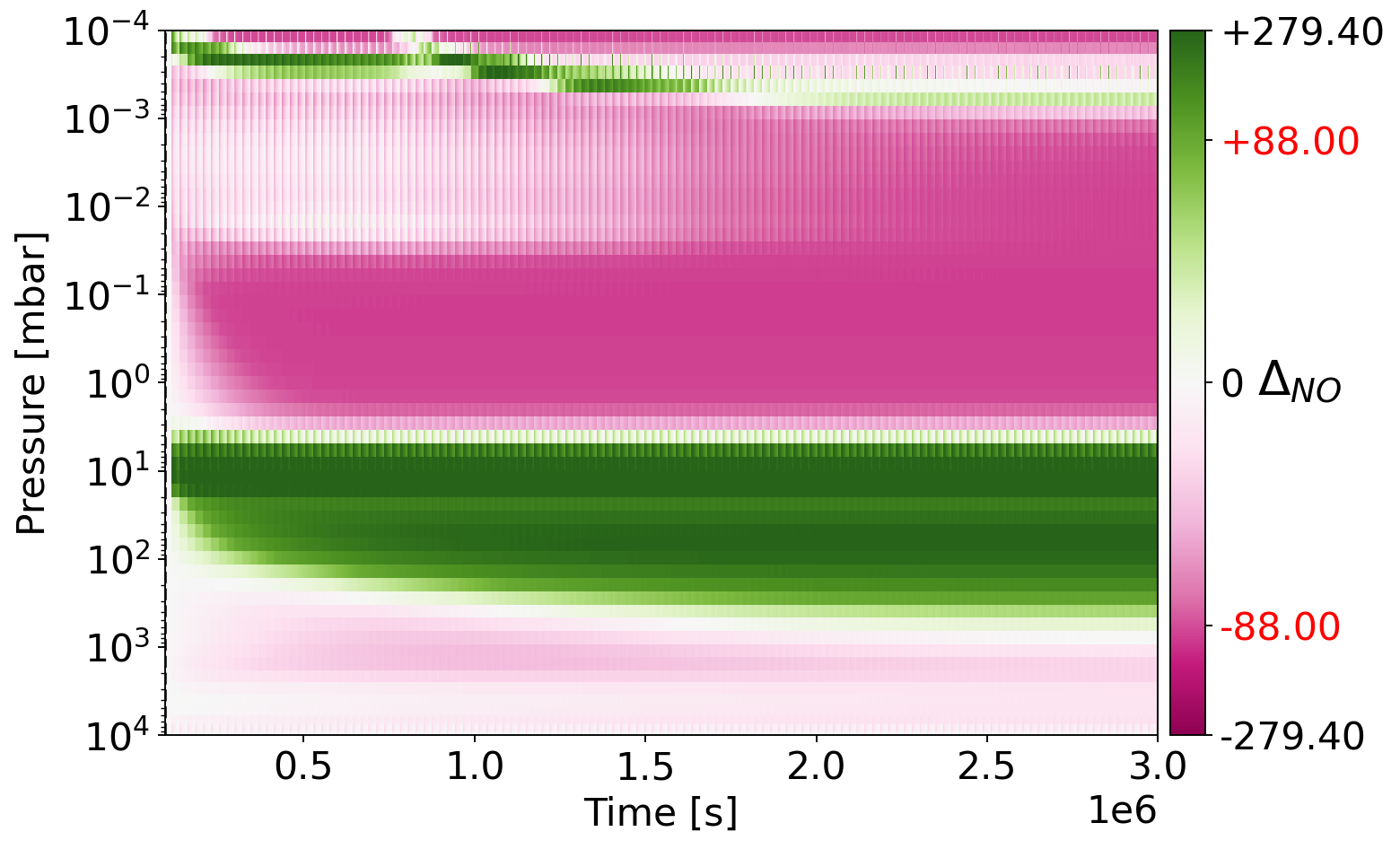}
        \captionof{figure}{$\Delta_{\text{NO},p}$ in the cold planet versus time. Solved self-consistently with the periodic flare model.}
        \label{fig:pcmCPer412NO}
    \end{figure}

 
    Our results (\Cref{fig:pcmCPer412NO}) show NO production at \SI{10}{\milli\bar} and removal at \SI{0.01}{\milli\bar}, similar to the results of V16. \ce{NO} produced at \SI{10}{\milli\bar} diffuses to adjacent regions at higher pressures.
    \par
    In the case of the hot planet, application of periodic flares causes the atmosphere to enter a strong resonant state with a period of approximately 26 times the period at which flares are applied. It is unlikely that this resonant mode would be driven by the more realistic stochastic flaring scheme. The model atmospheres of K22 --- with fixed $T(p)$ ---  entered into a range of oscillatory states with different average steady-state compositions depending on the rate of flare application. Our observation of a long-period resonance is not the same behaviour as seen in K22, and is instead only present when $T(p)$ is coupled self-consistently to the chemistry, so it can be attributed to a feedback between chemistry and the energy redistribution.
    \par 
    The introduction of self-consistent modelling has been shown in this section to influence: a) the inter-flare recovery of the atmosphere, b) the steady-state configuration of the atmosphere in cases where a steady-state exists, and c) the rate at which the atmosphere approaches steady-state. Therefore, it is important to carefully consider the resting period between flares. An assessment of the importance of self-consistent modelling in this context must be done with a reasonably accurate flare model, which can more realistically apply a series of flares over time. This is addressed in the following sections, where a stochastic flaring model adapted from \citet{Loyd2018} was used.
    
    \subsection{Stochastic flares}
    \label{ssec:flaresSto}
    Applying flares stochastically is more representative of the true behaviour of active M-dwarf stars. In response, the general behaviour of the atmospheres is different to the periodic cases, and has less dramatic compositional changes since individual flares are between $10^2$ and $10^6$ times less energetic than GF85 \citep{Loyd2018,Hawley1991}. The large number of species involved and their complex interactions require lengthy discussion, so species of particular interest are selected as case-studies. These species are selected because they demonstrate a variety of responses to flare activity, may be influential in radiative transfer, and may be relevant to observations \citep{Helling2020, Herbst2019}. Some species are hardly affected by the flares at all (e.g. \ce{N2}), while others are not sufficiently abundant to warrant analysis ($\max_p (\chi_s) < 10^{-12}$; e.g. \ce{C2H}). In this section, the time axis of the plots is typically limited to \SI{3e6}{\second} after the onset of flares for ease of comparison with the shortest simulations (which have reduced integration time due to computational limitations; see \Cref{sec:methods}).
    
    \subsubsection{Temperature anomaly}
    \label{sssec:stoTA}
\begin{figure}
        \includegraphics[width=0.95\columnwidth,keepaspectratio]{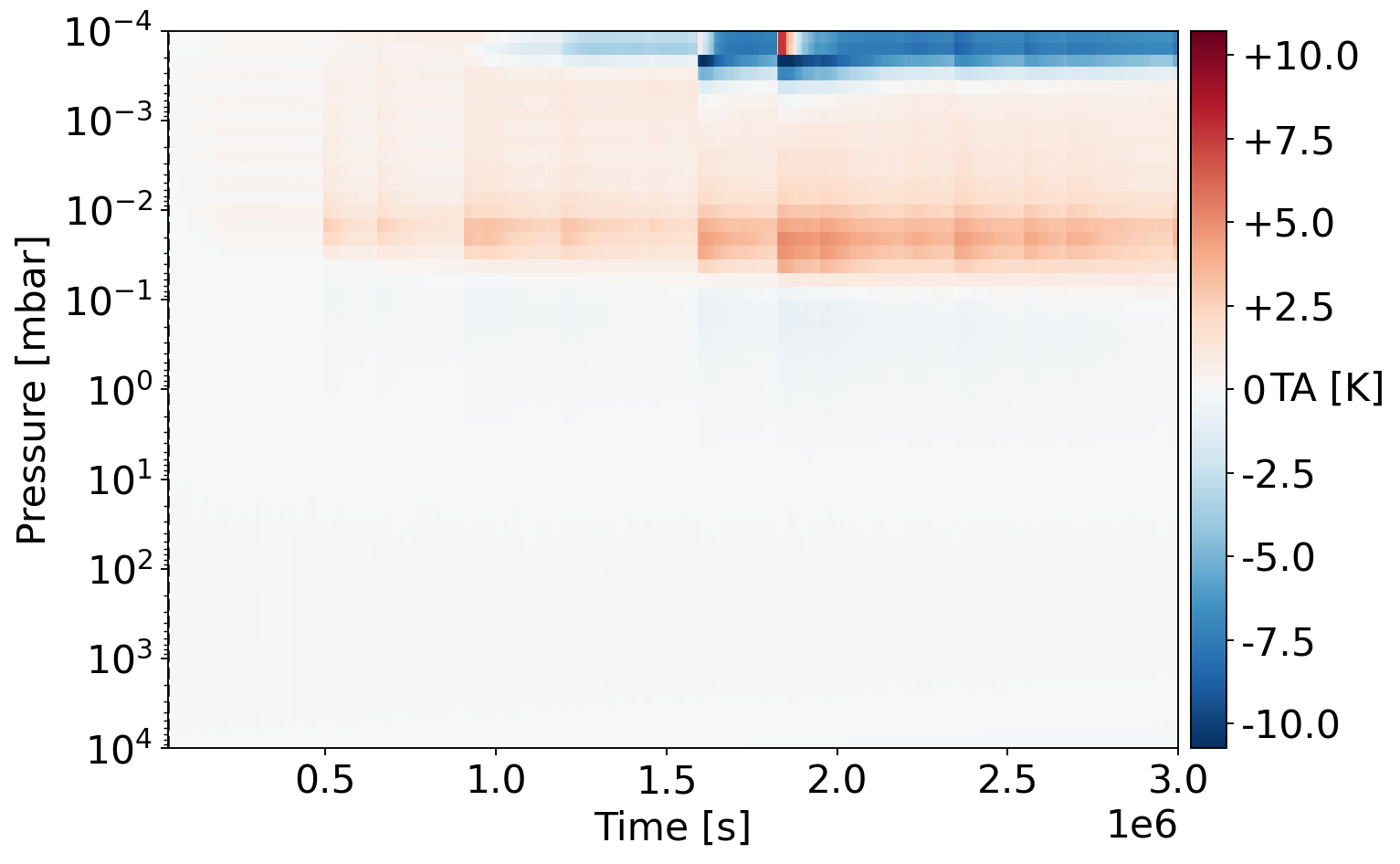}
        \captionof{figure}{TA of the cold planet versus time. Solved self-consistently with a stochastic flare model.}
        \label{fig:pcmCSto412T}
    \end{figure}

    The uppermost part of the atmosphere of the cold planet (\SI{e-4}{\milli\bar}) sees a negative TA, as defined by \Cref{eq:temperatureAnom}, which increases in magnitude over time to \SI{-10}{\kelvin} (\Cref{fig:pcmCSto412T}). There is also a region of $\text{TA} \sim +6 \text{ K}$ at higher pressures (\SI{2e-4}{\milli\bar}). The cooling indicates that the effect of UV absorption is primarily to drive chemistry, as photolysis of opaque species (e.g. \ce{H2O}, \ce{CO2}, \ce{CH4}, \ce{NH3}, \ce{C2H2}) by UV radiation reduces the amount of radiation absorbed in the uppermost part of the atmosphere. This radiation is instead absorbed at higher pressures, causing them to warm. Larger flares lead to larger changes in TA, and the temperature slowly begins to return to its quiescent value ($\text{TA} \rightarrow 0$) between flares and during less active intervals. 
    \par
\begin{figure*}
        \includegraphics[width=2.0\columnwidth,keepaspectratio]{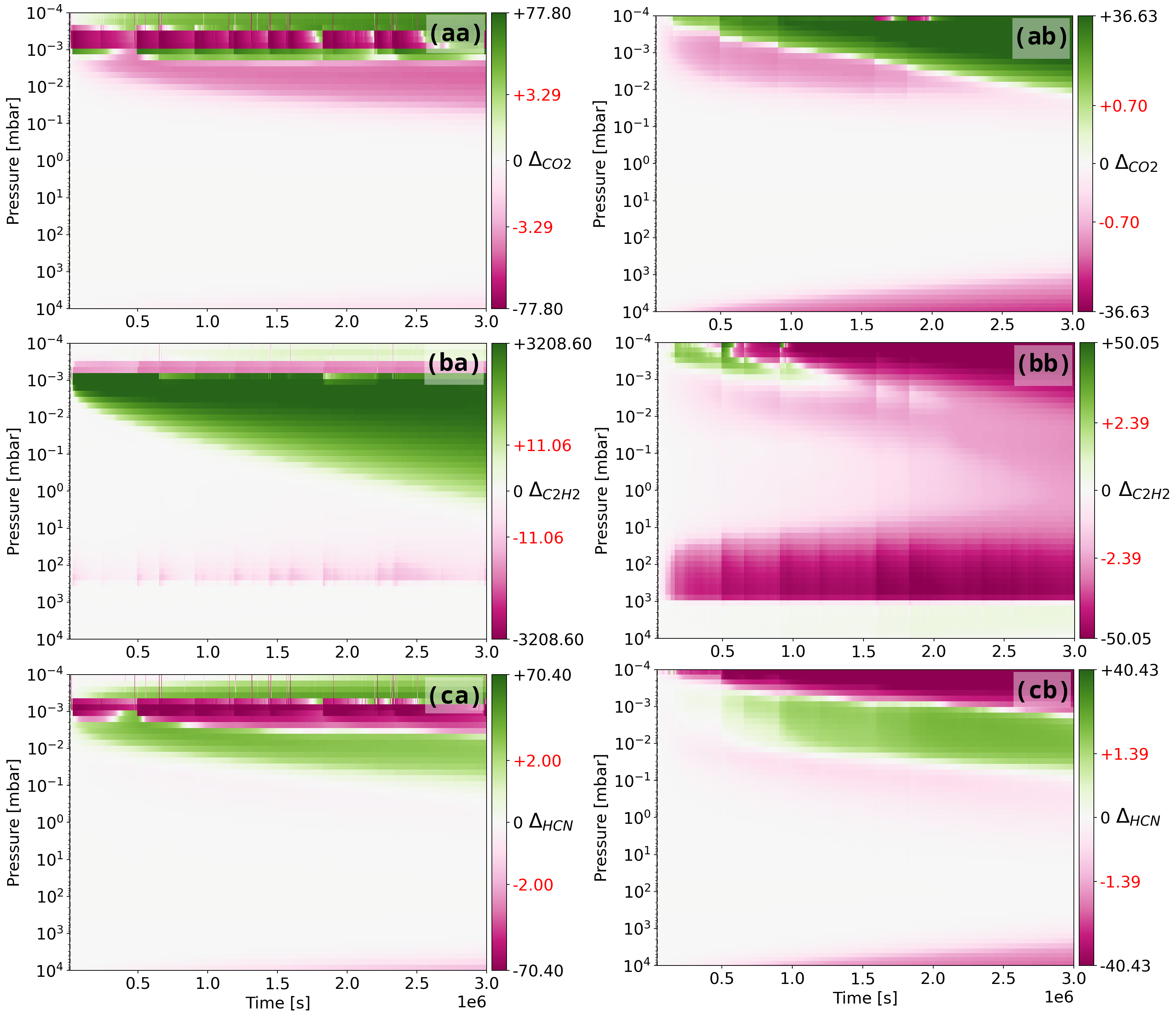}
        \captionof{figure}{$\Delta_{s}$ in the cold planet versus time for $s \in \{\ce{CO2},\ce{C2H2},\ce{HCN}\}$. These simulations used the stochastic flare model. Left column: NEQ; right column: CNEQ.}
        \label{fig:pcmCSto412-1}
    \end{figure*}

 
    In the atmosphere of the cold planet, the magnitude of the TA is generally much smaller across all pressures compared to the case where the periodic flare model was applied. This makes sense given that the flares in the stochastic flare model have less energy and thus perturb the chemistry of the atmosphere less. However, one significant difference between these two simulations is the pressures at which the heating and cooling occur. Shortly after the onset of flares the models behave similarly, with cooling at the top of the column and a region of heating below, but after some time in the periodic case a new region of heating is established at the top of the model (\Cref{ssec:flaresPerCneq}). Under the stochastic flare model, the TA of the atmosphere does approach the same tipping point, but does not undergo the same shift in behaviour. This difference is due to the fact that the flares drive chemistry less strongly and regularly under the stochastic flare model, which does not push the atmosphere towards the same tipping point in behaviour. Similar transitions are seen for some species in the atmosphere of the cold planet; e.g. \ce{C2H2} between \SI{e-4}{\milli\bar} and \SI{e-3}{\milli\bar} plotted in panel (bb) of \Cref{fig:pcmCSto412-1}. While the \SI{-10}{\kelvin} TA is small relative to the absolute temperature of the atmosphere, conservation of energy between flaring events can lead to significantly different chemical responses, which are discussed in the following sections.
    \par 
    The hot planet does not see any significant TA ($< 1 \text{ K}$), as was also true when the periodic flare model was applied (\Cref{ssec:flaresPerCneq}). 
    
    \subsubsection{Carbon dioxide}
    \label{sssec:stoCO2}
    \ce{CO2} is a large source of opacity across a range of wavelengths, so changes in its abundance were predicted to influence the temperature of atmosphere across a range of pressure levels \citep{Rothman1992,Venot2013}. The abundance of \ce{CO2} in the hot planet's atmosphere does not change significantly throughout the flare simulations.
    \par 
    \Cref{fig:pcmCSto412-1} shows the simulated change in \ce{CO2} abundance due to flares driving NEQ and CNEQ chemistry in the atmosphere of the cold planet; panels (aa) and (ab) respectively. Changes in \ce{CO2} abundance are larger when simulated with NEQ chemistry and are approximately in a steady-state by \SI{3e6}{\second}. In contrast, the self-consistent simulation shows that the \ce{CO2} produced at low pressure is mixed down to higher pressures through diffusion. This difference is primarily due to differing pre-flare abundances of \ce{CO2} near the top of the atmosphere between the NEQ and CNEQ cases (\Cref{fig:tilesQsc}). The evolutionary behaviour of the system at some time $t$ depends not only on the UV flux at that time, but also the temperature and composition of the atmosphere at that time $t$. This means that systems which are initialised in different pre-flare states may evolve differently even under otherwise identical conditions. By comparing the results of the NEQ and the CNEQ simulations against each other, for a given planet, we are seeing the influence of both the initial state and the response to the flares (both of which are effected by the use of self-consistent modelling). This logic does not allow us to disregard the role of self-consistency, but it does mean that differing behaviour between NEQ and CNEQ cases is not exclusively due to chemistry-temperature feedback being driven by enhanced UV flux.
    \par 
    In low-pressure regions where \ce{CO2} is removed, it is due to photolysis; the process \ce{CO2 ->[h\nu] CO + O} occurs in the NEQ case of the cold planet at a rate of \SI{6.4e5}{\per\cm\tothe3\per\second} at \SI{7e-4}{\milli\bar}. 
    \par
    Production of \ce{CO2} can be explained through a series of reactions which are enabled by the increased UV flux during active periods. The steps, enumerated below, operate on the understanding that a reaction \ce{A + B <=> C + D} at equilibrium will see an increase in its forward rate if the abundances of C or D are decreased 
    \begin{enumerate}
        \item The abundance of \ce{H2CO} is decreased due to photolysis of \ce{H2CO}.
        \item The rate of \ce{O2 + CH2CHO -> H2CO + OH + CO} is increased to compensate for the loss of \ce{H2CO} (\SI{3.1e2}{\per\cm\tothe3\per\second} at \SI{e-3}{\milli\bar} in the CNEQ case of the cold planet).
        \item The abundance of \ce{O2} is decreased as a result of the previous step.
        \item The rate of \ce{CH3COOO -> 2CH3 + O2 + 2CO2} is increased to compensate for the loss of \ce{O2} (\SI{1.2e1}{\per\cm\tothe3\per\second} at \SI{e-3}{\milli\bar}).
        \item The abundance of \ce{CO2} is increased as a result of the previous step --- note that two carbon dioxide molecules are produced for each oxygen molecule.
    \end{enumerate}
    This is somewhat analogous to Le Chatelier's principle at equilibrium.
    \par

    \subsubsection{Acetylene}
    \label{sssec:stoC2H2}
\begin{figure}
        \includegraphics[width=0.95\columnwidth,keepaspectratio]{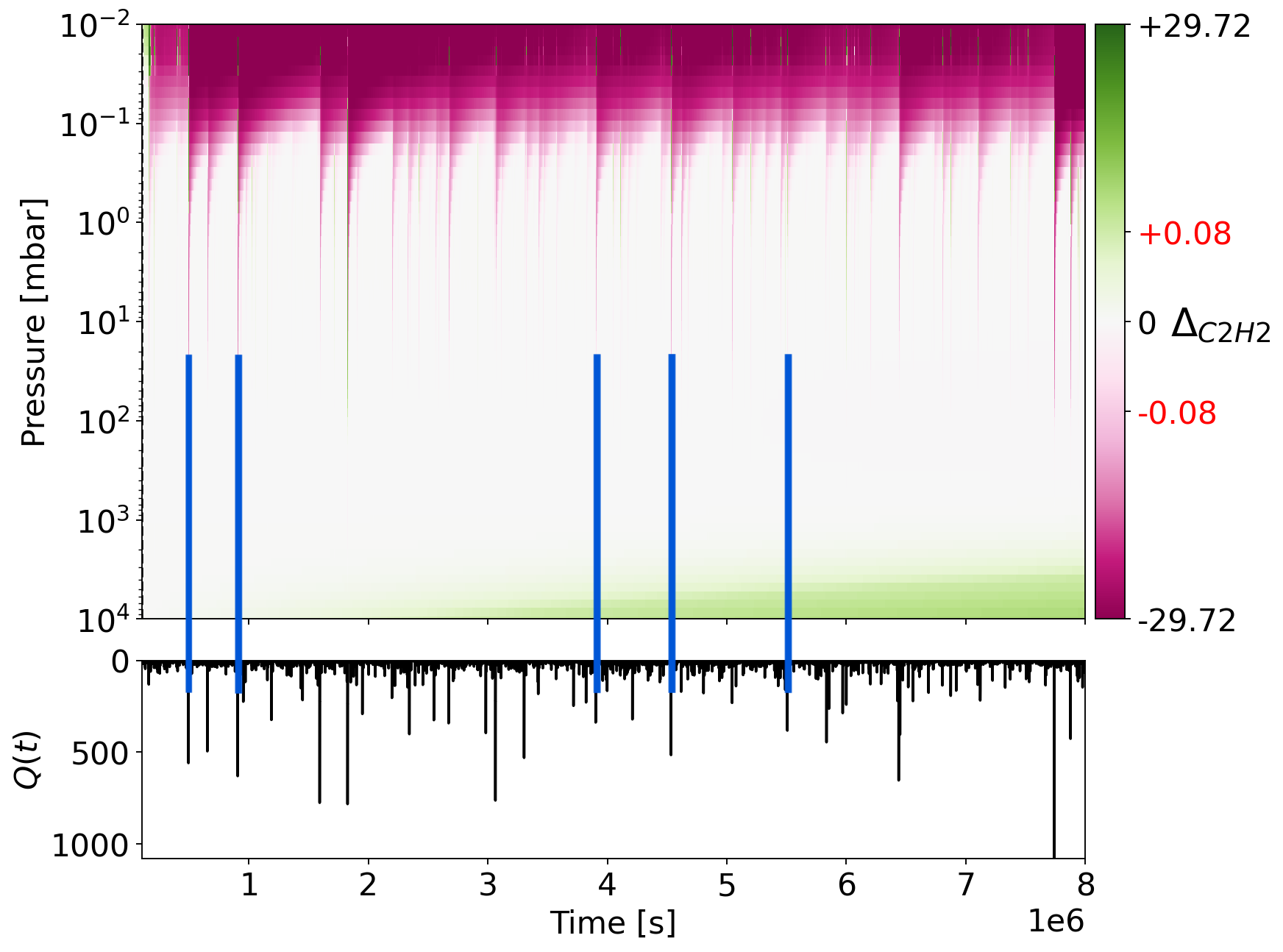}
        \captionof{figure}{$\Delta_{\ce{C2H2},p}$ in the hot planet versus time. Solved non-consistently with the stochastic flare model. Compositional changes are plotted alongside the quiescent flux scale factor $Q(t)$ to demonstrate the relationship between the chemistry and the UV flux. The vertical blue lines are included for ease of comparison so that the fingerprinting of $Q(t)$ on the composition may be visualised. Note that this figure extends the time-axis to \SI{8e6}{\second}.}
        \label{fig:pcmISto1632C2H2}
    \end{figure}

    \Cref{fig:pcmISto1632C2H2} demonstrates the relationship between the abundance of acetylene and UV flux  in the hot planet. From this plot it is possible to verify that the chemistry is driven most strongly when flares occur, which also follows from the single-flare responses discussed in \Cref{ssec:flaresSglNeq} --- the vertical blue lines indicate this for the cases of five flares. \Cref{fig:pcmISto1632C2H2} also shows that the abundance of \ce{C2H2} in the hot planet reaches an average steady-state very rapidly when simulated with NEQ chemistry. Once established, the atmosphere maintains this state for the duration of flare activity. Evolution about this average steady-state is on short time-scales as the atmosphere responds to individual flares. The application frequency of the flares is sufficient to prevent the abundance from returning to the PFS.
    \par 
    In the case of the cold planet (panel (ba) of \Cref{fig:pcmCSto412-1}), the abundance of acetylene does not tend towards a steady-state across much of the atmosphere when simulated with NEQ chemistry. This comparatively slow response is primarily because the chemical time-scales in the atmosphere of the colder planet is much longer compared to the hot planet due to the typical temperatures involved, although the opacity of the atmosphere also plays a role.
    
    \par 
    In the CNEQ case plotted in panel (bb) of \Cref{fig:pcmCSto412-1}, the atmosphere sees removal of \ce{C2H2} across all pressure levels between \SI{e3}{\milli\bar} and \SI{e-4}{\milli\bar}, with more than 50\% of the acetylene being removed. This is in contrast to the NEQ case (ba), where $\chi_{\ce{C2H2}}$ is enhanced between \SI{e-3}{\milli\bar} and \SI{e1}{\milli\bar} (with its abundance increasing by more than a factor of 37). 
    \par 
    Panel (ba) of \Cref{fig:pcmCSto412-1} shows that acetylene is initially produced near \mbox{$p=$\SI{e-3}{\milli\bar}}. This production then extends to deeper levels as various species (e.g. \ce{H2O}) are removed by the UV, which enables the UV radiation to penetrate deeper. The main pathway for formation of acetylene is \ce{C2H + H -> C2H2} which, as well as being the reverse pathway to the photolysis of acetylene, is enabled by the photolysis of various hydrogen-rich species such as \ce{NH3}, \ce{CH3}, and \ce{H2O}. Following the logic in \Cref{sec:introduction} and in \cite{Drummond2016}, these reactions explain why the NEQ case and CNEQ cases differ: species which play a significant role in the radiative transfer through the column are being affected by the flares, and without re-adjusting to RCE this is requiring that the chemistry to compensate in the NEQ case. \ce{C2H2} is also transported to pressures greater than \SI{e-3}{\milli\bar} by diffusion (occurring at a rate of \SI{-1.0e3}{\per\cm\tothe3\per\second} at \mbox{$p=$\SI{5e-3}{\milli\bar}}, a factor of 83 larger than the rate of \ce{C2H + H -> C2H2}).
    \par
    The CNEQ simulation initially sees production of acetylene in the upper atmosphere, as per the NEQ result. After some time, photolysis dominates and \ce{C2H2} is broken down by the UV radiation predominantly into \ce{C2H} and \ce{H} at a rate of \SI{+1.1e+2}{\per\cm\tothe3\per\second} at \mbox{$p=$\SI{2e-4}{\milli\bar}}. The difference between the NEQ and CNEQ cases is partially because the two cases do not have the same initial abundances of other related species, which are evolving alongside the photochemistry of \ce{C2H2}. Regions where evolution differs most between these two cases is also where the temperature is most affected in the CNEQ case.
    \par
    Although they appear differently upon inspection due to colour-bar normalisation, the regions where acetylene is removed in both the NEQ and CNEQ cases --- around \SI{2e2}{\milli\bar} --- show changes in abundance of a similar amplitude: $\Delta_{\ce{C2H2}}$ at \mbox{$t=$\SI{2.5e6}{\second}} being equal to -13 and -41 respectively. 
    
    \subsubsection{Hydrogen Cyanide}
    \label{sssec:stoHCN}
\begin{figure}
        \includegraphics[width=0.95\columnwidth,keepaspectratio]{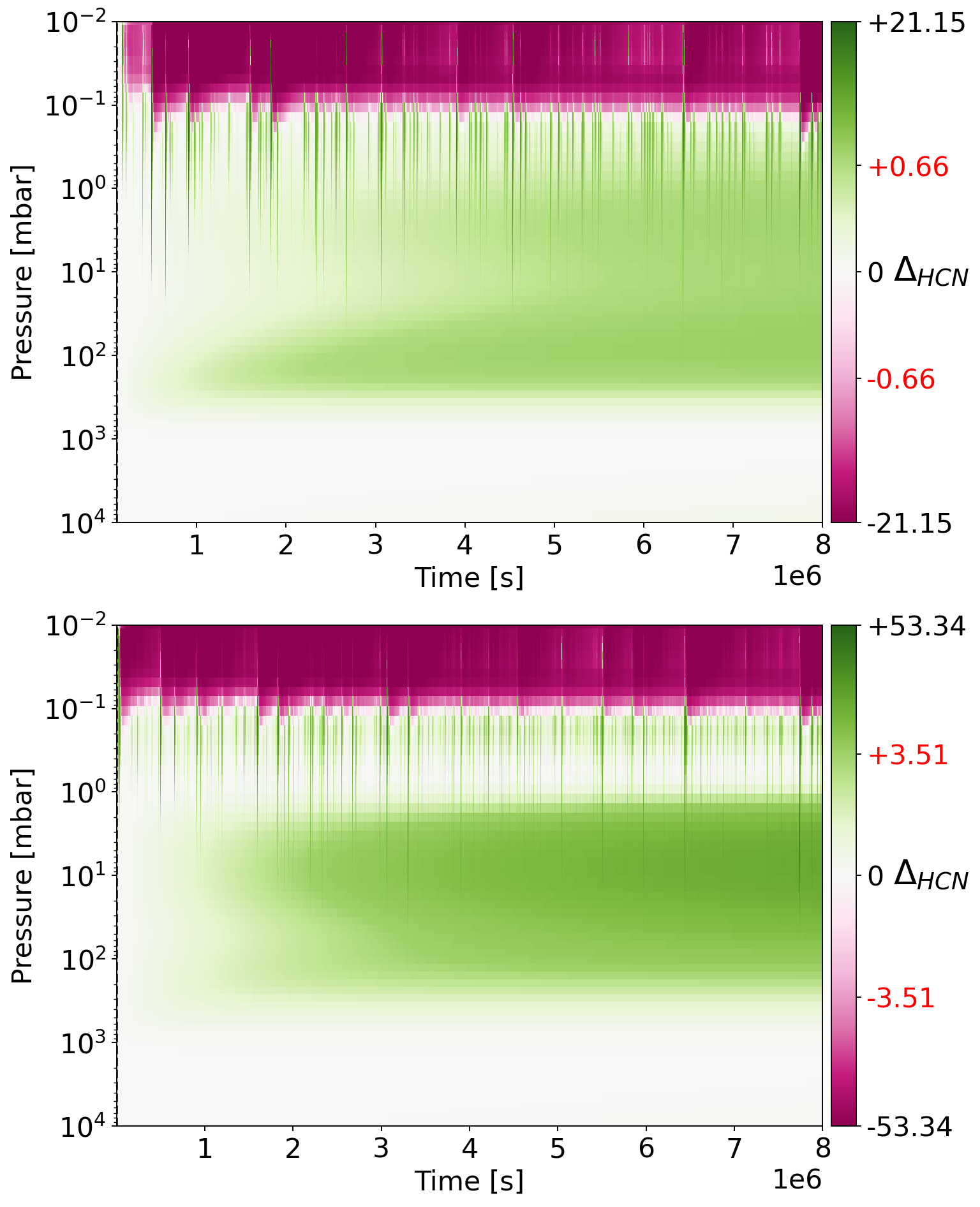}
        \captionof{figure}{$\Delta_{\ce{HCN},p}$ in the hot planet versus time, with a stochastic flare model. Top panel: NEQ; bottom panel: CNEQ.}
        \label{fig:pcmSSto1632HCN}
    \end{figure}

    In the case of the hot planet, \ce{HCN} is quickly depleted by photolysis for \mbox{$p<$\SI{1e-1}{\milli\bar}}, where it rapidly approaches an average steady-state. The abundance of \ce{HCN} in the NEQ simulation is plotted in the top panel of \Cref{fig:pcmSSto1632HCN} for the case of the hot planet. At higher pressures, \ce{HCN} sees a low-amplitude trend of increasing abundance as the simulation evolves. The trend does not correspond closely to flaring events, so it can be attributed to the depletion region in upper atmosphere allowing enhanced UV radiative-transfer to higher pressure levels. This behaviour in the high-pressure regime is common across many species in the case of the hot planet, but is not seen in the case of the cold planet. The hot planet is stable to convection in this subadiabatic regime (see \Cref{fig:tilesQsc}(d)) according to the Schwarzschild stability criterion, and the pattern is not typical of diffusive mixing, so this behaviour is most likely a chemical process. This can be verified by comparing the diffusive and chemical rates: at \SI{e1}{\milli\bar}, the rate of \ce{HCNH -> HCN + H} (which is the dominant reaction involving \ce{HCN}) is \SI{1.3e2}{\per\cm\tothe3\per\second} compared to \mbox{$\partial \phi/ \partial z =$\SI{-3.3e0}{\per\cm\tothe3\per\second}}. The trend of increasing \ce{HCN} abundance at high pressures exists for both CNEQ and NEQ simulations of the hot planet with approximately equal amplitudes.
    \par 
    In the case of the hot planet, the depletion region reaches a steady-state much faster and with greater $\Delta_{\ce{HCN}}$ when simulated self-consistently --- compare the top and bottom panels of \Cref{fig:pcmSSto1632HCN}. The high-pressure trend of \ce{HCN} production is also much stronger in the atmosphere of the hot planet compared to the cold one. 
    
    It was deduced in \ref{sssec:stoCO2} that the differing evolutionary behaviours of \ce{CO2} between NEQ and CNEQ simulations are in some cases partially determined before any flares are applied. Simulations of \ce{HCN} in the hot planet indicate that the differences between these two cases are indeed due to both the PFS of the atmosphere but also its response to the increased UV radiation. They cannot be exclusively a result of differing pre-flare states, as the quiescent states used for NEQ and CNEQ simulations are identical in the case of the hot planet.
    \par 
    Panel (ca) of \Cref{fig:pcmCSto412-1} shows that, for the cooler atmosphere, the impact of any particular flare does not play an important role in the evolution of the abundance of \ce{HCN}, and a trend persists smoothly. The abundance of \ce{HCN} is enhanced near \SI{7e-3}{\milli\bar}, and is transported via diffusion processes at a rate $\partial \phi/ \partial z$ comparable to the chemical kinetics. It is possible that the smooth and continuous trend of \ce{HCN} at high pressures would not exist in 2D or 3D models, according to the findings of K22.
    \par 
    For the evolution of \ce{HCN} in the cold planet, the NEQ and CNEQ simulations (panels (ca) and (cb) of \Cref{fig:pcmCSto412-1}) share similar features, although the position of the depletion region -- and hence production region below it -- differs as a result of differences in quiescent states, as discussed for \ce{CO2}. In the CNEQ case, the depletion and production regions propagate to increasingly deeper levels, as the system evolves, as a result of photochemistry. In the CNEQ case, diffusion acts to transport \ce{HCN} upwards, so these trends are entirely chemical. In the NEQ case, the number of levels which see production initially increases over time, but by \SI{3e6}{\second} is significantly slowed. These differences mean that the $1\sigma$ and $2\sigma$ changes to $\chi_{\ce{HCN}}$ are smaller in the CNEQ case, as the effects of photochemistry are distributed across more levels.
    \par 
    
    \begin{figure*}
        \includegraphics[width=2.0\columnwidth,keepaspectratio]{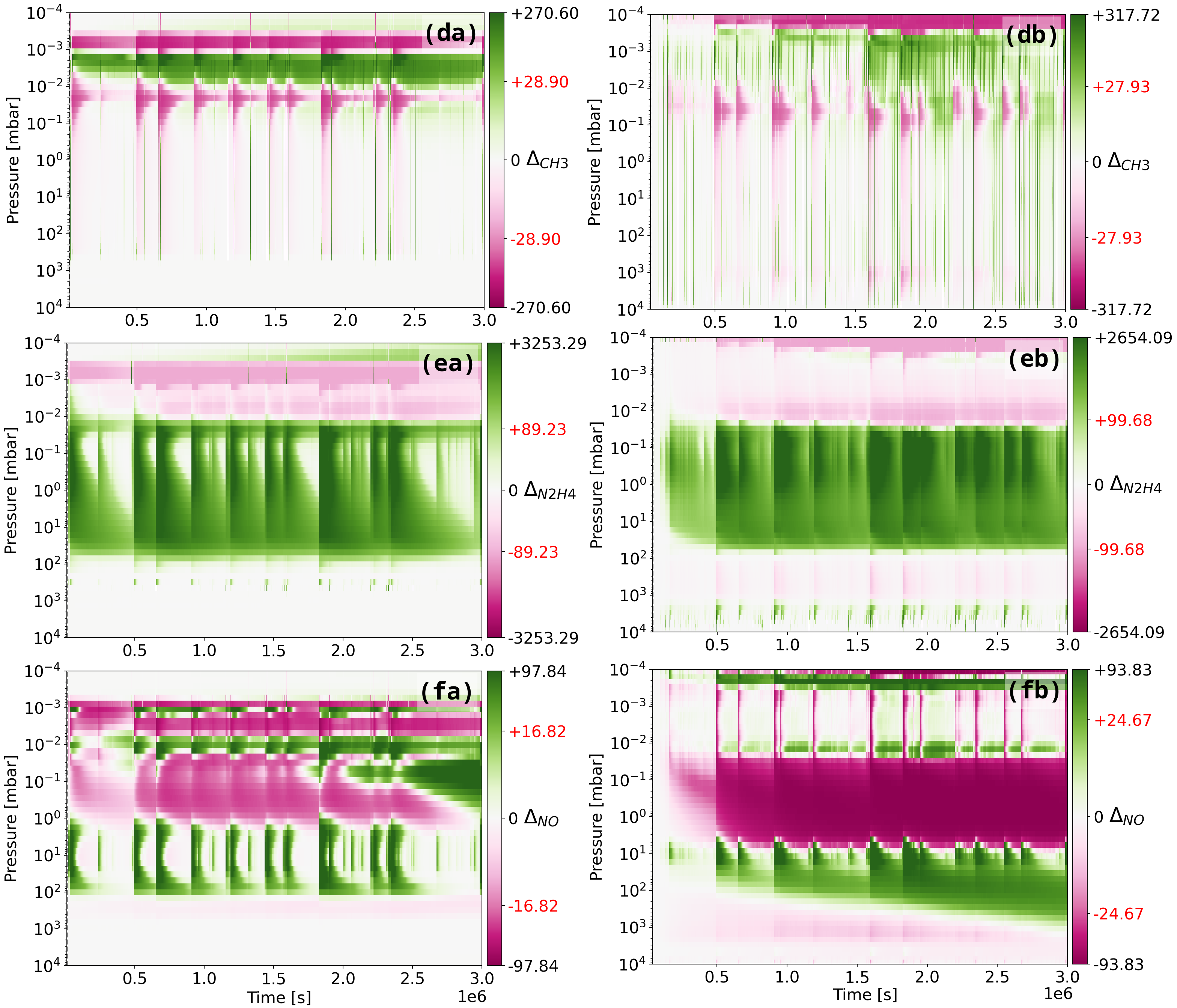}
        \captionof{figure}{Continuation of \Cref{fig:pcmCSto412-1} for species $s \in \{\ce{CH3},\ce{N2H4},\ce{NO}\}$.}
        \label{fig:pcmCSto412-2}
    \end{figure*}

    \subsubsection{Methyl group}
    \label{sssec:stoCH3}
    Primarily due to the photodissociation of \ce{CH4}, the abundance of \ce{CH3} in the cold planet is strongly affected by flares when simulated with NEQ chemistry (plotted in panel (da) of \Cref{fig:pcmCSto412-2}). The amount of \ce{CH3} produced or removed, pressure depending, is tied to the amplitude of the flares. Between high-energy flare events, the atmosphere begins to return to its PFS. This is in stark contrast to \ce{HCN} and \ce{C2H2} in the cold planet, which have consistent trends and no clear `fingerprinting' on $\chi$ by particular flares. This difference is best explained by the fact that reactions involving \ce{CH3} have much higher rates than those involving \ce{HCN} and \ce{C2H2}. Despite \ce{CH3} approaching its PFS between flares in panel (da) of \Cref{fig:pcmCSto412-2}, its abundance fluctuates with significant magnitude, temporarily increasing by factor of $\sim 3.7$ in the upper atmosphere.
    \par 
    Most species behave differently when simulated self-consistently. However, $\Delta_{\ce{CH3}}$ does not differ much between the NEQ and CNEQ cases: compare the panels (da) and (db) of \Cref{fig:pcmCSto412-2}. At \SI{2e-2}{\milli\bar}, \ce{CH3} is depleted at times corresponding to moderate and large flare events, with its abundance returning to its PFS value during less active periods. The dominant reaction pathway contributing to this behaviour is \ce{CH3 + H -> CH4} (at a rate of \SI{+1.1e+4}{\per\cm\tothe3\per\second}), which is enabled by the rapid photolysis of hydrogen-bearing species such as \ce{NH3 ->[h\nu]  NH2 + H} (at a rate of \SI{+2.3e+5}{\per\cm\tothe3\per\second}). Self-consistent modelling makes little difference to these chemical processes as they occur on much faster time-scales than energy transport.
    \par 

    \subsubsection{Hydrazine}
    \label{sssec:stoN2H4}
    Production of hydrazine (\ce{N2H4}) is significant at high pressures (see panel (ea) of \Cref{fig:pcmCSto412-2}), with the abundance of \ce{N2H4} increasing by more than a factor of 30 when UV activity peaks in the NEQ case. Although photolysis of hydrazine occurs at low-pressures, it sees a net production across many pressure levels due to photolysis products of other species reacting together to form hydrazine. At \mbox{$p=$\SI{3e-2}{\milli\bar}}, this is primarily via \ce{2NH2 -> N2H4} (at a rate of \SI{+1.3e+4}{\per\cm\tothe3\per\second}) enabled by \ce{NH3 ->[h\nu] NH2 + H} (at a rate of \SI{+1.3e+5}{\per\cm\tothe3\per\second}). Net production of hydrazine peaks when the most intense flares occur, and then between these events the abundance of \ce{NH3} begins to return to the PFS at a pressure-dependent rate. 
    \par 
    As with \ce{CH3}, restoration of \ce{N2H4} abundance towards its PFS between flares follows from the post-flare behaviour predicted in \Cref{sec:introduction} and discussed in \Cref{ssec:flaresSglNeq}. In comparison, self-consistent solving (plotted in panel (eb) of \Cref{fig:pcmCSto412-2}) slows the return to quiescent abundances, especially at pressures between  \SI{e-2}{\milli\bar} and \SI{e2}{\milli\bar}. Therefore, the effects of self-consistently solving for the abundance of \ce{N2H4} mean that:
    \begin{itemize}[labelwidth=1mm,itemindent=!]
        \label{itm:stoN2H4_effects}
        \item the abundance of \ce{N2H4} in the atmosphere stays semi-permanently in a state determined both by the flare intensity and occurrence frequency,
        \item these changes in abundance relative to the PFS occur near the pressure levels typically probed by transmission spectroscopy\footnote{see Figure 2 of \cite{Sing2015}},
        \item the effects of sequential flares compound to a greater degree in the CNEQ case, where the $1\sigma$ change in $\Delta_{\ce{N2H4}}$ is large.
    \end{itemize}
    \code{ATMO} does not include the opacity of hydrazine in its radiative transfer scheme. However, chemically-related species such as \ce{NH3} and \ce{H} do present features in the transmission spectra.
    \par 

    \subsubsection{Nitrogen monoxide}
    \label{sssec:stoNO}
    \Cref{fig:pcmCSto412-2} plots the evolution of the abundance of \ce{NO} in the atmosphere of the cold planet for the NEQ and CNEQ cases (panels (fa) and (fb) respectively). A prominent feature of both plots is the production region at approximately \SI{10}{\milli\bar}, which is as a result of the decomposition \ce{HNO -> H + NO} (\SI{1.8e+03}{\per\cm\tothe3\per\second} at \mbox{$p=$\SI{5e1}{\milli\bar}}), corresponding to depletion of \ce{HNO} near this pressure level. This production region begins to return to the abundances of the PFS much faster in the NEQ case, while the CNEQ case is much more `smeared' across time, similar to the behaviour seen in the production of \ce{N2H4} (\Cref{sssec:stoN2H4}). \ce{HNO} produced at this level is diffusively mixed to higher pressures at a rate $\partial \phi/\partial z$ approximately equal to half of the rate of its production $P$. The same can be said for the depletion region of \ce{NO} at approximately \SI{0.1}{\milli\bar}.
    \par
    The chemical inertia of the system can cause the atmosphere to overshoot its pre-flare abundance in less active periods as reaction rates readjust to the now reduced UV irradiation (c.f. \Cref{ssec:flaresSglNeq}). As a result, the $1 \sigma$ value of $\Delta_{\ce{NO}}$ is larger in the CNEQ case as flares compound upon each other more constructively. 
    \par 
    The $2 \sigma$ value of $\Delta_{\ce{NO}}$ represents the peak changes in composition. These intervals are brief, so the energy conservation afforded by self-consistent modelling plays a less important role. It follows that the $2 \sigma$ values are similar between the NEQ and CNEQ cases (panels (fa) and (fb) of \Cref{fig:pcmCSto412-2}).
    \par
\begin{figure}
        \includegraphics[width=0.95\columnwidth,keepaspectratio]{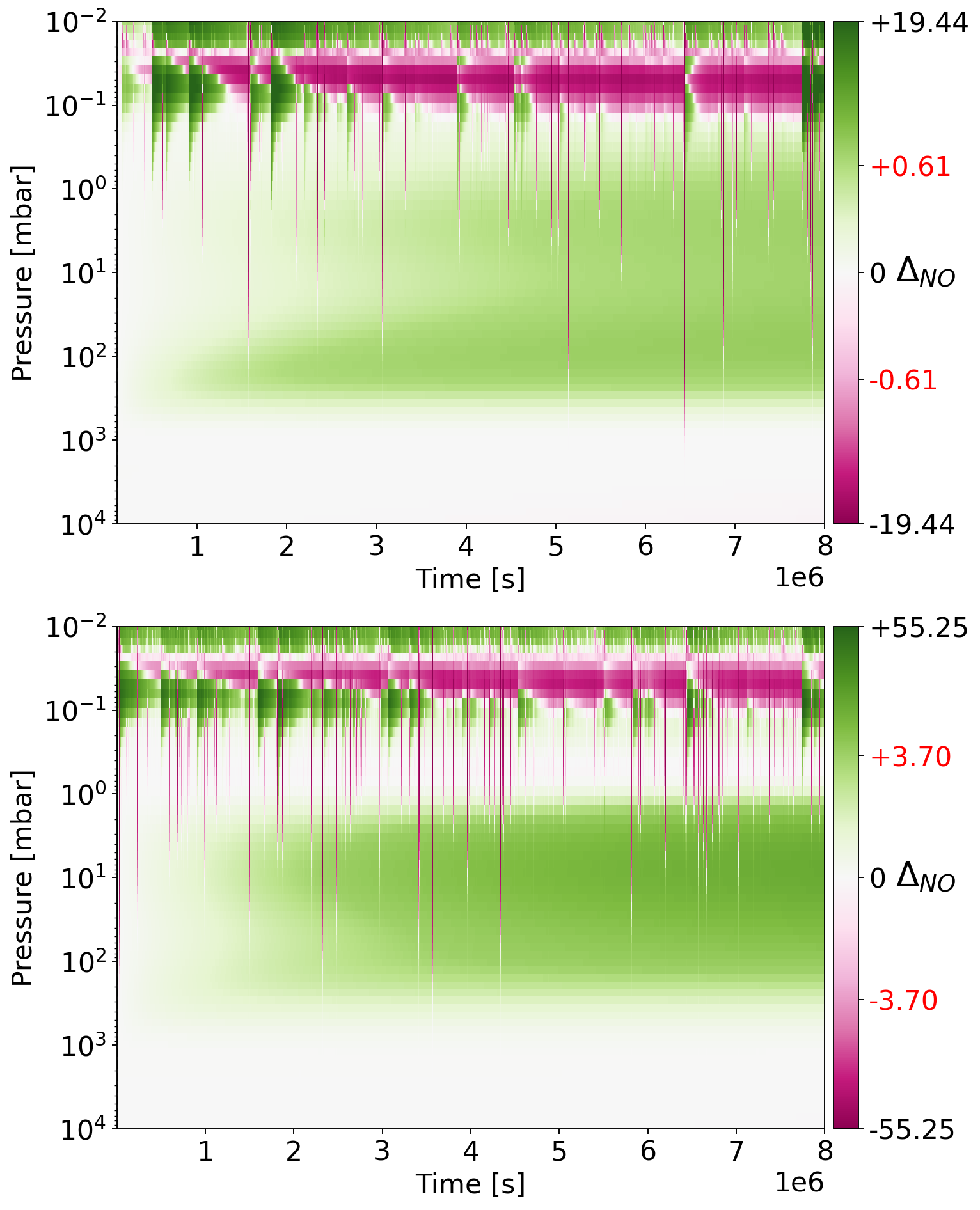}
        \captionof{figure}{$\Delta_{\ce{NO},p}$ in the hot planet versus time, with a stochastic flare model. Top panel: NEQ; bottom panel: CNEQ.}
        \label{fig:pcmSSto1632NO}
    \end{figure}

 
    For \ce{NO} in the atmosphere of the hot planet, the amplitudes of the changes in composition vary between the NEQ and CNEQ cases (top and bottom panels of \Cref{fig:pcmSSto1632NO} respectively). The distribution of $\Delta_{\ce{NO}}$ is similar across pressure-space between the simulations of the hot planet, in terms of where \ce{NO} sees production and depletion, but both the $1 \sigma$ and $2 \sigma$ values are larger when simulated self-consistently. 
    \par 
    
    \subsection{Observability of flare-induced changes}
    \label{ssec:synthObs}
    Previous sections presented and discussed the changes in atmospheric composition due to flares, and assessed the role of self-consistent modelling. For there to be differences in observables, it is not required that CNEQ chemistry yield larger changes in composition than NEQ chemistry, all other things being equal. Reduced spectral features are equally as interesting as enhanced ones; previous research indicates that self-consistently solving for the composition and temperature structure of an atmosphere is likely to reduce its transit depth \citep{Drummond2016}.
    \par 
    Equation \ref{eq:transit_duration} yields transit durations $\tau_c$ of 2.0510 and 0.5180 hours for the cold and hot planets respectively. These are much shorter than the 20 hours used in L22, which they selected for convenience. $\tau_c$ is required for calculating the relative change in transit depth due to flares ($C$, as per \Cref{eq:transDiff}) and is also used as an input for PandExo. 
    \par 
    The transmission spectra generated by \code{ATMO} are calculated using only 5000 correlated-$k$ bands across the wavelength range \mbox{(0.2 {\textmu}m $\le \lambda \le$ 1980.0 {\textmu}m)}, which leads to relatively low resolution when this is then truncated to the range of $\lambda$ which \textit{JWST} can detect. This reduced resolution prevents narrow features from being resolved, and so the effects of flares on transmission spectra are underestimated in our analysis. 
    \begin{figure}
        \includegraphics[width=\columnwidth,keepaspectratio]{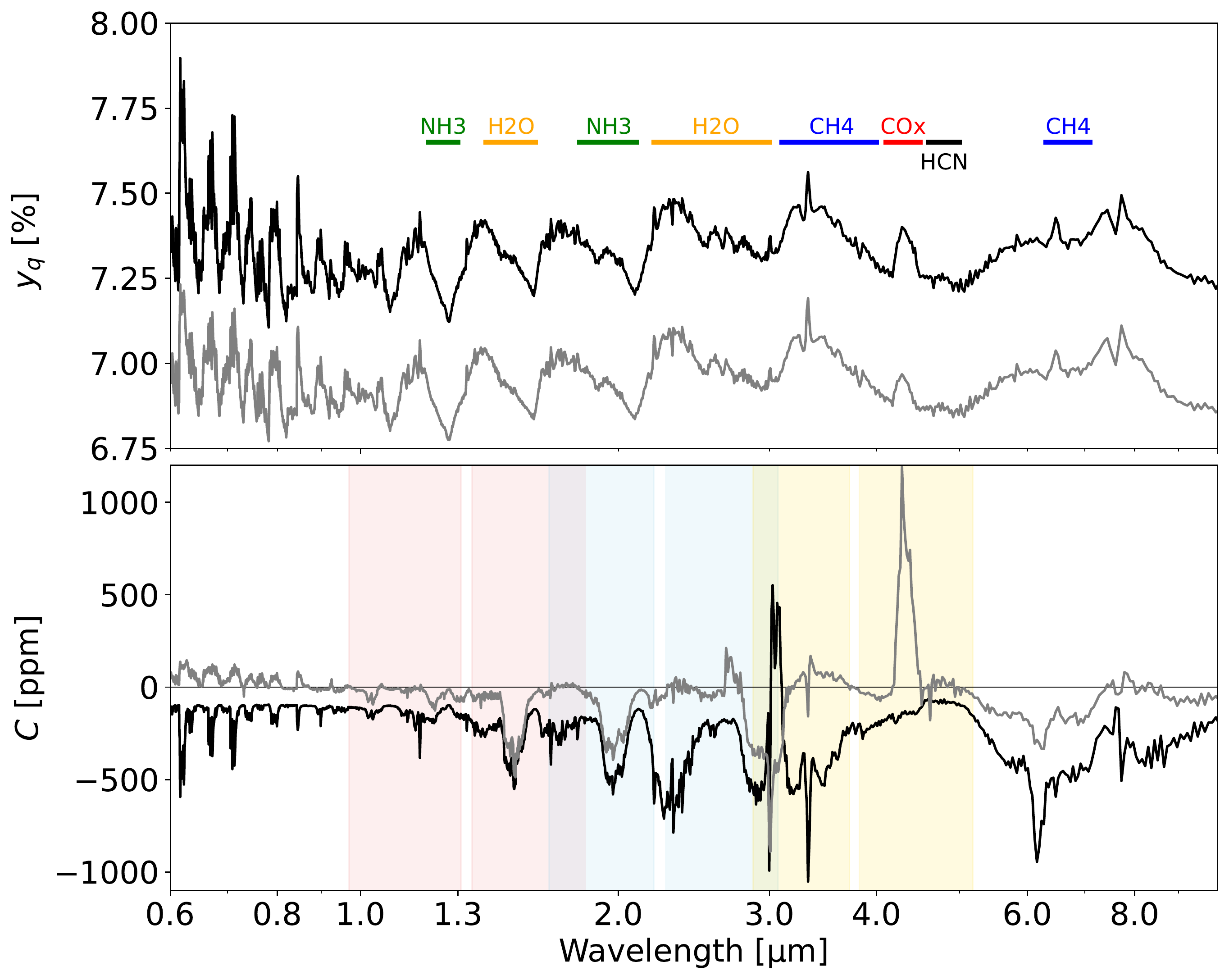}
        \captionof{figure}{Transmission spectra generated by \code{ATMO}. The top panel shows quiescent transit depth. On the same wavelength axis, the bottom panel plots the relative change in transit depth due to flares $C$ as per Equation \ref{eq:transDiff}. The black and grey lines correspond to spectra derived from NEQ and CNEQ simulations respectively. No artificial shift has been applied to data on the vertical axes. The wavelengths detectable by NIRSpec are highlighted in the bottom plot as filled regions: G140H/F100LP in red, G235H/F170LP in blue, and G395H/F290LP in yellow \citep{jdox2016}. Note that high-resolution observation modes have spectral gaps due to physical gaps between detectors inside NIRSpec \citep{jdox2016}. The top panel shows the transmission spectrum of the cold planet before flares were applied. $C$, shown in the bottom panel, is calculated using the pre-flare spectrum and the average spectrum after flares have applied at \mbox{$t=$\SI{4e6}{\second}}, as outlined in \Cref{ssec:transSpec}. Spectral features associated with various species and groups are labelled in in the top panel \citep{Gordon2022}. The COx feature (marked in red) has contributions from both \ce{CO} and \ce{CO2} molecules in the atmosphere.}
        \label{fig:trans_raw_412}
    \end{figure}
    \begin{figure}
        \includegraphics[width=\columnwidth,keepaspectratio]{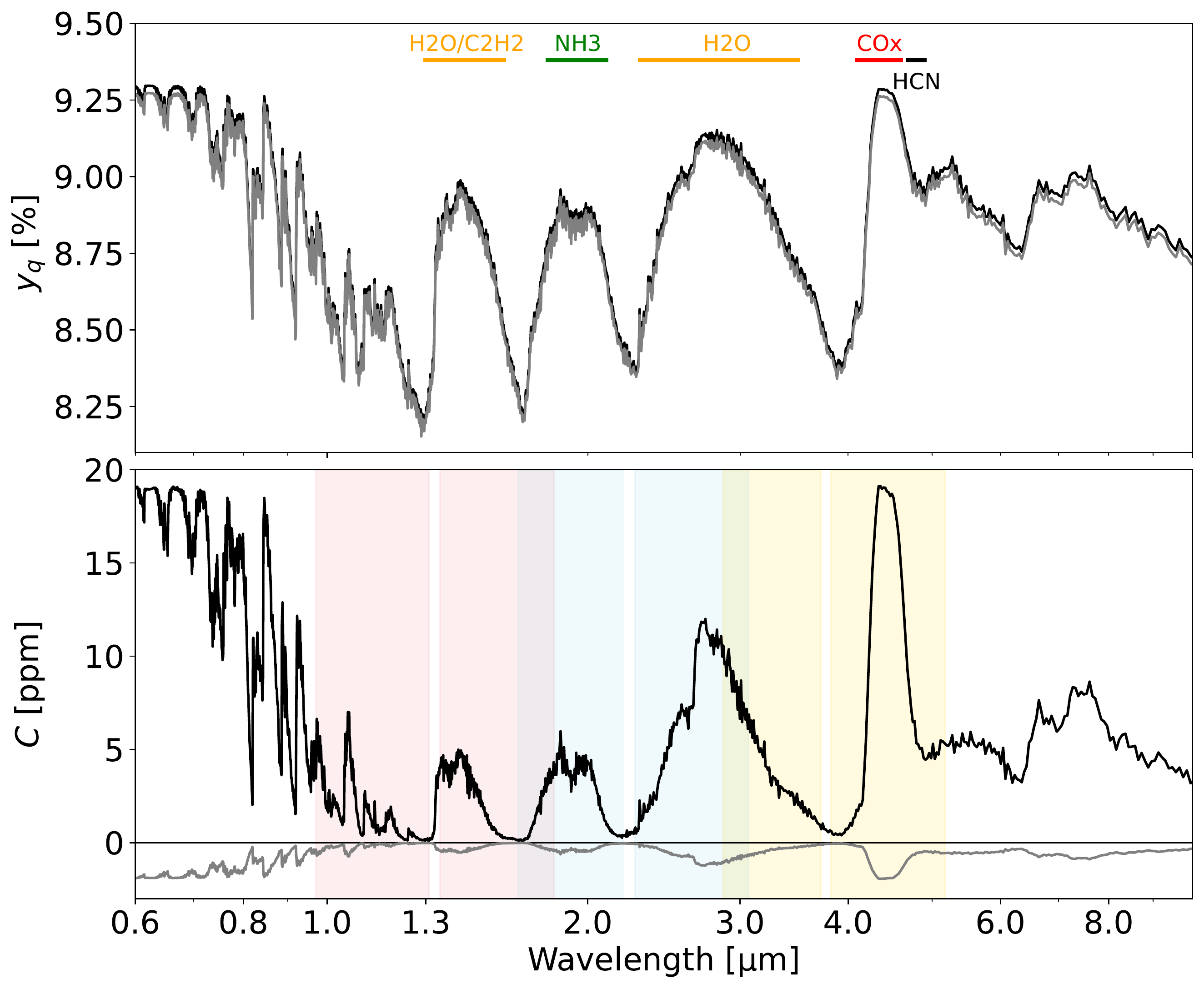}
        \captionof{figure}{Same as \Cref{fig:trans_raw_412}, but for the hot planet.}
        \label{fig:trans_raw_1632}
    \end{figure}
    \par
    V16 sees a maximum $|C|$ of $\sim 1200 \text{ ppm}$ after the atmosphere had adapted to the periodic application of flares, which is less than the maximum $|C|$ for our cold case ($\sim 1600 \text{ ppm}$), but more than maximum $|C|$ for our hot case ($\sim 20 \text{ ppm}$); this fits with their `observed' planet having a $T_{\text{eff}}$ which lies between our hot and cold cases.
    \par 
    It is clear from the plots in \Cref{fig:trans_raw_412,,fig:trans_raw_1632} that enhanced UV due to stellar flares induces compositional changes which manifest in transmission spectra. Features within these spectra can be associated with $\Delta_{s}$ from previous sections. For example, the COx feature near 4 {\textmu}m (identifiable in the transmission spectrum of both planets discussed in this work) is associated with both \ce{CO} and \ce{CO2} species in the atmosphere. The large COx feature follows from the fact that \ce{CO2} has a large opacity in the infrared, and is a strong greenhouse gas \citep{Wei2018}. The size of COx feature differs in amplitude when simulated self-consistently, compared to when the temperature profile is fixed, although this behaviour differs between the two planets explored in this work. In the case of the cold planet, this feature sees greater changes in magnitude when the atmosphere is evolved self-consistently (see bottom panel of \Cref{fig:trans_raw_412}), which is due to an increased \ce{CO2} abundance near \SI{e-2}{\milli\bar}, in contrast to the depletion of \ce{CO2} at these same pressures in the non-consistent case (see panels (aa) and (ab) of \Cref{fig:pcmCSto412-1}). This difference is primarily due to differing quiescent states (as a result of self-consistent solving) enabling differential evolution, rather than exclusively due to differential evolution during the course of flares, as previously discussed. The bottom panel of \Cref{fig:trans_raw_1632} shows that the same COx feature has opposite behaviour in response to flares in the case of the hot planet, which corresponds with \ce{CO2} evolving very differently in the hotter atmosphere. 
    
    \par 
    Spectra generated from self-consistent simulations have smaller transit depths than those generated from those with fixed $T(p)$ (c.f. the top panels of \Cref{fig:trans_raw_412,fig:trans_raw_1632}), which parallels \citet{Drummond2016} and follows from the related discussions in \Cref{sec:introduction,sec:methods}. As a result, $C$ is generally, although not exclusively, smaller for atmospheres simulated self-consistently compared to those in which $T(p)$ was fixed; the shapes and positions of spectral features are very similar between these two cases. Exceptions to this rule are features associated with \ce{CO} and \ce{CH4}, which both see larger $C$ in self-consistent simulations. 
    \par 
    While the quiescent transit depths are always larger in the case of the hotter planet, because it is closer to its parent star, the change in transit depth $C$ is much smaller. This follows from the fact that photochemistry is much less important in the case of the hot planet, and even less so when simulated self-consistently (see the related discussion in Section \ref{ssec:flaresSto}).
    \par 
    \par
\begin{figure}
        \includegraphics[width=0.8\columnwidth,keepaspectratio]{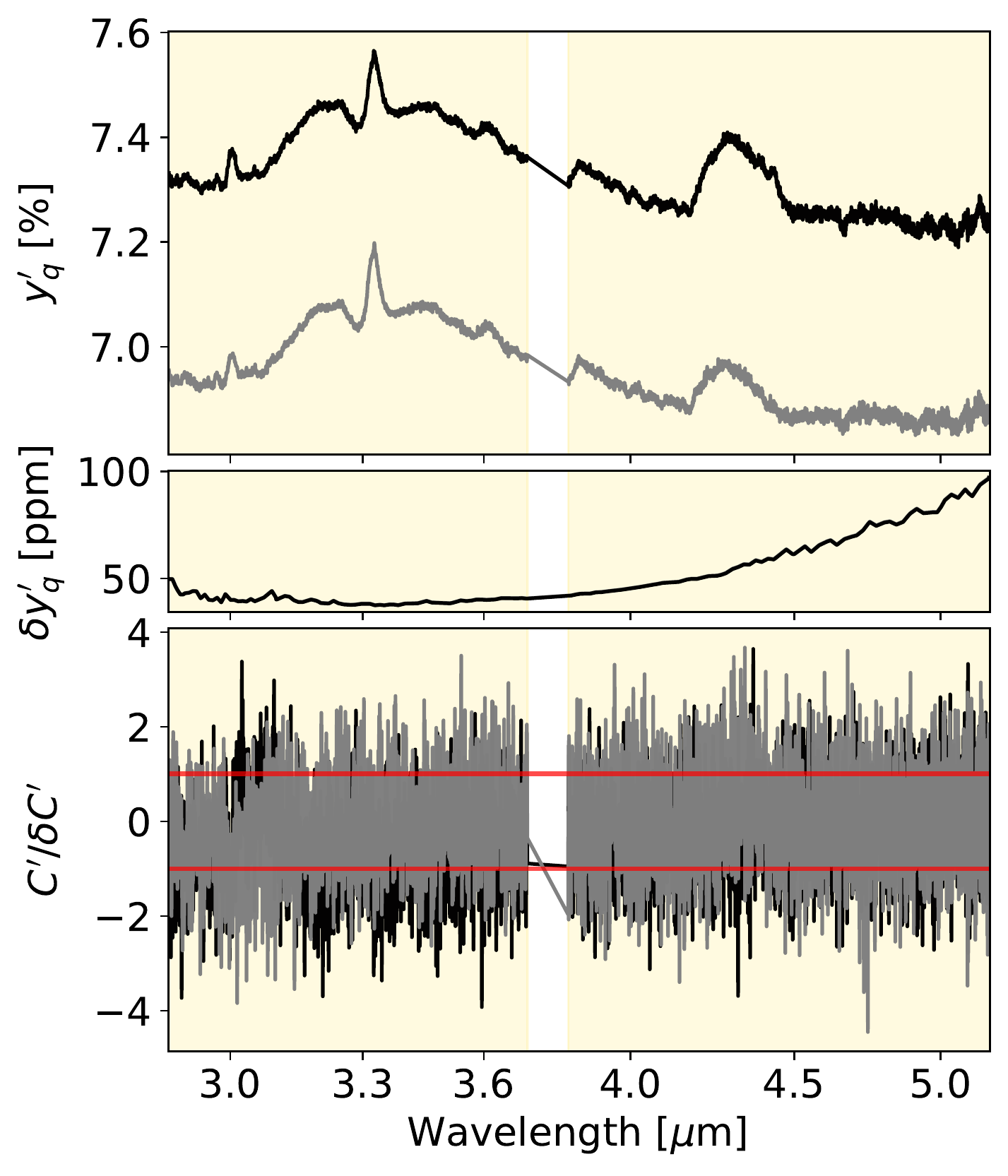}
        \captionof{figure}{Simulated two-transit observations (with NIRSpec using the G395H disperser) of the cold planet simulated self-consistently (grey) and with fixed $T(p)$ (black). Top panel: quiescent transit depth $y_q'$. Middle panel: error on the transit depth estimated by PandExo $\delta y_q'$. Bottom panel: SNR $C'/\delta C'$ for the relative change in the transit depth associated with flares. Horizontal red lines denote an SNR magnitude of unity.}
        \label{fig:pandexo_412_g395h_n2}
    \end{figure}

\begin{figure}
        \includegraphics[width=0.8\columnwidth,keepaspectratio]{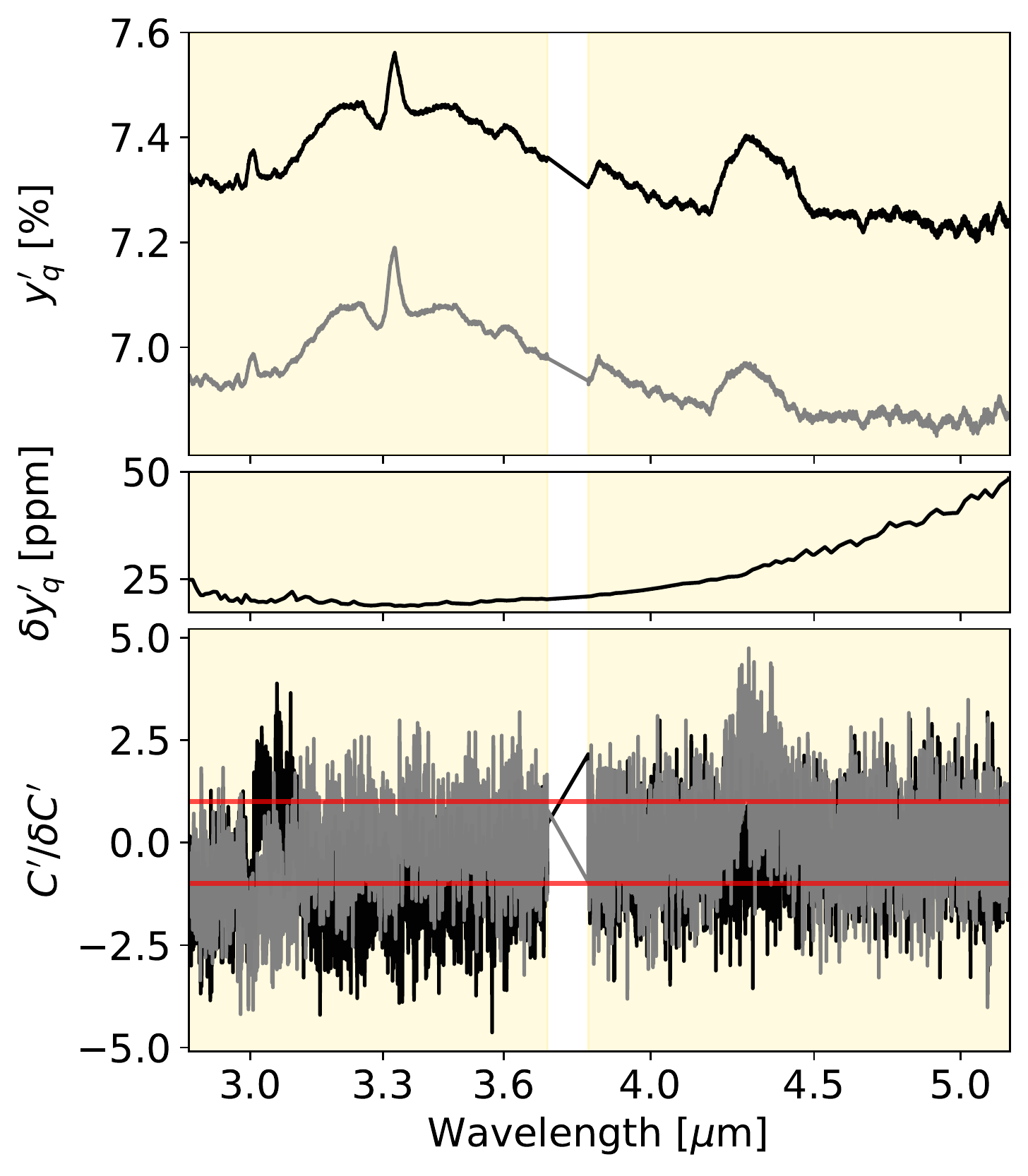}
        \captionof{figure}{Same as \Cref{fig:pandexo_412_g395h_n2} but integrated across eight transits, instead of two.}
        \label{fig:pandexo_412_g395h_n8}
    \end{figure}

 
    Figure \ref{fig:pandexo_412_g395h_n2} plots the simulated transit depth of the cold planet observed with NIRSpec G395H, assuming two transits per simulated observation. Spectral features identified in the raw transmission spectrum (\Cref{fig:trans_raw_412}) are present in the top panel of the figure, although with the additional noise added by PandExo. While the errors associated with the transit depth are small, once propagated through Equation \ref{eq:transDiff} the resultant error  $\delta C'$ is on a similar order to $C'$ itself. While it is clear from the bottom panel of \Cref{fig:pandexo_412_g395h_n2} that the SNR $C'/\delta C'$ has a magnitude greater than unity across most of the wavelength range, individual spectral features are difficult to identify for the NEQ and CNEQ cases in the figure. By increasing the number of integrated transits from two to eight (see \Cref{fig:pandexo_412_g395h_n8}), these features become readily identifiable with those indicated in the spectrum of $C$ before observational limitations are applied (\Cref{fig:trans_raw_412}). With eight transits, it is possible to distinguish differences between the NEQ and CNEQ cases in these synthetic observations of the cold planet. It would therefore be reasonable to conclude that the cumulative effects of flares on the atmosphere of the cold planet would be present in observations with a sufficient number of transits ($>2$), and that solving for this atmosphere self-consistently impacts observations made with NIRSpec. Observing the cold planet for eight transits would require at least 16.4 hours of integration time, which is more time than most targets will be allocated in the near future.
    \par 
    
    \par 
    The transit depth for observations of the hot planet are larger than those of the cold planet. However, the effects of flares on the atmosphere of the hot planet are much less observable. Even when observing the planet for 2000 transits, spectral features are not readily identifiable in plots of SNR versus wavelength. Evolution of spectral features with flares is not great enough such that they would manifest in observations with NIRSpec. This is true for fixed $T(p)$ and when $T(p)$ is solved self-consistently with the chemistry.
    \par 
    It is worth noting that \code{ATMO} does not account for the effects of aerosols when performing its radiative transfer calculations. It is understood that clouds can block spectral features arising from regions of the atmosphere below cloud-level, which could impact these synthetic observations \citep{Kreidberg2014}. We do not expect this effect to significantly impact the results of this work, as the majority of compositional changes induced by the flares occur at pressures lower than those where aerosol layers would develop.
    
    
    \section{Conclusion}
    \label{sec:conclusion} 
    The effects of flare-driven chemistry on exoplanet atmospheres have been investigated previously, but this work is the first assessment on the necessity of self-consistently solving for $T(p)$ alongside the chemistry. Simulating flares with coupled temperature and non-equilibrium chemistry shows different behaviour compared to when the PT profile is fixed. Many chemical species see significant changes in abundance due to the effects of flares; many changing in abundance by more than a factor of 100 versus the PFS. \ce{CH3} and \ce{NO} demonstrate complex behaviour across a range of pressure levels, the behaviours of \ce{C2H2} and \ce{N2H4} differ dramatically when simulated self-consistently, and an analysis of chemical pathways associated with \ce{CO2} demonstrates the critical role of minor species. Temperature changes associated with flares in the upper atmospheres of cool HJs are on the order of \SI{10}{\kelvin}, and are smaller for hotter HJs. 
    \par
    Applying flares using a realistic energy distribution is necessary when modelling the time-evolution of these atmospheres, as the atmosphere is sensitive to both the energy delivered by flares, as well as the rate at which flares occur. It is necessary to allow the atmosphere to evolve for a long period of time ($>$\SI{3e6}{\second}) from the onset of when flares are first applied, as an average steady-state is not always reached rapidly; this is especially important when modelling flares stochastically, as the model requires sufficient time to fully explore the flare frequency distribution. The cold planet assessed in this work was, for several species, still evolving when the simulations ended. It is therefore possible that compositional changes relative to the PFS would increase in magnitude with more simulation time. It would be interesting to investigate the significance of vertical mixing, especially as the parameter $K_{zz}$ is varied.
    
    \par 
    Flare-driven changes to planetary composition and temperature impact the transmission spectra of sufficiently cool HJ atmospheres. Synthetic observations with NIRSpec of a relatively cool HJ ($T_{\text{eff}} = 412 \text{ K}$) across only two transits tentatively show spectral features associated with flare-driven evolution. With more observation time \mbox{($\gtrsim 8$transits)} these features become increasingly apparent. Synthetic observations of this cold planet yield different spectral features in cases where it is evolved self-consistently as flares are applied, compared to when $T(p)$ is fixed. For hotter atmospheres (i.e. $T_{\text{eff}} = 1632 \text{ K}$), it is not possible to resolve spectral features associated with flare-driven chemistry, whether or not the atmosphere is evolved self-consistently. 
    \par 
    We conclude with the following statements:
    \begin{enumerate}
        \item modulation of the UV flux impinging on a HJ drives photochemistry in its atmosphere,
        \item photochemical processes affect the abundance of many different chemical species, including those which are not themselves sensitive to UV radiation,
        \item the temperature of such an atmosphere is marginally affected by these compositional changes,
        \item the effects on composition and temperature are typically more significant for cooler planets,
        \item it is possible to observe the effects of flare-driven changes to composition and temperature with \textit{JWST}-NIRSpec for relatively cool HJ atmospheres, 
        \item it is worth considering flare-driven behaviours in forward models of relatively cool planets around active stars,
        \item if flare-driven behaviours are included in forward models, we recommend that these models allow for self-consistent feedback between chemistry and temperature.
    \end{enumerate}
    
    \par 
    As \code{ATMO} is a 1D model it cannot account for much of the dynamics currently thought to take place in the atmospheres of gassy exoplanets such as horizontal advection, and geospatial differences in composition and temperature \citep{Showman2011, Koll2018, Hammond2021}. It is possible that this might affect inter-flare behaviours, as the night sides of planets are not exposed to increased UV irradiation but can contribute to chemistry via transport. \cite{Konings2022} and \cite{Chen2020} used higher-dimensional models to approach this problem but did not combine them with self-consistent temperature-chemistry feedback. In our work, the temperature changes induced by flares are too small to directly drive dynamical effects. However, it has been shown that a significant amount of energy can be transported from the dayside to the nightside of HJs by the corresponding dissociation and recombination of molecular hydrogen; it is possible that flare-driven photolysis of \ce{H2} could augment this otherwise thermolytic process, however this is difficult to account for with a 1D model \citep{Bell2018,Tan2019,Roth2021}. We have found that self-consistently solving for temperature and composition is important for accurately generating transmission spectra of flare-driven atmospheres, so it would be interesting to see this approach combined with a treatment of hazes and cloud, which themselves require higher-dimensional models to accurately describe. 
    \par 
    One important factor to consider alongside the effect of flares is the role of stellar activity alone. Recent work has shown that presence of phenomena such spots and plages in the atmosphere of a host star can interfere with observations of orbiting exoplanets \citep{Thompson2023,Rackham2019,Barstow2015}. Even if changes in composition due to flares were detectable by \textit{JWST}, discerning their spectral signatures from stellar activity could prove difficult.

    \section{Acknowledgements}
    \label{sec:acknowledgements}
    The authors would like to acknowledge the developers of NumPy, SciPy, and Matplotlib for their dedicated work to the open-source community \citep{Harris2020, Virtanen2020, Hunter2007}. The lead developer of \code{ATMO}, Dr P. Tremblin, is thanked for his contributions to the model. Feedback and suggestions from multiple anonymous reviewers significantly improved the quality of this work. The advice and support of Miss A. Goodsall is greatly appreciated. 
    
    \section{Data Availability}
    \label{sec:data_avail}
    The data underlying this article will be shared on reasonable request to the corresponding author.

    
    \bibliographystyle{mnras}
    \bibliography{mnras} 

\begin{thebibliography}{}
\makeatletter
\relax
\def\mn@urlcharsother{\let\do\@makeother \do\$\do\&\do\#\do\^\do\_\do\%\do\~}
\def\mn@doi{\begingroup\mn@urlcharsother \@ifnextchar [ {\mn@doi@}
  {\mn@doi@[]}}
\def\mn@doi@[#1]#2{\def\@tempa{#1}\ifx\@tempa\@empty \href
  {http://dx.doi.org/#2} {doi:#2}\else \href {http://dx.doi.org/#2} {#1}\fi
  \endgroup}
\def\mn@eprint#1#2{\mn@eprint@#1:#2::\@nil}
\def\mn@eprint@arXiv#1{\href {http://arxiv.org/abs/#1} {{\tt arXiv:#1}}}
\def\mn@eprint@dblp#1{\href {http://dblp.uni-trier.de/rec/bibtex/#1.xml}
  {dblp:#1}}
\def\mn@eprint@#1:#2:#3:#4\@nil{\def\@tempa {#1}\def\@tempb {#2}\def\@tempc
  {#3}\ifx \@tempc \@empty \let \@tempc \@tempb \let \@tempb \@tempa \fi \ifx
  \@tempb \@empty \def\@tempb {arXiv}\fi \@ifundefined
  {mn@eprint@\@tempb}{\@tempb:\@tempc}{\expandafter \expandafter \csname
  mn@eprint@\@tempb\endcsname \expandafter{\@tempc}}}

\bibitem[\protect\citeauthoryear{Adrian, Christensen  \& Liu}{Adrian
  et~al.}{2000}]{Adrian2000}
Adrian R.~J.,  Christensen K.~T.,   Liu Z.-C.,  2000, \mn@doi [Experiments in
  Fluids] {10.1007/s003489900087}, 29, 275

\bibitem[\protect\citeauthoryear{Alatalo et~al.,}{Alatalo
  et~al.}{2016}]{jdox2016}
Alatalo K.,  et~al., 2016, {JWST User Documentation (JDox)}, JWST User
  Documentation Website

\bibitem[\protect\citeauthoryear{Amundsen, Baraffe, Tremblin, Manners, Hayek,
  Mayne  \& Acreman}{Amundsen et~al.}{2014}]{Amundsen2014}
Amundsen D.,  Baraffe I.,  Tremblin P.,  Manners J.,  Hayek W.,  Mayne N.,
  Acreman D.,  2014, \mn@doi [A\&A] {10.1051/0004-6361/201323169}, 564, A59

\bibitem[\protect\citeauthoryear{Asplund, Grevesse, Sauval  \& Scott}{Asplund
  et~al.}{2009}]{Asplund2009}
Asplund M.,  Grevesse N.,  Sauval A.~J.,   Scott P.,  2009, \mn@doi [Annual
  Review of Astronomy and Astrophysics]
  {10.1146/annurev.astro.46.060407.145222}, 47, 481

\bibitem[\protect\citeauthoryear{Bailes et~al.,}{Bailes
  et~al.}{2011}]{Bailes2011}
Bailes M.,  et~al., 2011, \mn@doi [Science] {10.1126/science.1208890}, 333,
  1717

\bibitem[\protect\citeauthoryear{Barstow, Aigrain, Irwin, Kendrew  \&
  Fletcher}{Barstow et~al.}{2015}]{Barstow2015}
Barstow J.~K.,  Aigrain S.,  Irwin P. G.~J.,  Kendrew S.,   Fletcher L.~N.,
  2015, \mn@doi [Monthly Notices of the Royal Astronomical Society]
  {10.1093/mnras/stv186}, 448, 2546

\bibitem[\protect\citeauthoryear{Batalha et~al.,}{Batalha
  et~al.}{2017}]{Batalha2017}
Batalha N.~E.,  et~al., 2017, \mn@doi [Publications of the Astronomical Society
  of the Pacific] {10.1088/1538-3873/aa65b0}, 129, 064501

\bibitem[\protect\citeauthoryear{Baudino, Molliere, Venot, Tremblin, Bezard  \&
  Lagage}{Baudino et~al.}{2017}]{Baudino2017}
Baudino J.-L.,  Molliere P.,  Venot O.,  Tremblin P.,  Bezard B.,   Lagage
  P.-O.,  2017, \mn@doi [The Astrophysical Journal] {10.3847/1538-4357/aa95be},
  850, 150

\bibitem[\protect\citeauthoryear{Bell \& Cowan}{Bell \& Cowan}{2018}]{Bell2018}
Bell T.~J.,  Cowan N.~B.,  2018, \mn@doi [The Astrophysical Journal]
  {10.3847/2041-8213/aabcc8}, 857, L20

\bibitem[\protect\citeauthoryear{Caffau, Ludwig, Steffen, Freytag  \&
  Bonifacio}{Caffau et~al.}{2010}]{Caffau2010}
Caffau E.,  Ludwig H.-G.,  Steffen M.,  Freytag B.,   Bonifacio P.,  2010,
  \mn@doi [Solar Physics] {10.1007/s11207-010-9541-4}, 268, 255

\bibitem[\protect\citeauthoryear{Chadney, Koskinen, Galand, Unruh  \&
  Sanz-Forcada}{Chadney et~al.}{2017}]{Chadney2017}
Chadney J.,  Koskinen T.,  Galand M.,  Unruh Y.,   Sanz-Forcada J.,  2017,
  \mn@doi [A\&A] {10.1051/0004-6361/201731129}, 608, A75

\bibitem[\protect\citeauthoryear{{Chen}, {Zhan}, {Youngblood}, {Wolf},
  {Feinstein}  \& {Horton}}{{Chen} et~al.}{2020}]{Chen2020}
{Chen} H.,  {Zhan} Z.,  {Youngblood} A.,  {Wolf} E.~T.,  {Feinstein} A.~D.,
  {Horton} D.~E.,  2020, Nature Astronomy, pp 1--13

\bibitem[\protect\citeauthoryear{Czesla, Salz, Schneider  \& Schmitt}{Czesla
  et~al.}{2013}]{Czesla2013}
Czesla S.,  Salz M.,  Schneider P.~C.,   Schmitt J. H. M.~M.,  2013, \mn@doi
  [Astronomy \& Astrophysics] {10.1051/0004-6361/201322272}, 560, A17

\bibitem[\protect\citeauthoryear{D{\'{e}}sert, Vidal-Madjar, des Etangs, Sing,
  Ehrenreich, H{\'{e}}brard  \& Ferlet}{D{\'{e}}sert et~al.}{2008}]{Desert2008}
D{\'{e}}sert J.-M.,  Vidal-Madjar A.,  des Etangs A.~L.,  Sing D.,  Ehrenreich
  D.,  H{\'{e}}brard G.,   Ferlet R.,  2008, \mn@doi [Astronomy \&
  Astrophysics] {10.1051/0004-6361:200810355}, 492, 585

\bibitem[\protect\citeauthoryear{Drummond}{Drummond}{2017}]{Drummond2017}
Drummond B.,  2017, PhD thesis, University of Exeter

\bibitem[\protect\citeauthoryear{Drummond, Tremblin, Baraffe, Amundsen, Mayne,
  Venot  \& Goyal}{Drummond et~al.}{2016}]{Drummond2016}
Drummond B.,  Tremblin P.,  Baraffe I.,  Amundsen D.~S.,  Mayne N.~J.,  Venot
  O.,   Goyal J.,  2016, \mn@doi [Astronomy {\&} Astrophysics]
  {10.1051/0004-6361/201628799}, 594, A69

\bibitem[\protect\citeauthoryear{Fortney, Cooper, Showman, Marley  \&
  Freedman}{Fortney et~al.}{2006}]{Fortney2006}
Fortney J.~J.,  Cooper C.~S.,  Showman A.~P.,  Marley M.~S.,   Freedman R.~S.,
  2006, \mn@doi [The Astrophysical Journal] {10.1086/508442}, 652, 746

\bibitem[\protect\citeauthoryear{Fortney, Lodders, Marley  \& Freedman}{Fortney
  et~al.}{2008}]{Fortney2008}
Fortney J.~J.,  Lodders K.,  Marley M.~S.,   Freedman R.~S.,  2008, \mn@doi
  [The Astrophysical Journal] {10.1086/528370}, 678, 1419

\bibitem[\protect\citeauthoryear{Fortney, Mordasini, Nettelmann, Kempton,
  Greene  \& Zahnle}{Fortney et~al.}{2013}]{Fortney2013}
Fortney J.~J.,  Mordasini C.,  Nettelmann N.,  Kempton E. M.-R.,  Greene T.~P.,
    Zahnle K.,  2013, \mn@doi [The Astrophysical Journal]
  {10.1088/0004-637x/775/1/80}, 775, 80

\bibitem[\protect\citeauthoryear{Fortney, Dawson  \& Komacek}{Fortney
  et~al.}{2021}]{Fortney2021}
Fortney J.~J.,  Dawson R.~I.,   Komacek T.~D.,  2021, \mn@doi [Journal of
  Geophysical Research: Planets] {10.1029/2020je006629}, 126

\bibitem[\protect\citeauthoryear{Fossati, Young, Shulyak, Koskinen, Huang,
  Cubillos, France  \& Sreejith}{Fossati et~al.}{2021}]{Fossati2021}
Fossati L.,  Young M.~E.,  Shulyak D.,  Koskinen T.,  Huang C.,  Cubillos
  P.~E.,  France K.,   Sreejith A.~G.,  2021, \mn@doi [Astronomy {\&}
  Astrophysics] {10.1051/0004-6361/202140813}, 653, A52

\bibitem[\protect\citeauthoryear{Gordon \& McBride}{Gordon \&
  McBride}{1994}]{Gordon1994}
Gordon S.,  McBride B.,  1994, {NASA Reference Publication 1311}.
NASA

\bibitem[\protect\citeauthoryear{Gordon et~al.,}{Gordon
  et~al.}{2022}]{Gordon2022}
Gordon I.,  et~al., 2022, \mn@doi [Journal of Quantitative Spectroscopy and
  Radiative Transfer] {10.1016/j.jqsrt.2021.107949}, 277, 107949

\bibitem[\protect\citeauthoryear{Goyal et~al.,}{Goyal et~al.}{2017}]{Goyal2017}
Goyal J.,  et~al., 2017, \mn@doi [Monthly Notices of the Royal Astronomical
  Society] {10.1093/mnras/stx3015}, 474, 5158–5185

\bibitem[\protect\citeauthoryear{Grayver, Bower, Saur, Dorn  \& Morris}{Grayver
  et~al.}{2022}]{Grayver2022}
Grayver A.,  Bower D.~J.,  Saur J.,  Dorn C.,   Morris B.~M.,  2022, Interior
  heating of rocky exoplanets from stellar flares with application to
  TRAPPIST-1, \mn@doi{10.48550/ARXIV.2211.06140}, \url
  {https://arxiv.org/abs/2211.06140}

\bibitem[\protect\citeauthoryear{Günther et~al.,}{Günther
  et~al.}{2020}]{Guenther2020}
Günther M.,  et~al., 2020, \mn@doi [The Astronomical Journal]
  {10.3847/1538-3881/ab5d3a}, 159, 60

\bibitem[\protect\citeauthoryear{Hammond \& Lewis}{Hammond \&
  Lewis}{2021}]{Hammond2021}
Hammond M.,  Lewis N.~T.,  2021, \mn@doi [Proceedings of the National Academy
  of Sciences] {10.1073/pnas.2022705118}, 118

\bibitem[\protect\citeauthoryear{Harris et~al.,}{Harris
  et~al.}{2020}]{Harris2020}
Harris C.~R.,  et~al., 2020, \mn@doi [Nature] {10.1038/s41586-020-2649-2}, 585,
  357

\bibitem[\protect\citeauthoryear{Hawley \& Pettersen}{Hawley \&
  Pettersen}{1991}]{Hawley1991}
Hawley S.~L.,  Pettersen B.~R.,  1991, \mn@doi [The Astrophysical Journal]
  {10.1086/170474}, 378, 725

\bibitem[\protect\citeauthoryear{Hawley, Davenport, Kowalski, Wisniewski, Hebb,
  Deitrick  \& Hilton}{Hawley et~al.}{2014}]{Hawley2014}
Hawley S.~L.,  Davenport J. R.~A.,  Kowalski A.~F.,  Wisniewski J.~P.,  Hebb
  L.,  Deitrick R.,   Hilton E.~J.,  2014, \mn@doi [The Astrophysical Journal]
  {10.1088/0004-637x/797/2/121}, 797, 121

\bibitem[\protect\citeauthoryear{Helling, Kawashima, Graham, Samra, Chubb, Min,
  Waters  \& Parmentier}{Helling et~al.}{2020}]{Helling2020}
Helling C.,  Kawashima Y.,  Graham V.,  Samra D.,  Chubb K.~L.,  Min M.,
  Waters L. B. F.~M.,   Parmentier V.,  2020, \mn@doi [Astronomy \&
  Astrophysics] {10.1051/0004-6361/202037633}, 641, A178

\bibitem[\protect\citeauthoryear{Herbst et~al.,}{Herbst
  et~al.}{2019}]{Herbst2019}
Herbst K.,  et~al., 2019, \mn@doi [Astronomy {\&} Astrophysics]
  {10.1051/0004-6361/201935888}, 631, A101

\bibitem[\protect\citeauthoryear{Hilton, West, Hawley  \& Kowalski}{Hilton
  et~al.}{2010}]{Hilton2010}
Hilton E.~J.,  West A.~A.,  Hawley S.~L.,   Kowalski A.~F.,  2010, \mn@doi [The
  Astronomical Journal] {10.1088/0004-6256/140/5/1402}, 140, 1402

\bibitem[\protect\citeauthoryear{Hindmarsh}{Hindmarsh}{1983}]{Hindmarsh1983}
Hindmarsh A.,  1983, Scientific Computing: Applications of Mathematics and
  Computing to the Physical Sciences.
North-Holland Publishing Company, p.~55

\bibitem[\protect\citeauthoryear{Hunter}{Hunter}{2007}]{Hunter2007}
Hunter J.~D.,  2007, \mn@doi [Computing in Science \& Engineering]
  {10.1109/mcse.2007.55}, 9, 90

\bibitem[\protect\citeauthoryear{Kim, Cho, Mok, Yoo  \& Cho}{Kim
  et~al.}{2013}]{Kim2013}
Kim J.,  Cho H.-K.,  Mok J.,  Yoo H.~D.,   Cho N.,  2013, \mn@doi [Journal of
  Photochemistry and Photobiology B: Biology]
  {10.1016/j.jphotobiol.2012.11.007}, 119, 46

\bibitem[\protect\citeauthoryear{Koll \& Komacek}{Koll \&
  Komacek}{2018}]{Koll2018}
Koll D. D.~B.,  Komacek T.~D.,  2018, \mn@doi [The Astrophysical Journal]
  {10.3847/1538-4357/aaa3de}, 853, 133

\bibitem[\protect\citeauthoryear{Konings, Baeyens  \& Decin}{Konings
  et~al.}{2022}]{Konings2022}
Konings T.,  Baeyens R.,   Decin L.,  2022, \mn@doi [Astronomy \& Astrophysics]
  {10.1051/0004-6361/202243436}, 667, A15

\bibitem[\protect\citeauthoryear{Kreidberg et~al.,}{Kreidberg
  et~al.}{2014}]{Kreidberg2014}
Kreidberg L.,  et~al., 2014, \mn@doi [Nature] {10.1038/nature12888}, 505, 69

\bibitem[\protect\citeauthoryear{Lee, Fletcher  \& Irwin}{Lee
  et~al.}{2011}]{Lee2011}
Lee J.-M.,  Fletcher L.~N.,   Irwin P. G.~J.,  2011, \mn@doi [Monthly Notices
  of the Royal Astronomical Society] {10.1111/j.1365-2966.2011.20013.x}, 420,
  170

\bibitem[\protect\citeauthoryear{Lewis et~al.,}{Lewis et~al.}{2020}]{Lewis2020}
Lewis N.~K.,  et~al., 2020, \mn@doi [The Astrophysical Journal Letters]
  {10.3847/2041-8213/abb77f}, 902, L19

\bibitem[\protect\citeauthoryear{Liang, Parkinson, Lee, Yung  \& Seager}{Liang
  et~al.}{2003}]{Liang2003}
Liang M.-C.,  Parkinson C.~D.,  Lee A. Y.-T.,  Yung Y.~L.,   Seager S.,  2003,
  \mn@doi [The Astrophysical Journal] {10.1086/379314}, 596, L247

\bibitem[\protect\citeauthoryear{Lothringer, Barman  \& Koskinen}{Lothringer
  et~al.}{2018}]{Lothringer2018}
Lothringer J.~D.,  Barman T.,   Koskinen T.,  2018, \mn@doi [The Astrophysical
  Journal] {10.3847/1538-4357/aadd9e}, 866, 27

\bibitem[\protect\citeauthoryear{Louca, Yamila, Tsai, Froning, Parke~Loyd  \&
  France}{Louca et~al.}{2022}]{Louca2022}
Louca A.~J.,  Yamila M.,  Tsai S.-M.,  Froning C.~S.,  Parke~Loyd R.,   France
  K.,  2022, Monthly Notices of the Royal Astronomical Society

\bibitem[\protect\citeauthoryear{Loyd et~al.,}{Loyd et~al.}{2018}]{Loyd2018}
Loyd R. O.~P.,  et~al., 2018, \mn@doi [The Astrophysical Journal]
  {10.3847/1538-4357/aae2bd}, 867, 71

\bibitem[\protect\citeauthoryear{Melbourne et~al.,}{Melbourne
  et~al.}{2020}]{Melbourne2020}
Melbourne K.,  et~al., 2020, \mn@doi [The Astronomical Journal]
  {10.3847/1538-3881/abbf5c}, 160, 269

\bibitem[\protect\citeauthoryear{Mignon et~al.,}{Mignon
  et~al.}{2023}]{Mignon2023}
Mignon L.,  et~al., 2023, \mn@doi [Astronomy \& Astrophysics]
  {10.48550/ARXIV.2303.03998}

\bibitem[\protect\citeauthoryear{Mizutani, Mikuni, Takahasi  \& Noda}{Mizutani
  et~al.}{1975}]{Mizutani1975}
Mizutani H.,  Mikuni H.,  Takahasi M.,   Noda H.,  1975, \mn@doi [Origins of
  Life] {10.1007/bf00928899}, 6, 513

\bibitem[\protect\citeauthoryear{Mounzer et~al.,}{Mounzer
  et~al.}{2022}]{Mounzer2022}
Mounzer D.,  et~al., 2022, \mn@doi [Astronomy \& Astrophysics]
  {10.1051/0004-6361/202243998}, 668, A1

\bibitem[\protect\citeauthoryear{Murray-Clay, Chiang  \& Murray}{Murray-Clay
  et~al.}{2009}]{MurrayClay2009}
Murray-Clay R.~A.,  Chiang E.~I.,   Murray N.,  2009, \mn@doi [The
  Astrophysical Journal] {10.1088/0004-637x/693/1/23}, 693, 23

\bibitem[\protect\citeauthoryear{Odert et~al.,}{Odert et~al.}{2020}]{Odert2020}
Odert P.,  et~al., 2020, \mn@doi [Astronomy \& Astrophysics]
  {10.1051/0004-6361/201834814}, 638, A49

\bibitem[\protect\citeauthoryear{Ohmura \& Ohmura}{Ohmura \&
  Ohmura}{1961}]{Ohmura1961}
Ohmura T.,  Ohmura H.,  1961, \mn@doi [Physical Review]
  {10.1103/physrev.121.513}, 121, 513

\bibitem[\protect\citeauthoryear{Parmentier et~al.,}{Parmentier
  et~al.}{2018}]{Parmentier2018}
Parmentier V.,  et~al., 2018, \mn@doi [Astronomy \& Astrophysics]
  {10.1051/0004-6361/201833059}, 617, A110

\bibitem[\protect\citeauthoryear{Parmentier, Showman  \& Fortney}{Parmentier
  et~al.}{2020}]{Parmentier2020}
Parmentier V.,  Showman A.~P.,   Fortney J.~J.,  2020, \mn@doi [Monthly Notices
  of the Royal Astronomical Society] {10.1093/mnras/staa3418}, 501, 78

\bibitem[\protect\citeauthoryear{Pettersen \& Coleman}{Pettersen \&
  Coleman}{1981}]{Pettersen1981}
Pettersen B.~R.,  Coleman L.~A.,  1981, \mn@doi [The Astrophysical Journal]
  {10.1086/159500}, 251, 571

\bibitem[\protect\citeauthoryear{Phillips et~al.,}{Phillips
  et~al.}{2020}]{Phillips2020}
Phillips M.,  et~al., 2020, \mn@doi [A\&A] {10.1051/0004-6361/201937381}, 637,
  A38

\bibitem[\protect\citeauthoryear{Piette, Madhusudhan, McKemmish, Gandhi,
  Masseron  \& Welbanks}{Piette et~al.}{2020}]{Piette2020}
Piette A. A.~A.,  Madhusudhan N.,  McKemmish L.~K.,  Gandhi S.,  Masseron T.,
  Welbanks L.,  2020, \mn@doi [Monthly Notices of the Royal Astronomical
  Society] {10.1093/mnras/staa1592}, 496, 3870

\bibitem[\protect\citeauthoryear{Poling, Prausnitz  \& O'Connell}{Poling
  et~al.}{2001}]{Poling2001}
Poling B.,  Prausnitz J.,   O'Connell J.,  2001, The properties of gases and
  liquids, 5th edn.
McGraw-Hill, New York

\bibitem[\protect\citeauthoryear{Rackham, Apai  \& Giampapa}{Rackham
  et~al.}{2019}]{Rackham2019}
Rackham B.~V.,  Apai D.,   Giampapa M.~S.,  2019, \mn@doi [The Astronomical
  Journal] {10.3847/1538-3881/aaf892}, 157, 96

\bibitem[\protect\citeauthoryear{Reiners, Basri  \& Browning}{Reiners
  et~al.}{2009}]{Reiners2009}
Reiners A.,  Basri G.,   Browning M.,  2009, \mn@doi [The Astrophysical
  Journal] {10.1088/0004-637x/692/1/538}, 692, 538

\bibitem[\protect\citeauthoryear{Ridgway et~al.,}{Ridgway
  et~al.}{2022}]{Ridgway2022}
Ridgway R.~J.,  et~al., 2022, \mn@doi [Monthly Notices of the Royal
  Astronomical Society] {10.1093/mnras/stac3105}

\bibitem[\protect\citeauthoryear{Rodriguez-Lopez}{Rodriguez-Lopez}{2019}]{RodriguezLopez2019}
Rodriguez-Lopez C.,  2019, \mn@doi [Frontiers in Astronomy and Space Sciences]
  {10.3389/fspas.2019.00076}, 6

\bibitem[\protect\citeauthoryear{Roth, Drummond, H{\'{e}}brard, Tremblin, Goyal
   \& Mayne}{Roth et~al.}{2021}]{Roth2021}
Roth A.,  Drummond B.,  H{\'{e}}brard E.,  Tremblin P.,  Goyal J.,   Mayne N.,
  2021, \mn@doi [Monthly Notices of the Royal Astronomical Society]
  {10.1093/mnras/stab1256}, 505, 4515

\bibitem[\protect\citeauthoryear{Rothman, Hawkins, Wattson  \& Gamache}{Rothman
  et~al.}{1992}]{Rothman1992}
Rothman L.,  Hawkins R.,  Wattson R.,   Gamache R.,  1992, \mn@doi [Journal of
  Quantitative Spectroscopy and Radiative Transfer]
  {10.1016/0022-4073(92)90119-o}, 48, 537

\bibitem[\protect\citeauthoryear{Segura, Walkowicz, Meadows, Kasting  \&
  Hawley}{Segura et~al.}{2010}]{Segura2010}
Segura A.,  Walkowicz L.~M.,  Meadows V.,  Kasting J.,   Hawley S.,  2010,
  \mn@doi [Astrobiology] {10.1089/ast.2009.0376}, 10, 751

\bibitem[\protect\citeauthoryear{Sharp \& Burrows}{Sharp \&
  Burrows}{2007}]{Sharp2007}
Sharp C.~M.,  Burrows A.,  2007, \mn@doi [The Astrophysical Journal Supplement
  Series] {10.1086/508708}, 168, 140

\bibitem[\protect\citeauthoryear{Showman \& Polvani}{Showman \&
  Polvani}{2011}]{Showman2011}
Showman A.~P.,  Polvani L.~M.,  2011, \mn@doi [The Astrophysical Journal]
  {10.1088/0004-637x/738/1/71}, 738, 71

\bibitem[\protect\citeauthoryear{Sing et~al.,}{Sing et~al.}{2015}]{Sing2015}
Sing D.~K.,  et~al., 2015, \mn@doi [Nature] {10.1038/nature16068}, 529, 59

\bibitem[\protect\citeauthoryear{Stamnes, Thomas  \& Stamnes}{Stamnes
  et~al.}{2017}]{Stamnes2017}
Stamnes K.,  Thomas G.~E.,   Stamnes J.~J.,  2017, Radiative Transfer in the
  Atmosphere and Ocean.
Cambridge University Press, \mn@doi{10.1017/9781316148549}

\bibitem[\protect\citeauthoryear{Steinrueck, Showman, Lavvas, Koskinen, Tan  \&
  Zhang}{Steinrueck et~al.}{2021}]{Steinrueck2021}
Steinrueck M.~E.,  Showman A.~P.,  Lavvas P.,  Koskinen T.,  Tan X.,   Zhang
  X.,  2021, \mn@doi [Monthly Notices of the Royal Astronomical Society]
  {10.1093/mnras/stab1053}, 504, 2783

\bibitem[\protect\citeauthoryear{Tan \& Komacek}{Tan \&
  Komacek}{2019}]{Tan2019}
Tan X.,  Komacek T.~D.,  2019, \mn@doi [The Astrophysical Journal]
  {10.3847/1538-4357/ab4a76}, 886, 26

\bibitem[\protect\citeauthoryear{Thomas \& Stamnes}{Thomas \&
  Stamnes}{2002}]{Thomas2002}
Thomas G.~E.,  Stamnes K.,  2002, Radiative Transfer in the Atmosphere and
  Ocean.
Cambridge University Press

\bibitem[\protect\citeauthoryear{Thompson et~al.,}{Thompson
  et~al.}{2023}]{Thompson2023}
Thompson A.,  et~al., 2023, \mn@doi [The Astrophysical Journal]
  {10.48550/ARXIV.2302.04574}

\bibitem[\protect\citeauthoryear{Thorngren, Gao  \& Fortney}{Thorngren
  et~al.}{2019}]{Thorngren2019}
Thorngren D.,  Gao P.,   Fortney J.~J.,  2019, \mn@doi [The Astrophysical
  Journal] {10.3847/2041-8213/ab43d0}, 884, L6

\bibitem[\protect\citeauthoryear{Tremblin, Amundsen, Mourier, Baraffe,
  Chabrier, Drummond, Homeier  \& Venot}{Tremblin et~al.}{2015}]{Tremblin2015}
Tremblin P.,  Amundsen D.~S.,  Mourier P.,  Baraffe I.,  Chabrier G.,  Drummond
  B.,  Homeier D.,   Venot O.,  2015, \mn@doi [The Astrophysical Journal]
  {10.1088/2041-8205/804/1/l17}, 804, L17

\bibitem[\protect\citeauthoryear{Tremblin et~al.,}{Tremblin
  et~al.}{2019}]{Tremblin2019}
Tremblin P.,  et~al., 2019, \mn@doi [The Astrophysical Journal]
  {10.3847/1538-4357/ab05db}, 876, 144

\bibitem[\protect\citeauthoryear{Tsai et~al.,}{Tsai et~al.}{2022}]{tsai2022}
Tsai S.-M.,  et~al., 2022, Direct Evidence of Photochemistry in an Exoplanet
  Atmosphere, \mn@doi{10.48550/ARXIV.2211.10490}, \url
  {https://arxiv.org/abs/2211.10490}

\bibitem[\protect\citeauthoryear{Venot, H{\'{e}}brard, Ag{\'{u}}ndez,
  Dobrijevic, Selsis, Hersant, Iro  \& Bounaceur}{Venot
  et~al.}{2012}]{Venot2012}
Venot O.,  H{\'{e}}brard E.,  Ag{\'{u}}ndez M.,  Dobrijevic M.,  Selsis F.,
  Hersant F.,  Iro N.,   Bounaceur R.,  2012, \mn@doi [Astronomy {\&}
  Astrophysics] {10.1051/0004-6361/201219310}, 546, A43

\bibitem[\protect\citeauthoryear{Venot et~al.,}{Venot et~al.}{2013}]{Venot2013}
Venot O.,  et~al., 2013, \mn@doi [Astronomy \& Astrophysics]
  {10.1051/0004-6361/201220945}, 551, A131

\bibitem[\protect\citeauthoryear{Venot, Rocchetto, Carl, Hashim  \&
  Decin}{Venot et~al.}{2016}]{Venot2016}
Venot O.,  Rocchetto M.,  Carl S.,  Hashim A.~R.,   Decin L.,  2016, \mn@doi
  [The Astrophysical Journal] {10.3847/0004-637X/830/2/77}, 830, 77

\bibitem[\protect\citeauthoryear{Virtanen et~al.,}{Virtanen
  et~al.}{2020}]{Virtanen2020}
Virtanen P.,  et~al., 2020, \mn@doi [Nature Methods]
  {10.1038/s41592-019-0686-2}, 17, 261

\bibitem[\protect\citeauthoryear{Wakeford, Visscher, Lewis, Kataria, Marley,
  Fortney  \& Mandell}{Wakeford et~al.}{2016}]{Wakeford2016}
Wakeford H.~R.,  Visscher C.,  Lewis N.~K.,  Kataria T.,  Marley M.~S.,
  Fortney J.~J.,   Mandell A.~M.,  2016, \mn@doi [Monthly Notices of the Royal
  Astronomical Society] {10.1093/mnras/stw2639}, 464, 4247

\bibitem[\protect\citeauthoryear{Wei, Hsieh, Chiu, Yen, Lee, Tsai  \& Ting}{Wei
  et~al.}{2018}]{Wei2018}
Wei P.-S.,  Hsieh Y.-C.,  Chiu H.-H.,  Yen D.-L.,  Lee C.,  Tsai Y.-C.,   Ting
  T.-C.,  2018, \mn@doi [Heliyon] {10.1016/j.heliyon.2018.e00785}, 4, e00785

\bibitem[\protect\citeauthoryear{Weng, Li, Ald{\'{e}}n  \& Li}{Weng
  et~al.}{2021}]{Weng2021}
Weng W.,  Li S.,  Ald{\'{e}}n M.,   Li Z.,  2021, \mn@doi [Applied
  Spectroscopy] {10.1177/0003702821990445}, 75, 1168

\bibitem[\protect\citeauthoryear{Wiese \& Fuhr}{Wiese \&
  Fuhr}{2009}]{Wiese2009}
Wiese W.~L.,  Fuhr J.~R.,  2009, \mn@doi [Journal of Physical and Chemical
  Reference Data] {10.1063/1.3077727}, 38, 565

\bibitem[\protect\citeauthoryear{Zhang, West, Irwin, Nixon  \& Yung}{Zhang
  et~al.}{2015}]{Zhang2015}
Zhang X.,  West R.~A.,  Irwin P. G.~J.,  Nixon C.~A.,   Yung Y.~L.,  2015,
  \mn@doi [Nature Communications] {10.1038/ncomms10231}, 6

\bibitem[\protect\citeauthoryear{Zhuleku, Warnecke  \& Peter}{Zhuleku
  et~al.}{2020}]{Zhuleku2020}
Zhuleku J.,  Warnecke J.,   Peter H.,  2020, \mn@doi [Astronomy \&
  Astrophysics] {10.1051/0004-6361/202038022}, 640, A119

\bibitem[\protect\citeauthoryear{van Dishoeck, Jonkheid  \& van Hemert}{van
  Dishoeck et~al.}{2006}]{Dishoeck2006}
van Dishoeck E.~F.,  Jonkheid B.,   van Hemert M.~C.,  2006, \mn@doi [Faraday
  Discussions] {10.1039/b517564j}, 133, 231

\makeatother
\end{thebibliography}

    \appendix
    
    \section{Assessment of the integration error on the radiative-convective model across a flare time-series}
    \label{app:energyError}
    
    The UV flux $F_{\text{rad}}^0(\lambda,t)$ impinging on the top of the atmosphere is continuous over time. Throughout a flare, $F_{\text{rad}}^0$ increases rapidly and then decays back to its quiescent value. In the model used in this work, $F_{\text{rad}}^0$ versus time is calculated according to the method outlined in Section \ref{ssec:flaringModels}, using the model of \cite{Loyd2018} The chemical time-step $\delta t_{\text{chem}}$ is dynamically calculated throughout the simulation, meaning that it can readily encapsulate non-equilibrium processes responding to $F_{\text{rad}}^0$. However, the iteration frequency at which the system solved for RCE is less flexible: it is prescribed at the start of the simulation by the parameter $M$, and is held constant such that the time between RCE solutions is given by $\delta t_{\text{RCE}} \approx M \delta t_{\text{chem}}$. This approach begs the question: how frequently does the model need to re-calculate the temperature profile, by solving for RCE, such that it reasonably captures the impact of flares on the temperature profile?
    \par 
    One way to assess this is by simply comparing the RCE solution time-scale to the time-scale of flares. Given a chemical iteration $\delta t_{\text{chem}}$ on the order of 15 seconds and an $M$ of 15 iterations, the model will solve for RCE every 225 seconds, which is $\sim 1/13 \text{th}$ of the duration of GF85. This would indicate that $M = 15$ is sufficient, as the temperature profile can respond to each phase of the flare.
    \par 
    We can attempt to place an upper limit on the error associated with our choice of $M=15$. To do this, we may assume that:
    \begin{itemize}
        \item the photochemical timescale is shorter than the radiative timescale, 
        \item the chemical and thermal response to a flare is well behaved,
        \item the photochemistry responds instantaneously to the $F_{\text{rad}}^0$, 
    \end{itemize}
    Note that these assumptions are applied in this section only. The first two assumptions are reasonable for the atmospheres explored in this work. The third assumption would require that $M=1$ and that $\delta t_{\text{chem}}$ be very small in order to capture the effects of changing UV flux continuously. Finite reaction rates within the chemical network apply an inertia to the system such that the composition does not respond instantaneously to changes in $F_{\text{rad}}^0$. The inertia in the system introduced by the finite reaction means that $M=1$ is not a strictly necessary requirement of this system. 
    \par 
    The error associated with a given $M$ can be estimated by comparing the total energy delivered to the atmosphere when the flux is updated every $\delta t_{\text{RCE}}$ to the total energy delivered when flux is updated continuously. This method is outlined in the following paragraphs.
    \par 
    The total radiant exposure (which has dimensions of energy per unit area) of the atmosphere over the course of the simulation is given by
    \begin{equation}
        I = \frac{1}{hc} \int F_{\text{rad}}^0(\lambda,t) dt
        \label{eq:exposure}\
    \end{equation}
    Substituting Equation \ref{eq:qfsf} into Equation \ref{eq:exposure} and assigning integration limits from $t=0$ to $t=t_f$ yields
    \begin{equation}
        I = \frac{F_{\text{rad}}^0 (\lambda, 0)}{hc}  \Big( \int_0^{t_f} Q(t) dt + t_f \Big)
        \label{eq:exposure2}
    \end{equation}
    The error in the evaluation of radiant exposure can be quantified as $\varepsilon = I_d / I_c - 1$, where $I_d$ corresponds to the implemented case where the model solves for RCE every $M$ chemical iterations, and $I_c$ corresponds to the hypothetical case where the radiative-convective responds continuously to flares. Together, we are left with
    \begin{equation}
        \varepsilon = \frac{\sum_{i=0}^{N-1} Q(i \delta t_{\text{RCE}}) \delta t_{RCE} - t_f} { \int_0^{t_f} Q(t) dt - t_f} - 1
        \label{eq:errExposure}
    \end{equation}
    where the number of steps $N$ is defined such that $t_f = N * \delta t_{\text{RCE}} $. The integral in the denominator is calculated using Simpson's rule with a high resolution time-series for $Q(t)$.
    \par
\begin{figure}
        \includegraphics[width=0.85\columnwidth,keepaspectratio]{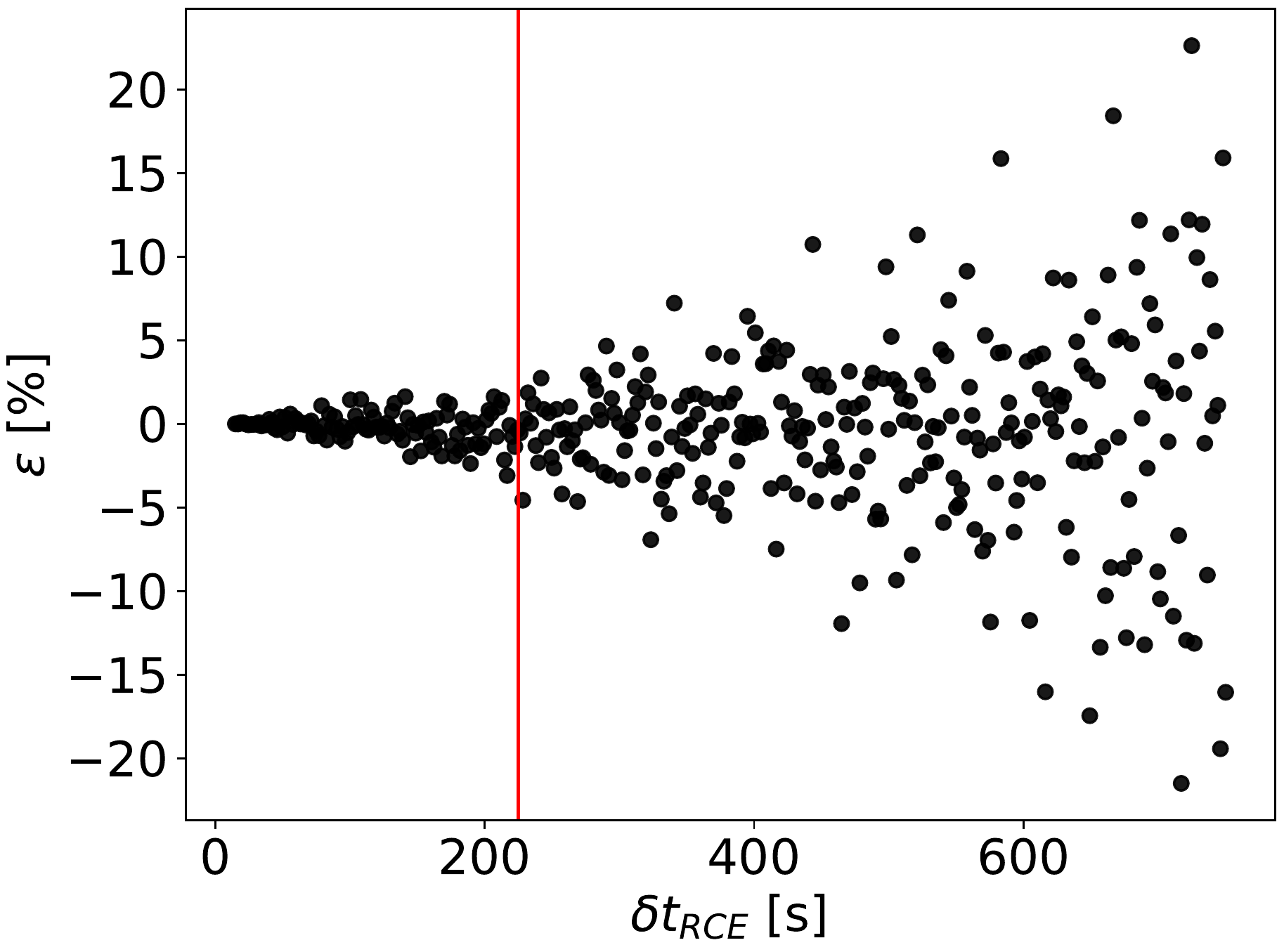}
        \captionof{figure}{Scatter plot of integration error $\varepsilon$ versus RCE convergence frequency $\delta t_{\text{RCE}}$. The recurrent and stochastic nature of the flares means that the error swings between positive and negative values chaotically, but its magnitude generally increases with increasing $M$. The red line marks a RCE convergence frequency of 225 seconds. The chemical time-step is taken to be 15 seconds. The integration time $t_f$ was taken to be \SI{2e6}{\second}.}
        \label{fig:modptProof}
    \end{figure}

 
    Figure \ref{fig:modptProof} shows that the error $\varepsilon$ increases with $\delta t_{\text{RCE}}$, as expected. For the value of $M=15$ used for the simulations in this work, $\delta t_{\text{RCE}}$ is approximately 225 seconds, which corresponds to $\varepsilon < 5\%$. Given that $\varepsilon$ quantifies an upper-limit on the error associated with iterative RCE solving, it would be reasonable to conclude that $M=15$ is satisfactory for the systems simulated in this work. Decreasing the value of $M$ would reduce this error somewhat, but would require compromising on computational performance, which limits either the model resolution or the total integration time of the simulation.
    
    \section{Elemental abundances}
    \label{app:elemList}
    
    Table \ref{tab:elems} enumerates several measures of bulk elemental composition for both planets explored in this work. \code{ATMO} calculates the bulk composition by enhancing solar elemental abundances according to [M/H]. Solar abundances are derived from \cite{Caffau2010} and \cite{Asplund2009}.
    
    \begin{table}
        \begin{center}
            \begin{tabular}{p{2cm}|p{2cm}}
                Measure & Value \\
                \hline
                [M/H]   &  1.7000 \\
                He/H    &  0.0955 \\
                Fe/H    &  0.0017 \\
                C/O     &  0.5500 \\
                X       &  0.4340 \\
                Y       &  0.1650 \\
                Z       &  0.4010
            \end{tabular}
        \end{center}
        \caption{Measures of the bulk elemental abundance of the two planets explored in this work. Values given to four decimal places.}
        \label{tab:elems}
    \end{table}
    
    \section{List of chemical species}
    \label{app:chemList}
    
    \ce{H2},
    \ce{O(^{3}P)},
    \ce{O(^{1}D)},
    CO,
    C,
    CH,
    \ce{3CH2},
    \ce{1CH2},
    \ce{H2O},
    \ce{O2},
    \ce{H2O2},
    \ce{CH4},
    \ce{H2CO},
    \ce{CH3OH},
    \ce{CO2},
    \ce{CH3OOH},
    \ce{C2H2},
    \ce{C2H4},
    \ce{C2H6},
    \ce{CH2CO},
    \ce{CH3CHO},
    \ce{C2H5OH},
    \ce{C2H5OOH},
    \ce{CH3COOOH},
    \ce{C3H8},
    \ce{C4H8Y},
    \ce{C4H10},
    \ce{C2H5CHO},
    \ce{C3H7OH},
    \ce{C2H6CO},
    \ce{C3H8CO},
    \ce{C2H3CHOZ},
    \ce{C2H4O},
    H,
    C7H8,
    OH,
    OOH,
    \ce{CH3},
    HCO,
    \ce{CH2OH},
    \ce{CH3O},
    \ce{CH3OO},
    \ce{C2H},
    \ce{C2H3},
    \ce{C2H5},
    \ce{CHCO},
    \ce{CH2CHO},
    \ce{CH3CO},
    \ce{C2H5O},
    \ce{C2H4OOH},
    \ce{C2H5OO},
    \ce{CH3COOO},
    \ce{1C3H7},
    \ce{1C4H9},
    \ce{CH3OCO},
    \ce{CO2H},
    \ce{2C2H4OH},
    \ce{1C2H4OH},
    \ce{2C3H7},
    \ce{2C4H9},
    \ce{N2},
    He,
    Ar,
    \ce{N(^{4}S)},
    \ce{N(^{2}D)},
    NH,
    \ce{NH2},
    \ce{NH3},
    NNH,
    NO,
    \ce{NO2},
    \ce{N2O},
    NCN,
    HNO,
    CN,
    HCN,
    \ce{H2CN},
    HCNN,
    HCNO,
    HOCN,
    HNCO,
    HON,
    NCO,
    \ce{HNO2},
    HONO,
    \ce{NO3},
    \ce{HONO2},
    \ce{CH3ONO},
    \ce{CH3NO2},
    \ce{CH3NO},
    \ce{C3H7O},
    \ce{C4H9O},
    \ce{C6H6},
    \ce{N2O3},
    \ce{NH2OH},
    \ce{N2O4},
    \ce{N2H2},
    \ce{N2H3},
    \ce{N2H4},
    HNNO,
    HNOH,
    \ce{HNO3},
    \ce{H2NO},
    CNN,
    \ce{H2CNO},
    \ce{C2N2},
    \ce{HCNH},
    Na,
    NaH,
    NaO,
    NaOH,
    NaCl,
    K,
    KH,
    KO,
    KOH,
    KCl,
    \ce{HO2},
    SO,
    \ce{SO2},
    Cl,
    HCl,
    ClO,
    \ce{Cl2},
    Ti,
    TiO,
    V,
    VO,
    Si,
    SiH,
    S,
    SH,
    \ce{H2S},
    Mg,
    MgH,
    MgS,
    Al,
    AlH,
    Fe,
    FeH,
    Cr,
    CrN,
    CrO,
    Ca,
    F,
    HF,
    Li,
    LiCl,
    LiH,
    LiF,
    Cs,
    CsCl,
    CsH,
    CsF,
    Rb,
    RbCl,
    RbH,
    RbF,
    P,
    PH,
    PH3,
    PO,
    P2,
    PS,
    \ce{PH2},
    \ce{P4O6}

    

    

    
    \bsp	
    \label{lastpage}
\end{document}